\documentclass[10pt,epsf]{article}
\usepackage{graphicx}
\usepackage[english]{babel}
\usepackage{simplewick}
\usepackage{braket}
\usepackage{amssymb}
\usepackage{dutchcal}
\usepackage{imakeidx}  
 \usepackage[margin=1cm]{caption} 
\usepackage[titletoc,title]{appendix}
\newcommand{\beq} {\begin{equation}}
\newcommand{\eeq} {\end{equation}}
\newcommand{\beqnarray} {\begin{eqnarray}}
\newcommand{\eeqnarray} {\end{eqnarray}}
\newcommand{\beqn} {\begin{eqnarray}}
\newcommand{\eeqn} {\end{eqnarray}}

\newcommand{\ih}{\frac{i}{\hbar}}      
\newcommand{\sumint}{\sum \!\!\!\!\!\!\!\!\int } 
\newcommand{\vol}{{\cal V}}                       
\newcommand{\kfermi}{{k_{\rm F}}}             
\newcommand{\intr}{{\hat {\cal U}}}                       
\newcommand{\tg}{{\tilde G}}                       
\newcommand{\br}{{\bf r}}

\newcommand{\bx}{{\bf x}}

\newcommand{\by}{{\bf y}}
\newcommand{\bk}{{\bf k}}

\newcommand{\hpsi}{{\hat \psi}}
\newcommand{\ha}{{\hat a}}

\newcommand{\hh}{{\hat h}}

\newcommand{\hB}{{\hat B}}
\newcommand{\hcB}{{\hat {\cal B}}}

\newcommand{\hF}{{\hat F}}

\newcommand{\hH}{{\hat H}}
\newcommand{\hI} {\hat {\mathbb {I} } }
\newcommand{\hN}{ \hat {\mathbb{N}  } }
\newcommand{\hO}{{\hat O}}
\newcommand{\hP}{{\hat P}}
\newcommand{\hQ}{{\hat Q}}

\newcommand{\hT}{\hat {\mathbb{T}}}
\newcommand{\hU}{{\hat U}}
\newcommand{\hV}{{\hat V}}
\newcommand{\hW}{{\hat W}}
\newcommand{\halpha}{\hat {\alpha}}

\newcommand{\tene} {{\cal E}} 
\newcommand{\rG}{{\rm G}}

\newcommand{\yspin}{{\cal Y}}

\newcommand{\barv}{\overline{V}}
\newcommand{\deno}{{\mathbb D}}
\newcommand{\rx}{{\rm x}}
\newcommand{\ry}{{\rm y}}

\newcommand{\half}{\frac{1}{2}}

\begin{document}
\begin{center}
{\Huge \bf Introducing the Random Phase Approximation Theory} \\
\vskip 1.cm 
{\Large Giampaolo Co'} \\
\vskip 0.5 cm
{\large Dipartimento di Matematica e Fisica, "Ennio De Giorgi" 
 \\ Universit\`a del Salento}\\ {and} \\
{\large INFN, Sezione di Lecce}  
\end{center}
%
\thispagestyle{empty}
\vskip 0.5 cm

\abstract{
 Random Phase Approximation (RPA) is the theory most commonly used to describe the 
excitations of many-body systems. In this article, the secular equations of the theory are obtained by 
using three different approaches: the equation of motion method, the Green function perturbation
theory and the time-dependent Hartree--Fock theory. Each approach emphasizes specific aspects of 
the theory overlooked by the other methods. Extensions
of the RPA secular equations to treat the continuum part of the excitation spectrum and also the
pairing between the particles composing the system are presented. 
Theoretical approaches which overcome the intrinsic approximations of RPA are outlined. 
}

\section{Introduction}
\label{sec:intro}
The aim of the Random Phase Approximation (RPA)  theory 
is the description of harmonic 
excitations of quantum many-body systems. This theory
was formulated by David Bohm and David Pines in the early 1950s  at the end of a set of articles dedicated to the 
description of collective oscillations of  electron gas \cite{boh51,pin52,boh53}. 
The approximation is well defined in the first of these articles \cite{boh51}, where it 
is used to eliminate the random movement of single electrons out of phase with
respect to the oscillations of the external probe exciting the system. The theory 
is presented only in the third of these articles \cite{boh53} and does not contain any 
random phase to be approximated. However,  the authors used the term  
{\emph{Random Phase Approximation}}  to identify the theory and it is 
 by this name that it is nowadays commonly known. 

The applications of  RPA in the 1950s and 1960s  were 
focused on the description of infinite, homogeneous and translationally invariant
systems, such as electron gas. A detailed historical overview of the works
of these early years is given in Ref.~\cite{pin16}.
Advances in the computing technologies 
allowed the application of RPA also to finite systems
such as atoms and especially nuclei. During the 1970s and 1980s, RPA
was the main theoretical tool used to investigate nuclear excitations of 
various types (see, for example, Refs. \cite{spe91,sch21} for a review). 
More recently, RPA has been applied to atomic
and molecular systems \cite{ren12}.
Nowadays, RPA calculations are rather standard and relatively
simple to carry out, so that they are, improperly, classified
as mean-field calculations. 

 RPA belongs to the category  of  effective theories. These 
theories use particle--particle interactions which do not have a strongly repulsive core 
at small inter-particle distances, a feature characterizing instead
the microscopic interactions which are tailored to describe two-particle data. 
 Hartree--Fock (HF) and Density Functional Theory (DFT)
are also effective theories. They are conceived to  describe the ground state of 
many-body systems, while RPA starts from the 
assumption that the ground state is known and considers the problem
of describing the excitation modes.

The validity of RPA is restricted to situations where the excitation energies 
are relatively small as compared to the global binding energies of the system.
This means that  RPA is not suitable for describing situations where the system 
 undergoes deep modifications of its structure, such as fission in nuclei or 
phase transitions in fluid. 

In the energy regime adequate to be described by RPA, 
it is plausible to separate the role of the external probe, 
which excites the system, from its response. 
Each probe, photon, electron,
neutrino, hadron, electric and magnetic field, sound wave, etc.,  
is described by a specific set of operators depending on the type of interaction
with the system. The response of the system depends only on the 
interactions between its fundamental components. For this reason,  
the many-body response is universal, independent of the specific
probe that induces it.  RPA evaluates this universal response.

Regarding the theoretical aspects of the theory, I like to quote what 
David Pines and Philippe Nozi\`eres write in
Chapter 5.2 of their book on quantum liquids \cite{pin66}:
\begin{quote}
\emph{{``The development, frequent independent rediscovery and gradual appreciation of the 
 Random Phase Approximation offers a useful lesson to theoretical physicist. 
 First, it illustrates the splendid variety of ways that can
 be developed for saying the same thing. Second, it suggests the usefulness of learning 
 different languages of theoretical physics and of attempting the reconciliation of seemingly
 different, but obviously related results."}} 
\end{quote}

Despite this clear statement, RPA is commonly presented in the context of specific
theoretical frameworks in order to attack some well identified  problem. 
In this article, I want to focus  attention on the theory in itself and I present
three different ways of  obtaining the secular RPA equations. 
In my opinion, this allows a richer comprehension of  the theory,
since each method emphasizes aspects overlooked by the other ones. 
The present article is not a review of the recent advances in  the use of RPA theory, 
but it aims to be a guide to understand it by pointing out its underlying assumptions,
its merits and its faults and by indicating how to improve it. 


The starting point of every many-body theory is the Independent Particle Model (IPM)
and in Section~\ref{sec:IPM}, I recall some aspects of this model which are important for 
the RPA theory. 
 RPA secular equations are derived in Sections~\ref{sec:EOM}--\ref{sec:TDHF} by using, 
 respectively, the method of the equations of motion, 
the perturbation calculation of the two-body Green function and the 
harmonic approximation of the time-dependent evolution of the HF equations.

The following two sections are dedicated to specific aspects which can be
considered by RPA. In Section~\ref{sec:CRPA}, I present how to describe
the fact that one particle can be emitted from the system,
and in Section \ref{sec:QRPA} how to treat pairing effects between the particles. 
Some issues related to the pragmatic application of RPA in actual calculations
are presented in Section~\ref{sec:applications}. 

Approaches that extend the usual RPA formulations are outlined in Section~\ref{sec:HRPA},
and the formulation of an RPA-like theory able to handle microscopic interactions is 
presented in Section~\ref{sec:CCRPA}.

Despite  my good intentions, I used numerous acronyms and to facilitate 
the reading I list them in Tab. \ref{tab:acronyms}.

\section{Independent Particle Models}
\label{sec:IPM}

The starting point of all the many-body theories is the Independent Particle Model (IPM).
In this model, each particle moves independently
of the presence of the other particles.This allows the definition of single-particle
(s.p.) energies and wave functions identified by a set of quantum numbers.
This is the basic language necessary to build any theory where the particles
interact among them. 

\subsection{Mean-Field model}
\label{sec:MF}
A very general expression of the hamiltonian describing the 
many-body system is
\beq
\hH= \sum_{i=1}^A \left( -\frac{\hbar^2} {2m_i} \nabla^2_i  + \hV_0(i) \right) 
+ \frac 1 2 \sum_{i,j=1}^A \hV(i,j)  + \cdots
\label{eq:mf.ham1}
\,\,\,,
\eeq
where $A$ is the number of  particles, each of them with mass $m_i$. 
In the expression (\ref{eq:mf.ham1}), the term containing the Laplace
operator $ \nabla^2_i$ represents the kinetic energy,
$\hV_0(i)$ is a generic potential acting on each particle and
$\hV(i,j)$ is the interaction between two particles. The
dots indicate the, eventual, presence of more complex terms of the interaction,
such as three-body forces. Henceforth, we shall not consider these latter terms. 

By adding to and subtracting from the expression ({\ref{eq:mf.ham1}) an average 
potential $\hU(i)$ acting on one particle at a time, we obtain:
\beq
\hH= 
{\underbrace 
{\sum_i^A \left( -\frac{\hbar^2} {2m_i} \nabla^2_i  + \hV_0(i) + \hU(i) \right)}_{\hH_0} }
+ 
{\underbrace {\frac {1} {2} \sum_{i,j}^A \hV(i,j) - \sum_i^A \hU(i) }_{\hH_1}}
\label{eq:mf.ham2}
\,\,\,.
\eeq
The part indicated by $\hH_0$ is a sum of terms acting on one particle,
the $i$-th particle, at a time. We can define each term of this sum as 
s.p. hamiltonian $\hh(i)$,
\beq
\hH_0 = \sum_i \hh(i) 
= \sum_i^A \left( -\frac{\hbar^2} {2m_i} \nabla^2_i  + \hV_0(i) + \hU(i) \right)
\label{eq:mf.hzero}
\,\,\,.
\eeq

The basic approximation of the Mean-Field (MF) model consists in neglecting,
in the expression (\ref{eq:mf.ham2}), 
the term $\hH_1$ called  \emph{residual interaction}. 
In this way, the many-body problem is transformed into a sum of many, independent,
one-body problems, which can be solved one at a time. 
The MF model is an IPM since the particles described by 
$\hH_0$ do not interact among them. 

The fact that the hamiltonian $\hH_0$ is a sum of independent terms
implies that its eigenstates can be built as a product of the eigenstates
of $\hh(i)$
\beq
\hh(i) | \phi_i \rangle = \epsilon_i | \phi_i \rangle
\,\,\,,
\label{eq:mf.hsp}
\eeq
therefore
\beq
 \hH_0 | \Phi \rangle =  \left( \sum_i \hh(i) \right) | \Phi \rangle = 
 \tene | \Phi \rangle
\,\,\,,
\eeq
where 
\beq
| \Phi \rangle =  | \phi_1 \rangle | \phi_2 \rangle \cdots | \phi_A \rangle
\,\,\,.
\eeq

For fermions, the antisymmetry of the global wave function under the
exchange of two particles implies that the wave function 
$| \Phi \rangle$ has to be described as the sum of antisymmetrized products 
of one-particle wave functions. This solution is known in the literature
as Slater determinant \cite{sla29}
\beq
| \Phi \rangle = \frac{1}{\sqrt{A !}} \, \det \{ \ket{\phi_i} \}
\,\,\,.
\eeq

Systems with global dimensions comparable to the average distances 
of two interacting particles are conveniently described by exploiting the spherical
symmetry. We are talking about nuclei, atoms and small molecules. 
After choosing the center of the coordinate system, 
it is convenient to use polar spherical coordinates. 

The single-particle wave function
can be expressed as a product of a radial part, depending only on  the distance  $r \equiv |\br|$ from 
the coordinate center, with a term dependent on the angular coordinates $\theta$ and $\phi$
and, eventually, the spin of the particle. The angular part 
has a well known analytic expression. For example,
in cases of an MF potential containing a spin-orbit term the s.p. wave functions  
are conveniently expressed  as:
\beq
\phi_{n l j m}(\br) = R_{n l j}(r) 
\sum_{\mu \sigma} \langle l\,\mu\,\half \, \sigma | j\, m \rangle 
Y_{l \mu}(\theta,\phi) \chi_\sigma =
R_{n l j}(r) \yspin_{ljm} (\theta,\phi)
\label{eq:mf.spwf2}
\,\,\,,
\eeq
where the spherical harmonics $Y_{l \mu}$ and the Pauli spinors $ \chi_\sigma$
are connected by the Clebsch--Gordan coefficients and form the so-called
  spin spherical harmonics \cite{edm57}. 

Systems with dimensions much larger than average distances between
two interacting particles are conveniently described by exploiting the translational 
invariance. In condensed matter conglomerates, the translational symmetry dominates.
A basic structure of the system is periodically repeated in three cartesian directions
and it is not possible to find a central point.

The basic MF model for this type of system considers the 
potential $\hU$ to be  constant. This fermionic system is commonly called 
\emph{Fermi gas}. It is a toy model, homogeneous, with infinite volume, 
composed by an infinite number of fermions which do not interact with 
each other. Since the energy scale is arbitrary, it is possible to select
$\hU=0$ without loosing generality. In this case, the
one-body Schr\"odinger equation is
\beq
-\frac{\hbar^2} {2m_j} \nabla_j^2 \phi_j(\br) = \epsilon_j \phi_j(\br) 
\label{eq:mf.pwwf}
\,\,\,.
\eeq

By defining 
\beq
\epsilon_j = \frac{\hbar^2 \bk^2_j}{ 2 m_j}
\label{eq:mf.enefg}
,
\eeq
the eigenfunction of Equation~(\ref{eq:mf.pwwf}) can be written as
\beq
\phi_j(\br) = \frac{1}{\sqrt{\vol}}\,e^{i \bk_j \cdot \br} \chi_\sigma \chi_\tau
\label{eq:mf.pw}
\,\,\,,
\eeq
where ${\cal V}$ is the volume of the system and $\chi$ are the Pauli spinors related to the
spin of the fermion and, eventually, to its isospin. The third components of 
spin and isospin are indicated as $\sigma$ and $\tau$, respectively.
The physical quantities of interest are those independent of ${\cal V}$ 
whose value, at the end of the calculations, is taken to be infinite.

The solution of the Fermi gas model provides a set of continuum single 
particle energies. Each energy is characterized by $k\equiv|\bk|$, as indicated 
by Equation~(\ref{eq:mf.enefg}). In the ground state of the system, all the 
s.p. states with $k$ smaller than a value $\kfermi$, called Fermi
momentum, are fully occupied and those with  $k > \kfermi$  are empty.
Each state has a degeneracy of 2 in cases of electron gas and of 4 for nuclear
matter where each nucleon is characterized also by the isospin third component.

\subsection{Hartree--Fock Theory}
\label{sec:HF}

The theoretical foundation of the MF model is provided by the Hartree--Fock (HF) theory, 
which is based on the application of the variational principle,
one of the most used methods to solve
the Schr\"odinger equation in an approximated manner. The basic idea is that
the wave function which minimizes the energy, considered as functional of the
many-body wave function, is the correct eigenfunction of the hamiltonian. 
This statement is correct when the search for the minimum is carried out by considering the 
full Hilbert space. In reality, the problem is simplified by assuming a specific
expression of the wave function and the search for the minimum is carried out in 
the subspace spanned by all the wave functions which have the chosen 
expression. The energy value obtained in this manner is an upper bound of 
the correct energy eigenvalue of the hamiltonian. 
The formal properties of the variational principle are discussed in   
quantum mechanics textbooks. 

For a fermion system,
the HF equations are obtained by considering  trial many-body wave functions  
which are expressed as a single Slater determinant. This implies the existence 
of an orthonormal basis of s.p. wave functions.  The requirement 
that the s.p. wave functions are orthonormalized is a condition 
inserted in the variational equations in terms of Lagrange multipliers. 

We continue this discussion by using  Occupation Number Representation (ONR)
formalism, which describes the operators acting on the Hilbert space in terms of creation $\ha^+_\nu$
and destruction $\ha_\nu$ operators. Concise
presentations of this formalism are given in various
textbooks, for example, in  Appendix 2A of \cite{boh69}, in  Appendix C of \cite{row70}, 
in  Appendix C of \cite{rin80}, in Chapter 4 of \cite{suh07} and in Chapter 1 of \cite{bru16}.

In Appendix \ref{sec:app:HF} we show that the hamiltonian of the many-body 
system, if only two-body interactions are considered, can be written as
\beq
\hH = \sum_\nu \epsilon_\nu \ha^+_\nu \ha_{\nu}
- \half \sum_{ij} \barv_{i j i j}
+
\frac 1 4 \sum_{\mu \mu' \nu \nu'} \barv_{\nu \mu \nu' \mu'}
\hN[\ha^+_\nu \ha^+_\mu \, \ha_{\mu'} \, \ha_{\nu'}] 
= \hH_0 +\hV_{\rm res}
\label{eq:hf.ham4}
\,\,\,,
\eeq
where $\hH_0$ is the sum of the first two terms, while $\hV_{\rm res}$ is the last term.
We use the common convention of indicating with the latin letters $h,i,j,k,l$  s.p. states 
below the Fermi surface (hole states) and with the $m,n,p,q,r$ letters the s.p. states
above the Fermi energies (particle states). Greek letters indicate indexes which have to be defined; 
therefore, in the above equation, their sums run on all the set of s.p. states. 
In Equation (\ref{eq:hf.ham4}), $\epsilon_\nu$ is the energy of the s.p. state characterized
by the $\nu$ quantum numbers and $\barv$ is the antisymmetrized  
matrix element of the interaction defined as 
\beq
\barv_{\nu \mu \nu' \mu'}  \equiv \langle \nu \mu | \hV | \nu' \mu' \rangle
-  \langle \nu \mu | \hV | \mu' \nu' \rangle
\label{eq:v.avu}
\,\,\,.
\eeq

With the symbol $\hN$, we indicate the normal order operator which, by definition, 
arranges the set of creation and destruction operators in the brackets such that  
their expectation value on the ground state is zero. 
By considering this property of  $\hN$,
the expectation value of the hamiltonian between two Slater determinants 
assumes the expression
\beqn
\nonumber
&~&
\langle \Phi_0 | \hH | \Phi_0 \rangle
= \langle \Phi_0 | \hH_0 | \Phi_0 \rangle  
+ \langle \Phi_0 | \hV_{\rm res} | \Phi_0 \rangle \\
\nonumber
&=&
\sum_\nu \epsilon_\nu \langle \Phi_0 | \ha^+_\nu \ha_{\nu}  | \Phi_0 \rangle 
- \half \sum_{ij} \barv_{i j i j} \langle \Phi_0 | \Phi_0 \rangle \\
\nonumber
&+& 
\frac 1 4 \sum_{\mu \mu' \nu \nu'} \barv_{\nu \mu \nu' \mu'}
\langle \Phi_0 |  \hN[\ha^+_\nu \ha^+_\mu \, \ha_{\mu'} \, \ha_{\nu'}] | \Phi_0 \rangle \\
&=& \sum_i \epsilon_i - \half \sum_{ij} \barv_{i j i j}  \equiv \tene_0 [\Phi_0]
\label{eq:v.ezero}
\,\,\,,
\eeqn
which clearly indicates that the contribution of the residual interaction is zero 
and the only part of the interaction which is considered is the one-body term $\hH_0$.
This is a consequence of considering a single Slater determinant to describe the
system ground state.

In Equation~(\ref{eq:v.ezero}), we expressed the energy $\tene_0$ as a functional of the Slater determinant 
$\Phi_0$. The search for the minimum of the energy functional is carried out 
in the Hilbert subspace spanned by Slater determinants. The quantities to be varied are the s.p. 
wave functions forming these determinants. These s.p. wave functions must be orthonormalized and
this is an additional condition which has to be imposed in doing the variations.
Therefore, the problem to be solved is the search for a constrained minimum 
and it is tackled by using the Lagrange multipliers  technique. 

The calculation is well known in the literature (see, for example, chapter XVIII-9 of \cite{mes61} or
Chapter 8.4 of \cite{bra03}). The final result is a set of non-linear integro-differential equations 
providing the s.p. wave functions $\phi_k$ and the values of the Lagrange multipliers $\epsilon_k$.
In coordinate space, these equations can be expressed as
\beq
\hh \phi_k(\br) = 
-\frac{\hbar^2 \nabla^2}{2m} \phi_k(\br) 
+ {\underbrace {\hU(\br) \phi_k(\br)}_{\rm Hartree}}
- {\underbrace{\int d^3 r' \hW(\br,\br') \phi_k(\br')}_{\rm Fock--Dirac}} 
= \epsilon_k \phi_k(\br)
\label{eq:v.hf4}
\,\,\,.
\eeq
where the Hartree average potential is defined as
\beq
\hU(\br) \equiv  \sum_{j} \int d^3 r' \phi^*_j(\br') \hV(\br,\br') \phi_j(\br')
\label{eq:v.hartree}
\,\,\,,
\eeq
and the non-local Fock--Dirac term is
\beq
\hW(\br,\br') \equiv  \sum_{j} \phi^*_j(\br') \hV(\br,\br') \phi_j(\br)
\label{eq:v.fock}
\,\,\,.
\eeq

At this stage, the $\epsilon_k$ are the values of the Lagrange multipliers. A theorem, called 
 Koopmans \cite{koo34}, shows that these quantities are the differences 
between the energies of systems with $A+1$ and $A$ particles; therefore, they are identified
as s.p. energies. 

By neglecting the Fock--Dirac term, we obtain a differential equation of 
MF type. The Fock--Dirac term, also called the exchange term,
changes the bare mean-field equation by inserting the effect of the 
Pauli exclusion principle. 

The differential Equation (\ref{eq:v.hf4}) is solved numerically by using an
iterative procedure. One starts with a set of trial wave functions $\phi_k^{(1)}$ 
built with MF methods. With these trial wave functions, the  Hartree 
(\ref{eq:v.hartree}) and  Fock--Dirac (\ref{eq:v.fock}) terms are calculated 
and included in Equation~(\ref{eq:v.hf4}) which is solved with standard numerical methods. 
In this way, a new set of s.p. wave functions  $\phi_k^{(2)}$ is obtained and it is
used to calculate new $\hU$ and $\hW$ potentials. The process continues up to convergence. 
 
As already pointed out in the introduction, 
the interactions used in the HF calculations are not the microscopic interactions 
built to reproduce the experimental data of the two-particle  systems. 
These microscopic interactions
contain a strongly repulsive core and, if inserted in the integrals of Equations (\ref{eq:v.hf4}) 
and (\ref{eq:v.hartree}), they would produce terms much larger than $\epsilon_k$. 
This would 
attempt  calculating a relatively small number by summing and subtracting 
relatively  large numbers. 
HF calculations require interactions which have already tamed the strongly repulsive
core (an early discussion of this problem can be found in Chapter 13 
of \cite{row70}). 

 \subsection{Density Functional Theory}
\label{sec:var.DFT}
\index{Density Functional Theory (DFT)}
The HF theory is widely utilized in nuclear and atomic physics, but there are two 
problems concerning its use. A first one is related to the formal development of the 
theory and it shows up mainly in the nuclear physics framework where the 
commonly used effective interactions have a phenomenological 
input containing also terms explicitly dependent on the density of the system.
Without these terms, the HF calculations do not reproduce binding energies 
and densities of nuclei. The addition of these terms allows the construction of interactions
able to produce high quality results all through the nuclide table.
The physics simulated by these density dependent terms is still a matter of study. 
Formally, the variational principle used to derive the HF equation 
is not valid when the interaction depends explicitly on the density.

The second problem is of pragmatic type and it is related to the difficulty 
in evaluating the Fock--Dirac term of Equation~(\ref{eq:v.hf4}) for complicated systems
which do not show a well defined symmetry, for example, complex molecules.

The Density Functional Theory (DFT) solves both problems. 
This theory  is based on a theorem of Hohenberg and Kohn \cite{hoe64},
formulated in the 1960s. 

 Let us express the hamiltonian of a system of $A$ fermions of mass $m$ as:
\beq
\hH={\hat T}+\hU_{ext}+\hV
\label{dft:eq:h1}
\,\,,
\eeq
with
\beq
{\hat T}=\sum_{i=1}^A - \hbar^2\frac{\bigtriangledown_i^2}{2m}\quad ,\quad 
\hU_{ext}=\sum_{i=1}^A {\hat u}_{ext}(i)\quad ,\quad 
\hV=\frac{1}{2}\sum_{i,j=1}^A{\hat v}(i,j)
\label{dft:eq:h2}
\,\,,
\eeq
The kinetic energy term, $\hat T$ and the external potential $\hU_{ext}$, 
are one-body operators, while the interaction term $\hV$ is a two-body potential. 
The kinetic energy term plus $\hV$
are characteristic of the many-fermion system, while $\hU_{ext}$ depends on external 
situations and therefore, in principle, can be modified. 

The Hohenberg--Kohn theorem states that there is a bijective correspondence between the 
external potential  $\hU_{ext}$,  the ground state $| \Psi_0 \rangle$ and the 
number density 
\beq
\rho_0(\br)=
\langle \Psi_0 | \sum_{i=1}^A\delta(\br-\br_i) | \Psi_0 \rangle 
,
\label{dft:eq:rho}
\eeq 
of the system. 

The theorem has the following implications.
\begin{itemize}
\item[(a)]
Because of the bijective mapping 
\beq
 \hU_{ext} \Longleftrightarrow | \Psi_0 \rangle \Longleftrightarrow \rho_0
\,\,.
\eeq
we can consider the states $| \Psi_0 \rangle$ as functionals of the density $\rho_0$.

\item[(b)]
Because of (a), every observable is also a functional of $\rho_0$.
Specifically, this is true for the energy of the system
\beq
E[\rho_0] = \langle \Psi[\rho_0] | \hH | \Psi[\rho_0]\rangle = 
F[\rho_0] + \int d^3 r  \,  \hU_{ext}(\br) \, \rho_0(\br) 
,
\eeq
where the universal part, the part independent of the external potential, is defined as
\beq
F[\rho_0] \equiv \langle \Psi [\rho_0] | \left( {\hat T} + \hV \right) | \Psi [\rho_0] \rangle 
.
\eeq

\item[(c)]
The variational principle implies that  for each  $\rho \ne \rho_0$ the following relation holds:
\beq
E_0 \equiv E[\rho_0] < E[\rho]
.
\eeq
\end{itemize}

The focus of the theory has moved from the many-body wave function $\ket{\Psi_0}$ to the much 
simpler one-body density $\rho_0$. The idea of Kohn and Sham \cite{koh65}
is to reproduce the ground state density $\rho_0$ 
of a system of interacting fermions by using a fictitious system of non-interacting fermions.
This is done by changing the external part of the hamiltonian. In this view, the density 
(\ref{dft:eq:rho}) is expressed as a sum of orthonormalized s.p. wave functions
\beq
\rho_0(\br) = \sum_{i < \epsilon_{\rm F}} |\phi^{\rm KS}_i(\br) |^2
,
\label{dft:eq:densks}
\eeq
where $\epsilon_{\rm F}$ is the Fermi energy and KS indicates Kohn and Sham. 
The density  (\ref{dft:eq:densks}) is generated by a one-body hamiltonian whose
eigenstate is a Slater determinant $|\Phi^{\rm KS} \rangle$. 
The energy functional built in the  Kohn and Sham approach is usually expressed as:
\beq
E[\rho_0] = T^{\rm KS} [\rho_0] + E_{\rm H}^{\rm KS} [\rho_0] 
+ E_{\rm ext}^{\rm KS} [\rho_0] + E_{\rm xc}^{\rm KS} [\rho_0]
,
\label{dft:eq:eKS}
\eeq
where there is a kinetic energy term,
\beq
T^{\rm KS} [\rho_0] = \langle \Phi^{\rm KS}| {\hat T} |\Phi^{\rm KS} \rangle = 
\int\,d^3r \, \sum_i \phi^{* {\rm KS}}_i(\br) \left(- \frac{\hbar^2 \nabla^2}{2 m} \right) 
\phi^{\rm KS}_i(\br)
,
\eeq
a Hartree term,
\beq
E_{\rm H}^{\rm KS} [\rho_0] = \int\,d^3r_i \int\,d^3r_j  
\rho_0(\br_i) {\hat v} (\br_i,\br_j) \rho_0(\br_j) 
,
\eeq
and an external mean-field term
\beq
E_{\rm ext}^{\rm KS} [\rho_0] =  \int\,d^3r  \rho_0(\br_i) 
\hU^{\rm KS}_{\rm ext}(\br_i)
.
\eeq
The additional term, $E_{\rm xc}^{\rm KS}$, is said to be of   exchange and correlation.

The variational principle is applied to the energy functional (\ref{dft:eq:eKS})
and the final result is, again, a set of non-linear integro-differential 
equations, which allows the evaluation of  the Kohn and Sham s.p. wave functions
\beq
\left\{
- \frac{ \hbar^2 \nabla^2}{2 m} + \int d^3r_j  {\hat v}(\br,\br_j) \rho_0(\br_j)  + 
\hU^{\rm KS}_{\rm ext}(\br) + \hU^{\rm KS}_{\rm xc}(\br)
\right\} \phi^{\rm KS}_i(\br) = \epsilon_i \phi^{\rm KS}_i(\br)
\label{dft:eq:KS}
\,\,.
\eeq
This set of equations is solved numerically with iterative
techniques analogous to those used in the HF case.
In Equation (\ref{dft:eq:KS}), only local terms appear, contrary to the HF equations
which contain the non-local Fock--Dirac term. This makes the numerical solution of 
the KS equations much simpler than that of the HF equations and allows 
an application of the theory to systems difficult to treat with HF. 

While the only input of the HF theory is the effective interaction $\hV$, 
in the DFT one has, in addition, to define the exchange 
and correlation term $\hU_{\rm xc}^{\rm KS}$. The strategy for choosing 
this term is an open problem of investigation in the field.

Formally speaking, the s.p. wave functions $\phi^{\rm KS}$ and the Lagrange multipliers
$\epsilon_k$ of Equation~(\ref{dft:eq:KS}) do not have a well defined
physical interpretation. From the pragmatical
point of view, the values of these latter quantities are very close to the s.p. energies of the HF
theory defined by  Koopmans' theorem.


\subsection{Excited States in the Independent Particle Model}
\label{sec:IPM.excitation}

The IPM is quite successful in describing the ground state properties of the fermion systems. This 
is also due to the fact that  effective interactions are tailored to make this work. 
A good example of this is provided by the AMEDEE compilation of Hartree--Fock--Bogolioubov 
results concerning the ground states of nuclear isotope chains from $Z=6$ up to
 $Z=130$ \cite{cea}. Experimental values of binding energies and charge density radii
 are described with excellent accuracy by using a unique and universal effective 
 nucleon--nucleon interaction. 
 The situation changes immediately as soon as one tries to apply the same theoretical
 scheme to describe excited states.

The basic ansatz of the IPM is that a many fermion system can be described by a single 
Slater determinant $\ket{ \Phi }$. The Slater determinant describing the ground state,
$\ket{ \Phi_0}$, has all the s.p. states below the Fermi energy (hole states) fully occupied,
while those above it (particle states) are completely empty. 
In this picture, excited states are obtained
by promoting particles from states below the Fermi surface to states above it. 
By using the ONR,   this procedure can be formally described as
\beq
\ket{\Phi_N} = \ha^+_{p_1} \cdots \ha^+_{p_N}  \ha_{h_1} \cdots \ha_{h_N}  \ket{\Phi_0}
,
\eeq
where the $p$'s indicate particle states and the $h$'s the hole states.
The number $N$ of creation or destruction operators is obviously smaller than $A$, 
the number of fermions.
The state $\ket{\Phi_N}$ is a Slater determinant where $N$ hole states have been changed
with $N$ particle states and it is the eigenstate of the IPM hamiltonian
\beq
\hH_0 \ket{\Phi_N} = \tene_N \ket{ \Phi_N }
.
\eeq

The excitation energy of this system is given by the difference between the s.p. energies
of the particle states and that of the hole states
\beq
\omega^{\rm IPM}_N \equiv \tene_N - \tene_0 =
   \epsilon_{p_1} + \epsilon_{p_2} + \cdots + \epsilon_{p_N}  
 - ( \epsilon_{h_1}  + \epsilon_{h_2} + \cdots +  \epsilon_{h_N} )
 .
\eeq

A good example of the failure of this approach in describing the excitations
of a many-body systems is provided by the case of the $^{208}$Pb nucleus. 
We show in Figure \ref{fig:pb208a} the scheme of the s.p. levels 
 around the Fermi energy of this nucleus. 
 The energies of these levels have been obtained by exploiting
   Koopmans' theorem, i.e., by subtracting the experimental binding energies of the nuclei
 with one nucleon more or less, with respect to $^{208}$Pb. These nuclei are $^{207}$Tl, $^{209}$Bi
 and the two lead isotopes $^{207}$Pb and $^{209}$Pb. From the experimental values of the 
 angular momenta of these odd--even nuclei, we identified the quantum numbers of the s.p. 
 levels. 
 \vskip -2.0 cm 
\begin{figure}[ht] 
\begin{center}
\captionsetup{margin=2cm}
\includegraphics [scale=0.6,angle=90]{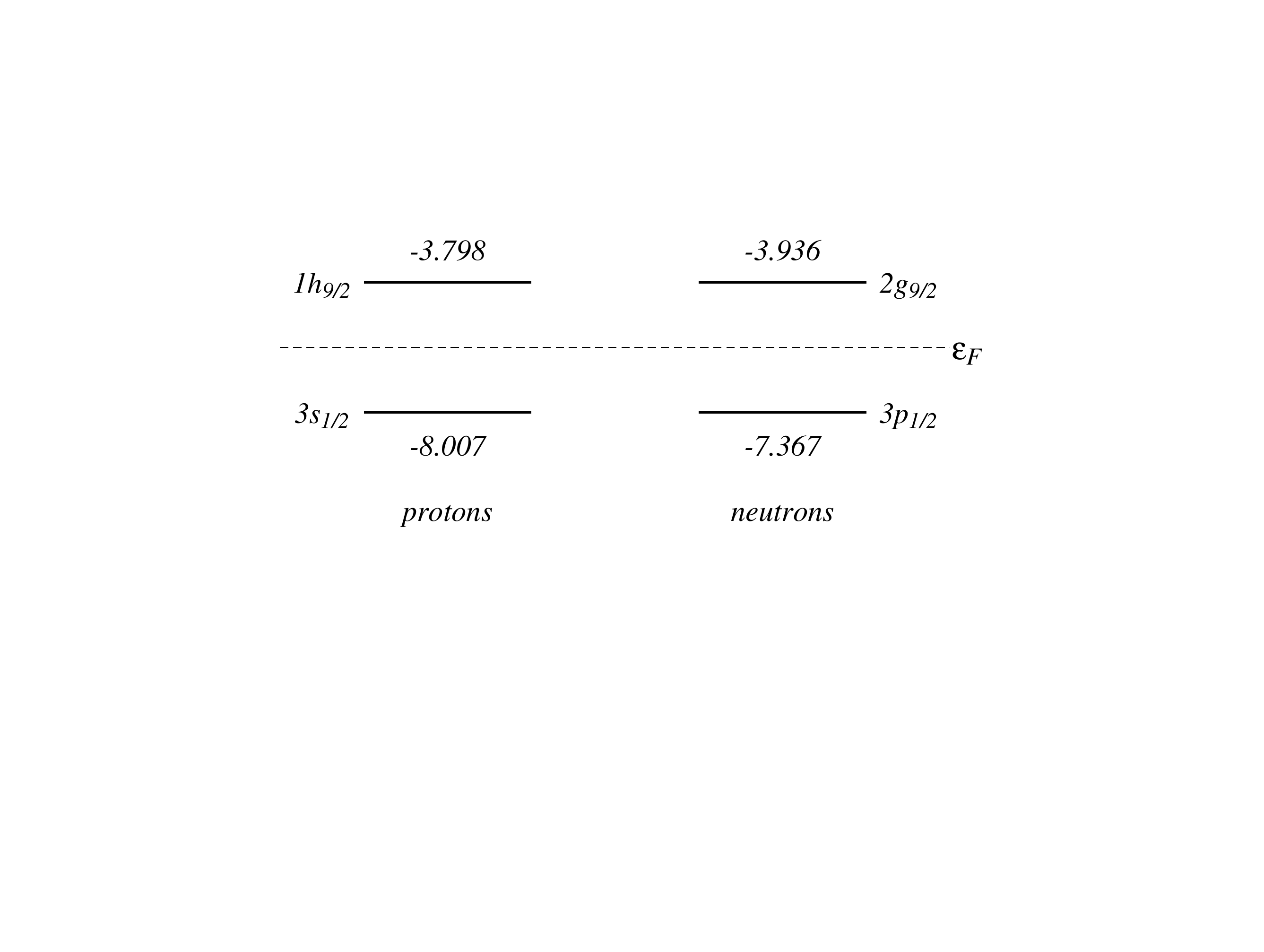} 
\vskip -4.0 cm 
\caption{\small 
Sketch of the s.p. levels around the Fermi surface in $^{208}$Pb. The numbers indicate, in MeV units,
the s.p. energies obtained as differences between the experimental binding energies of the 
nuclei with one nucleon more or less than $^{208}$Pb. 
}  
\label{fig:pb208a} 
\end{center}
\end{figure}

 The first excited state in the IPM framework is that obtained by promoting the nucleon lying
 on the s.p. state just below the Fermi surface to the state just above it. In the present case, this 
 one-particle one-hole ($1p-1h$) 
 excitation for the protons will be produced by the transition from the $3s_{1/2}$ state
 to the $1h_{9/2}$ state. The excitation energy of this transition is  4.209 MeV, the parity 
 is negative and the total angular momentum is 4 or 5. The
 analogous transition for the neutrons also implies  a negative  parity value and
  excitation energies of 3.431 MeV and also in this case the angular momentum values
 of the excited state can be 4 or 5. Measurements indicate that 
 the first excited state of the $^{208}$Pb has an excitation energy of
 2.614 MeV with angular momentum 3 and  negative parity. Evidently, the IPM is unable
 to predict the presence of this state. The part of the hamiltonian disregarded by the IPM, 
 the residual interaction, plays an important role. 
  RPA considers the presence of the residual interaction in the description 
 of the excitations of a many-body system. 

\section{RPA with the Equation of Motion Method}
\label{sec:EOM}

The first approach I present in order to obtain the RPA secular equations 
is the Equation of Motion (EOM) method inspired by the Heisenberg  
picture of quantum mechanics. 

Let us define an operator, $\hQ^+_\nu$, whose action on the ground state 
of the system defines its excited states
\beq
\hQ^+_\nu \ket{\Psi_0} = \ket{\Psi_\nu}  
,
\label{eq:rpa:defq}
\eeq
which satisfy  the eigenvalue equation
\beq
\hH \ket{\Psi_\nu} = E_\nu \ket{\Psi_\nu}
.
\eeq
In the above equations, the index $\nu$ indicates all the quantum numbers
characterizing the excited state. For example, in a finite fermion system, 
they are 
the excitation energy, the total angular momentum and the parity. 
The choice of  $\hQ^+_\nu$ defines completely the problem to be solved,
and also the ground state of the system through the equation
\beq
\hQ_\nu \ket{\Psi_0} = 0
.
\label{eq:rpa:defqa}
\eeq

It is worth  remarking that the states $\ket{\Psi_\nu}$ are not eigenstates of the full hamiltonian $\hH$
but, depending on the choice of $\hQ^+_\nu$, they are eigenstates of only a part of the hamiltonian. 
For example, if $\hQ^+_\nu = \ha^+_p \ha_h$, ground and excited states are Slater determinants
of the IPM  described
in Section \ref{sec:IPM.excitation}. As  has been already pointed out by discussing Equation~(\ref{eq:v.ezero}),
this choice does not consider the contribution of the residual interaction.

Let us calculate the commutator of the 
$\hQ^+_\nu$ operator with the hamiltonian
\beqn
\nonumber
\left[\hH,\hQ^+_\nu \right] \ket{\Psi_0} &=& \left( \hH \hQ^+_\nu - \hQ^+_\nu \hH \right) \ket{\Psi_0} 
= \hH \ket{\Psi_\nu} - \hQ^+_\nu E_0 \ket{\Psi_0} \\
&=&   E_\nu \ket{\Psi_\nu} - \hQ^+_\nu E_0 \ket{\Psi_0} 
   = \left( E_\nu - E_0 \right) \hQ^+_\nu \ket{\Psi_0}
   ,
\label{eq:rpa:moto1}
\eeqn
and for the operator $\hQ_\nu$,  we obtain
\beq
\left[\hH,\hQ_\nu \right] \ket{\Psi_0} = \left( \hH \hQ_\nu - \hQ_\nu \hH \right) \ket{\Psi_0} 
= \hH \hQ_\nu \ket{\Psi_0} -  E_0 \hQ_\nu \ket{\Psi_0} = 0
,
\label{eq:rpa:moto2}
\eeq
because of Equation~(\ref{eq:rpa:defqa}).

We multiply Equation (\ref{eq:rpa:moto1}) by a generic operator $\hat {\cal O}$ and by
$\bra{\Psi_0}$ and we subtract the complex conjugate. 
For\,Equations\,(\ref{eq:rpa:moto1})\,and\,(\ref{eq:rpa:moto2})}, we obtain
\beq
\braket{\Psi_0| \left[ {\hat {\cal O}} , [\hH,\hQ^+_\nu] \right] |\Psi_0} =
\left( E_\nu - E_0 \right) \braket{\Psi_0 | {\hat {\cal O}} \hQ^+_\nu | \Psi_0} 
= \left( E_\nu - E_0 \right) \braket{\Psi_0 | \left[ {\hat {\cal O}}, \hQ^+_\nu \right] | \Psi_0} 
.
\label{eq:rpa:moto}
\eeq
since $\bra{\Psi_0} \hQ^+_\nu = 0$.

This result is independent of the expression of the operator ${\hat {\cal O}}$.
In the construction of the various theories describing the system excited states, 
the ${\hat {\cal O}}$ operator is substituted by the $\delta \hQ_\nu$ operator representing
an infinitesimal variation of the excitation operator defined by Equation  (\ref{eq:rpa:defq}).

\subsection{Tamm--Dankoff Approximation}
\label{sec:rpa:TDA}
\index{Tamm--Dankoff approximation}

A first choice of the $\hQ^+_\nu$ consists in considering the excited state as a linear
combination of particle--hole excitations. This means that the excited state is not any
more a single Slater determinant as in the IPM, but it is described by a sum
of them. This choice of $\hQ^+_\nu$, leading to the so-called
Tamm--Dankoff approximation (TDA), is
\beq
\hQ^+_\nu \equiv \sum_{m\,i} X_{m i}^\nu \ha^+_m \ha_i 
,
\label{eq:rpa.q}
\eeq
where $X^\nu_{m i}$ is a real number and the usual convention
of indicating the hole states with the letters $h,i,j,k,l$ and the particle states with
$m, n, p, q, r$ has been adopted.

The definition (\ref{eq:rpa.q}) of the $\hQ^+_\nu$ operator implies that 
the ground state $\ket{\Psi_0}$ satisfying Equations (\ref{eq:rpa:moto1}) 
and (\ref{eq:rpa:moto2}) is the IPM ground state
$\ket{\Phi_0}$. In effect
\beq
\hQ_\nu \ket{\Phi_0} = \sum_{m\,i} X_{m i}^\nu \ha^+_i \ha_m  \ket{\Phi_0} = 0 
,
\label{eq:rpa:tdazero}
\eeq
since it is not possible to remove particles above the Fermi surface or to put particles
below it.

An infinitesimal variation of the $\hQ_\nu$ operator can be expressed as
\beq
\delta \hQ_\nu = \sum_{m\,i}  \ha^+_i \ha_m \delta {X_{m i}^{* \nu} } 
,
\label{eq:rpa:tdaq}
\eeq
since only the amplitudes $X_{m i}^{* \nu}$ can change. 
By substituting ${\hat {\cal O}}$ with $\delta \hQ_\nu$ in Equation (\ref{eq:rpa:moto}), we obtain
\beqn
\nonumber &~&
\braket{\Phi_0 | \Big[
\sum_{m\,i}  \ha^+_i \ha_m \delta {X_{m i}^{* \nu} } ,
\big[ \hH, \sum_{n\,j} X_{n j}^\nu \ha^+_{n} \ha_{j}  \big]
 \Big] | \Phi_0}
\\ &=& 
 \left(E_\nu - E_0 \right) 
 \braket{\Phi_0 | \Big[
\sum_{m\,i}  \ha^+_i \ha_m \delta {X_{m i}^{* \nu} } ,
 \sum_{n\,j} X_{n j}^\nu \ha^+_{n} \ha_{j}   
 \Big] | \Phi_0}
 .
\eeqn
Every variation $\delta X^{*\nu}_{ph}$ is independent of the other ones. For this reason, the
above equation is a sum of terms independent of each other. The equation is satisfied 
if all the terms related to the same variation of  $X^{\nu}_{ph}$ satisfy the relation. 
We can formally express this concept by considering a single term of the sum and by dividing it 
by $\delta X^{*\nu}_{ph}$ which is, by our choice, different from zero
\beq
\braket{\Phi_0 | \Big[ \ha^+_i \ha_m ,
[ \hH, \sum_{n\,j} X_{n j}^\nu \ha^+_{n} \ha_{j}  ] \Big] | \Phi_0}
=
 \left(E_\nu - E_0 \right) 
\sum_{n\,j} X^\nu_{n j}
 \braket{\Phi_0 | \left[ a^+_i a_m , a^+_n a_j   \right] | \Phi_0}
.
\label{eq:rpa:tda1}
\eeq

Let us calculate the right hand side of Equation~(\ref{eq:rpa:tda1}):
\beq
 \braket{\Phi_0 | \left[ \ha^+_i \ha_m , \ha^+_n \ha_j   \right] | \Phi_0} = 
  \braket{\Phi_0 |  \ha^+_i \ha_m  \ha^+_n \ha_j  | \Phi_0}
  -  \braket{\Phi_0 | \ha^+_n \ha_j , \ha^+_i \ha_m | \Phi_0}
  .
\label{eq:rpa:tda1a}
\eeq
We apply  Wick's theorem (see, for example, Ref.~\cite{fet71}) to the first term
\beq
\contraction {\Phi_0 | }{  \ha^+_i} {\ha_m  \ha^+_n }{\ha_j | \Phi_0  }
\contraction {\Phi_0 | \ha^+_i }{\ha_m} {} {\ha^+_n } 
\braket{\Phi_0 |  \ha^+_i \ha_m  \ha^+_n \ha_j  | \Phi_0} = \delta_{m n} \delta_{i j}
\label{eq:rpa.wick}
,
\eeq
where the lines indicate the operators to be contracted. 

The second term of Equation~(\ref{eq:rpa:tda1a}) is zero since
$
\ha_m \ket{\Phi_0} = 0 \;.
$
By using this result in Equation~(\ref{eq:rpa:tda1}), we obtain
\beq
\braket{\Phi_0 | \Big[ \ha^+_i \ha_m ,
[ \hH, \sum_{n\,j} X_{n j}^\nu \ha^+_{n} \ha_{j}  ]
 \Big] | \Phi_0}
=
 \left(E_\nu - E_0 \right) X^\nu_{mi} 
.
\label{eq:rpa:tda2}
\eeq

The evaluation of the double commutator of the left hand side of Equation~(\ref{eq:rpa:tda2}) 
is explicitly presented in Appendix \ref{sec:app.doublec}. We insert the results of Equations
(\ref{eq:app.pdfilon}) and (\ref{eq:app:em1}) into Equation~(\ref{eq:rpa:tda2}) and we consider 
the symmetry properties of the antisymmetrized matrix element of the interaction  
$\barv_{\alpha,\beta,\alpha',\beta'}$, Equation~(\ref{eq:v.avu}). 
Finally, we obtain the TDA equations:
\beq
\sum_{nj} X^\nu_{nj} 
\left[(\epsilon_n - \epsilon_j) \delta_{mn} \delta_{ij} + \barv_{mjin} \right]
= (E_\nu - E_0) X^\nu_{mi}
.
\label{eq:rpa:tda}
\eeq

The expression (\ref{eq:rpa:tda}) represents a homogenous system of linear 
equations whose unknowns are the $X^\nu_{mi}$. The number of unknowns, 
and therefore of  solutions, is given by the number of  particle--hole pairs
which truncates the sum.

The normalization condition of the excited state induces a relation between the 
$X^\nu_{mi}$ amplitudes:
\beqn
\nonumber
1 &=& \braket{\Psi_\nu | \Psi_\nu} =
\braket{\Phi_0 | \hQ_\nu \hQ^+_\nu | \Phi_0}
=\braket{\Phi_0 |  \sum_{p\,h}  \ha^+_h \ha_p X_{ph}^{* \nu}  
\sum_{p'\,h'} X_{p'h'}^\nu \ha^+_{p'} \ha_{h'} | \Phi_0}
\\
&=& \sum_{p\,h} \sum_{p'\,h'} X_{ph}^{* \nu}  X_{p'h'}^\nu 
\contraction {\Phi_0 |}{\ha^+_h} {\ha_p  \ha^+_{p'} }{\ha_{h'} }
\contraction {\Phi_0 | \ha^+_h }{\ha_p} {} {\ha^+_{p'} } 
\braket{\Phi_0 |   \ha^+_h \ha_p \ha^+_{p'} \ha_{h'} | \Phi_0}
= \sum_{p\,h}  |  X_{ph}^\nu |^2
,
\label{eq:rpa:tdanormx}
\eeqn
which defines without ambiguity the values of the $X^\nu_{ph}$ and suggests
their probabilistic interpretation. 

The TDA theory describes not only the energy spectrum of the system, but also for each 
excited state it provides the many-body wave function written in terms of single-particle
states. This allows the calculation of the transition probability from the ground state
to an excited state.

Let us assume that the action of the external field which excites the system is described
by a one-body operator
\beq
{\hat F} = \sum_{\mu \mu'} \braket{\mu| {\hat f} | \mu'} \ha^+_\mu \ha_{\mu'} 
\equiv \sum_{\mu \mu'}  f_{\mu  \mu'} \ha^+_\mu \ha_{\mu'} 
.
\label{eq:rpa:opext}
\eeq
The transition probability from the ground state to a TDA excited state is
\beqn
\nonumber
\braket{\Psi_\nu | {\hat F} | \Psi_0} &=& \braket{\Phi_0 | \hQ_\nu {\hat F} | \Phi_0} 
\\ \nonumber &=& 
\braket{\Phi_0 | \sum_{mi} X^{*\nu}_{mi} 
\ha^+_i \ha_m \sum_{\mu \mu'}  f_{\mu  \mu'} \ha^+_\mu \ha_{\mu'}  | \Phi_0} 
\\ \nonumber &=& 
 \sum_{mi} X^{*\nu}_{mi}  \sum_{\mu \mu'}  f_{\mu  \mu'} \braket{\Phi_0 |  \ha^+_i \ha_m  \ha^+_\mu \ha_{\mu'}  | \Phi_0} 
 \\ &=&
  \sum_{mi} X^{*\nu}_{mi}  \sum_{\mu \mu'}  f_{\mu  \mu'} \delta_{i \mu'} \delta_{m \mu} 
  = \sum_{mi} X^{*\nu}_{mi} \, f_{mi} 
.
\eeqn
where we used  Wick's theorem as in Equation~(\ref{eq:rpa.wick}). 
The many-body transition probabilities are described in terms of single-particle transition
probabilities.

\subsection{Random Phase Approximation}
\label{sec:rpa:RPA}

\subsubsection{Limits of the TDA}
The comparison between the TDA results and the experimental data is not satisfactory, 
especially in nuclear physics. For this reason, since the second half of the 1960s, the assumptions related to the TDA theory have been carefully analyzed.
These assumptions are related to the choice of the expression (\ref{eq:rpa:tdaq}) of the 
$\hQ_\nu$ operator. From these studies, it appeared clear that this choice is inconsistent
with the equations of motion (\ref{eq:rpa:moto}). 

This inconsistency can be seen in the following manner. 
The equation of motions
(\ref{eq:rpa:moto}) were obtained without making any assumption on the operator
${\hat {\cal O}}$. For the operator  ${\hat {\cal O} } = \ha^+_m \ha_i$, the equations of motion are:
\beq
\braket{\Psi_0| \left[ \ha^+_m \ha_i , [\hH,\hQ^+_\nu] \right] |\Psi_0} =
\left( E_\nu - E_0 \right) \braket{\Psi_0 | \ha^+_m \ha_i \hQ^+_\nu | \Psi_0} 
= \left( E_\nu - E_0 \right) \braket{\Psi_0 | \left[ \ha^+_m \ha_i, \hQ^+_\nu \right] |\Psi_0} 
.
\label{eq:rpa:nocon1}
\eeq
By inserting the expression of the TDA operator (\ref{eq:rpa:tdaq}) in the right hand
side of the above equation, we obtain
\beq
\sum_{nj} X^\nu_{nj} 
\braket{\Phi_0 | \left[ \ha^+_m \ha_i, \ha^+_n \ha_j \right] |\Phi_0} 
 = 
\sum_{nj} X^\nu_{nj} \left\{ 
\braket{\Phi_0 | \ha^+_m \ha_i \ha^+_n \ha_j  | \Phi_0}  
- \braket{\Phi_0 | \ha^+_n \ha_j \ha^+_m \ha_i  | \Phi_0} 
\right\} = 0 
.
\eeq
This result requires that also the left hand side of  Equation (\ref{eq:rpa:nocon1}) 
  be zero. The one-body term of the hamiltonian has a double commutator 
equal to zero
\[
\sum_{\alpha \beta} h_{\alpha,\beta} 
\braket{\Phi_0 | \left[ \ha^+_m \ha_i , (\ha^+_\alpha \ha_j  \delta_{n \beta} - 
\ha^+_n \ha_\beta \delta_{j \alpha}) \right]| \Phi_0} = 0
,
\]
but the double commutator of the interaction term is not equal to zero.
\[
\sum_{\alpha,\beta,\alpha',\beta'} \barv_{\alpha,\beta,\alpha',\beta'} 
\braket{\Phi_0 | \Big[ \ha^+_m \ha_i , 
\big[ \hN[ \ha^+_\alpha \ha^+_\beta \ha_{\beta'} \ha_{\alpha'}] , \ha^+_n \ha_j \big] 
\Big] | \Phi_0}
\neq 0
.
\]

\subsubsection{ RPA Equations}
\label{sec:RPA.rpa}

The most straightforward way of extending the TDA is to consider 
the RPA excitation operator (\ref{eq:rpa:tdaq}) defined as
\beq
\hQ^+_\nu \equiv \sum_{p\,h} X^\nu_{ph} \ha^+_p \ha_h - \sum_{p\,h} Y^\nu_{ph} \ha^+_h \ha_p 
,
\label{eq:rpa:rpaq}
\eeq
where both $X_{ph}^\nu$ and $Y_{ph}^\nu$ are numbers.

 RPA ground state is defined by the equation $\hQ_\nu \ket{\nu_0}= 0$. Evidently 
$\ket{\nu_0}$  is not an IPM ground state, i.e., a single Slater determinant.  
In this last case, we would have
\[
\hQ_\nu \ket{\Phi_0}
=  \sum_{p\,h} X^{*\nu}_{ph} \ha^+_h \ha_p \ket{\Phi_0}
 - \sum_{p\,h} Y^{* \nu}_{ph} \ha^+_p \ha_h \ket{\Phi_0}
 \ne 0
.
\] 
The first term is certainly zero, while the second one is not zero.
 RPA ground state $\ket{\nu_0}$ is more complex than the IPM ground state and
it contains effects beyond it. These effects, called generically correlations, are here described in terms
of hole--particle excitations, as we shall discuss in Section \ref{sec:RPA.ground}.

From the definition (\ref{eq:rpa:rpaq}) of RPA amplitudes, we obtain $\delta \hQ_\nu$ 
and by inserting it as ${\hat {\cal O}} = \delta \hQ_\nu$ in the equations of motion (\ref{eq:rpa:moto}) 
we obtain
\beqn
\nonumber
&~& \sum_{mi} \delta X^\nu_{mi}
\braket{\nu_0 | \Big[  \ha^+_i \ha_m  , \big[ \hH, \hQ^+_\nu \big]  \Big] | \nu_0}
- \sum_{mi} \delta Y^\nu_{mi} 
\braket{\nu_0 | \Big[  \ha^+_m \ha_i  , \big[ \hH, \hQ^+_\nu \big]  \Big]  | \nu_0}
  \\ 
&=&
(E_\nu - E_0)
\left\{
\sum_{mi} \delta X^\nu_{mi}
\braket{\nu_0 | \Big[  \ha^+_i \ha_m  ,  \hQ^+_\nu  \Big] | \nu_0}
- \sum_{mi} \delta Y^\nu_{mi} 
\braket{\nu_0 | \Big[  \ha^+_m \ha_i  ,  \hQ^+_\nu  \Big]  | \nu_0}
\right\}
.
\eeqn

As in the TDA case, the above equation represents a sum of independent terms since
each variation is independent of the other ones. By making equal the terms related to the
same variation, we obtain the following relations
\beqn
\braket{\nu_0 | \Big[  \ha^+_i \ha_m  , \big[ \hH, \hQ^+_\nu \big]  \Big] | \nu_0} &=&
(E_\nu - E_0) \braket{\nu_0 | \Big[  \ha^+_i \ha_m  ,  \hQ^+_\nu  \Big] | \nu_0}
\label{eq:rpa:rpa1a}
 \\
\braket{\nu_0 | \Big[ \ha^+_m \ha_i  , \big[ \hH, \hQ^+_\nu \big]  \Big]  | \nu_0} &=&
(E_\nu - E_0) \braket{\nu_0 | \Big[  \ha^+_m \ha_i  ,  \hQ^+_\nu  \Big]  | \nu_0}
\label{eq:rpa:rpa1b}
.
\eeqn

Let us consider the left hand side of Equation (\ref{eq:rpa:rpa1a}) 
\beqn
\nonumber
&~&
\braket{\nu_0 | \Big[  \ha^+_i \ha_m  , \big[ \hH, \hQ^+_\nu \big]  \Big] | \nu_0} \\
\nonumber 
&=&
\sum_{nj} X^\nu_{nj} 
\braket{\nu_0 | \Big[  \ha^+_i \ha_m  , \big[ \hH, \ha^+_n \ha_j \big]  \Big] | \nu_0} 
- \sum_{nj} Y^\nu_{nj} 
\braket{\nu_0 | \Big[  \ha^+_i \ha_m  , \big[ \hH, \ha^+_j \ha_n \big]  \Big] | \nu_0} \\
&\equiv&
\sum_{nj} X^\nu_{nj}  A_{minj} + \sum_{nj} Y^\nu_{nj} B_{minj}
.
\label{eq:rpa:AB1}
\eeqn
These equations define the elements of the $A$ and $B$ matrices. 

We calculate the right hand side of Equation (\ref{eq:rpa:rpa1a}) by using an approximation known
in the literature as \emph{Quasi-Boson-Approximation} (QBA) consisting in assuming
that the expectation value of a commutator between RPA ground states has the same value of 
the commutator between IPM states $\ket{\Phi_0}$. 
In the specific case under study, we have that 
\beq
\braket{\nu_0 | \Big[  \ha^+_i \ha_m  , \hQ^+_\nu \Big] | \nu_0} 
\simeq \braket{\Phi_0 | \Big[  \ha^+_i \ha_m  , \hQ^+_\nu \Big] | \Phi_0} 
\label{eq:rpa.qba}
.
\eeq
It  is worth  remarking that the QBA can be applied only for expectation values
of commutators. The idea is that pairs of creation
and destruction operators follow the rule
\[
[\ha^+_i \ha_m, \ha^+_n \ha_j] \simeq \delta_{mn} \delta_{ij}
,
\]
which means that the operators ${\hat {\cal O}}_{im} \equiv \ha^+_i \ha_m$ and
${\hat {\cal O}}^+_{jn} \equiv a^+_n a_j$ behave as boson operators.

By using the QBA, we can write
\beqn
\nonumber
&~& \braket{\nu_0 | \Big[  \ha^+_i \ha_m  , \hQ^+_\nu \Big] | \nu_0}  \\
\nonumber
&\simeq& \sum_{nj} X^\nu_{nj} 
\braket{\Phi_0 | [\ha^+_i \ha_m, \ha^+_n \ha_j]  | \Phi_0}  
- \sum_{nj} Y^\nu_{nj} 
\braket{\Phi_0 | [\ha^+_i \ha_m, \ha^+_j \ha_n]  | \Phi_0}  \\
\nonumber &=& 
\sum_{nj} X^\nu_{nj} \Big\{ \braket{\Phi_0 | \ha^+_i \ha_m \ha^+_n \ha_j | \Phi_0}  
- \braket{\Phi_0 | \ha^+_n \ha_j \ha^+_i \ha_m | \Phi_0}  \Big\} \\
\nonumber 
&-&  \sum_{nj} Y^\nu_{nj} \Big\{ \braket{\Phi_0 | \ha^+_i \ha_m \ha^+_j \ha_n  | \Phi_0} 
-  \braket{\Phi_0 | \ha^+_j \ha_n \ha^+_i \ha_m  | \Phi_0}  \Big\} \\
&=& \sum_{nj} X^\nu_{nj} \braket{\Phi_0 | \ha^+_i \ha_m \ha^+_n \ha_j | \Phi_0}  
= X^\nu_{mi} \delta_{mn} \delta_{ij} 
,
\label{eq:rpa:right1}
\eeqn
where we have taken into account  
that the terms multiplying $Y^\nu_{nj}$ do not conserve the 
particle number and, furthermore, that $a_m \ket{\Phi_0}=0$. 
Equation (\ref{eq:rpa:rpa1a}) becomes
\beq
\sum_{nj} X^\nu_{nj}  A_{minj} + \sum_{nj} Y^\nu_{nj} B_{minj} 
= (E_\nu - E_0)   X^\nu_{mi}
.
\label{eq:rpa:rpa1}
\eeq

For the calculation of the left hand side of Equation  (\ref{eq:rpa:rpa1b}),  
we consider that:
\beq
[\hH, \ha^+_n \ha_j]^+ = (\hH \ha^+_n \ha_j -  \ha^+_n \ha_j \hH)^+
= \ha^+_j \ha_n \hH - \hH \ha^+_j \ha_n = - [\hH, \ha^+_j \ha_n]
,
\eeq
since $\hH = \hH^+$ and then
\beq
\Big[ \ha^+_i \ha_m , [\hH, \ha^+_j \ha_n] \Big]^+ 
= -  \Big[ \ha^+_m \ha_i , -  [\hH, \ha^+_n \ha_j] \Big] 
=   \Big[ \ha^+_m \ha_i ,  [\hH, \ha^+_n \ha_j] \Big] 
.
\eeq
The double commutator becomes
\beqn
\nonumber
&~& \braket{\nu_0 | \Big[  a^+_m a_i  , \big[ \hH, Q^+_\nu \big]  \Big]  | \nu_0} 
\\ \nonumber &=&
\sum X^\nu_{nj} \braket{\nu_0 | \Big[  a^+_m a_i  , \big[ \hH, a^+_n a_j \big]  \Big]  | \nu_0} 
- \sum Y^\nu_{nj} \braket{\nu_0 | \Big[  a^+_m a_i  , \big[ \hH, a^+_j a_n \big]  \Big]  | \nu_0} 
\\ \nonumber &=&
\sum X^\nu_{nj} \braket{\nu_0 | \Big[  a^+_i a_m  , \big[ \hH, a^+_j a_n \big]  \Big]^+  | \nu_0} 
- \sum Y^\nu_{nj} \braket{\nu_0 | \Big[  a^+_i a_m  , \big[ \hH, a^+_n a_j \big] \Big]^+  | \nu_0} 
\\  &=&
\sum_{nj} X^\nu_{nj} (-  B^*_{minj} )+ \sum_{nj} Y^\nu_{nj} ( - A^*_{minj} ) 
,
\label{eq:rpa:AB2}
\eeqn
where we considered the definitions of the matrix elements $A$ and $B$ in Equation (\ref{eq:rpa:AB1}).

For the calculation of the right hand side of Equation (\ref{eq:rpa:rpa1b}) by using the QBA, 
we have
\beq
\braket{\nu_0 | \Big[  \ha^+_m \ha_i  ,  Q^+_\nu  \Big]  | \nu_0} \rightarrow \;({\rm QBA}) \rightarrow
- \sum_{nj} Y^\nu_{nj}  \braket{\Phi_0 | \Big[  \ha^+_m \ha_i  ,  \ha^+_j \ha_n  \Big]  | \Phi_0} =
Y^\nu_{mi} \delta_{ij} \delta_{mn}
;
\eeq
therefore, Equation (\ref{eq:rpa:rpa1b}) becomes
\beq
\sum_{nj} X^\nu_{nj} (-  B^*_{minj} )+ \sum_{nj} Y^\nu_{nj} ( - A^*_{minj} ) 
= (E_\nu - E_0) Y^\nu_{mi}
.
\label{eq:rpa:rpa2}
\eeq
 Equations (\ref{eq:rpa:rpa1}) and (\ref{eq:rpa:rpa2}) represent a homogenous system of 
linear equations whose unknowns are RPA amplitudes $X^\nu_{ph}$ and $Y^\nu_{ph}$. 
Usually, this system is presented as
\beq
\left(
\begin{array}{cc}
A & B \\
   & \\
B^*  & A^*
\end{array}
\right)
\left(  
\begin{array}{c}
X^\nu \\
\\
Y^\nu
\end{array}
\right) = 
(E_\nu - E_0)
\left(
\begin{array}{cc}
I & 0 \\
   & \\
0  & -I
\end{array}
\right)
\left(  
\begin{array}{c}
X^\nu \\
\\
Y^\nu
\end{array}
\right) = 
(E_\nu - E_0)
\left(  
\begin{array}{c}
X^\nu \\
\\
- Y^\nu
\end{array}
\right) 
\label{eq:rpa:matrix}
,
\eeq
where $A$ and $B$ are square matrices whose dimensions are those of the number of 
the particle--hole pairs describing the excitation, and $X$ and $Y$ are vectors of the same 
dimensions.

The expressions of the matrix elements of $A$ and $B$ in terms of effective interaction
between two interacting particles are obtained as in Appendix \ref{sec:app.doublec}
and they are:
\beqn
A_{minj} & \rightarrow \;({\rm QBA}) \rightarrow & 
\braket{\Phi_0 | \Big[  \ha^+_i \ha_m  , \big[ \hH, \ha^+_n \ha_j \big]  \Big] | \Phi_0} 
= (\epsilon_m - \epsilon_i ) \delta_{mn} \delta_{ij} + \barv_{mjin} 
,
\label{eq:rpa:amnij} \\
B_{minj} & \rightarrow \;({\rm QBA}) \rightarrow & 
- \braket{\Phi_0 | \Big[  \ha^+_i \ha_m  , \big[ \hH, \ha^+_j \ha_n \big]  \Big] | \Phi_0} 
= \barv_{mnij} 
.
\label{eq:rpa:bmnij} 
\eeqn
%
%

\subsubsection{Properties of RPA Equations}
\label{sec:propeRPA}
We consider RPA equations in the form
\[
\left(
\begin{array}{cc}
A & B \\
   & \\
B^*  & A^*
\end{array}
\right)
\left(  
\begin{array}{c}
X^\nu \\
\\
Y^\nu
\end{array}
\right) = 
\omega_\nu
\left(  
\begin{array}{c}
X^\nu \\
\\
- Y^\nu
\end{array}
\right) 
,
\]
where $\omega_\nu=E_\nu - E_0$ is the excitation energy.
\begin{itemize}
\item If $B=0$, we obtain the TDA equations.
\item
We take the complex conjugate of the above equations and obtain
\beq
\left(
\begin{array}{cc}
A & B \\
   & \\
B^*  & A^*
\end{array}
\right)
\left(  
\begin{array}{c}
Y^{*\nu} \\
\\
X^{*\nu}
\end{array}
\right) = 
- \omega_\nu
\left(  
\begin{array}{c}
Y^{*\nu} \\
\\
- X^{*\nu}
\end{array}
\right) 
.
\label{eq:rpa:cmplx} 
\eeq
This indicates that RPA equations are satisfied by 
positive and negative eigenvalues with the 
same absolute value.

\item 
Eigenvectors corresponding to different eigenvalues are orthogonal.
\[
\left(
\begin{array}{cc}
A & B \\
   & \\
B^*  & A^*
\end{array}
\right)
\left(  
\begin{array}{c}
X^{\nu} \\
\\
Y^{\nu}
\end{array}
\right) = 
\omega_\nu
\left(  
\begin{array}{c}
X^{\nu} \\
\\
- Y^{\nu}
\end{array}
\right) 
\;\;;\;\;
\left(
\begin{array}{cc}
A & B \\
   & \\
B^*  & A^*
\end{array}
\right)
\left(  
\begin{array}{c}
X^{\mu} \\
\\
Y^{\mu}
\end{array}
\right) = 
\omega_\mu
\left(  
\begin{array}{c}
X^{\mu} \\
\\
- Y^{\mu}
\end{array}
\right) 
.
\]
Let us calculate the hermitian conjugate of the second equation
\[
({X^\mu}^+ , {Y^\mu}^+) 
\left( \begin{array}{cc}
A & B \\
   & \\
B^*  & A^*
\end{array} \right)
= 
({X^\mu}^+ , -{Y^\mu})  \omega_\mu
.
\]
We multiply the first equation by  $({X^\mu}^+ , {Y^\mu}^+)$ on the left hand side, 
and the second equation on the right hand side by 
\[
\left(  \begin{array}{c}
X^{\nu} \\
\\
- Y^{\nu}
\end{array} \right) 
,
\]
and we obtain
\beqn
\nonumber 
({X^\mu}^+ , {Y^\mu}^+)
\left(
\begin{array}{cc}
A & B \\
   & \\
B^*  & A^*
\end{array}
\right)
\left(  
\begin{array}{c}
X^{\nu} \\
\\
Y^{\nu}
\end{array}
\right) 
&=& 
\omega_\nu
({X^\mu}^+ , {Y^\mu}^+)
\left(  \begin{array}{c}
X^{\nu} \\
\\
- Y^{\nu}
\end{array} \right) 
\\ \nonumber &~&
\\ \nonumber
({X^\mu}^+ , {Y^\mu}^+) 
\left( \begin{array}{cc}
A & B \\
   & \\
B^*  & A^*
\end{array} \right)
\left(  \begin{array}{c}
X^{\nu} \\
\\
 Y^{\nu}
\end{array} \right) 
&=& 
 \omega_\mu
({X^\mu}^+ , -{Y^\mu}) 
\left(  \begin{array}{c}
X^{\nu} \\
\\
Y^{\nu}
\end{array} \right) 
.
\eeqn
By subtracting the two equations, we have
\[
0 = (\omega_\nu - \omega_\mu) ({X^\mu}^+ X^\nu - {Y^\mu}^+ Y^\nu)
.
\]
Since we  assumed $\omega_\nu \ne \omega_\mu$, we obtain
\[
({X^\mu}^+ X^\nu - {Y^\mu}^+ Y^\nu)=0
.
\]

\item The normalization between two excited states requires

\beqn
\nonumber
\delta_{\nu \nu'} &=& \braket{\nu | \nu'} =
\braket{\nu_0 | \hQ_\nu \hQ^+_{\nu'} | \nu_0} = \braket{\nu_0 | [\hQ_\nu , \hQ^+_{\nu'} ]| \nu_0} 
\rightarrow \;{\rm QBA} \rightarrow
\braket{\Phi_0 | [\hQ_\nu, \hQ^+_{\nu'} ]| \Phi_0} \\
&=&
\sum_{mi} \left(X^\nu_{mi} X^{\nu'}_{mi} -   Y^\nu_{mi} Y^{\nu'}_{mi} \right)  
,
\label{eq:RPA:normalization}
\eeqn
where we used the fact that $\hQ_\nu \ket{\nu_0}=0$ to express the operator as commutator
in order to use the QBA.
\end{itemize}

\subsubsection{Transition Probabilities in RPA}
\label{sec.eom.trans}

In analogy with the TDA case, we assume that the action of the external field exciting
the system is described by a one-body operator expressed as in Equation (\ref{eq:rpa:opext}). The 
transition probability between RPA ground state and excited state is described by
\beq
\braket{\nu | {\hat F} | \nu_0} = \braket{\nu_0 | \hQ_\nu {\hat F} | \nu_0} 
= \braket{\nu_0 | [ \hQ_\nu, {\hat F} ] | \nu_0} 
,
\eeq
where we used, again, the fact that $\hQ_\nu \ket{\nu_0}=0$. Since the equation is expressed in terms
of commutator we can use the QBA
\beqn
\nonumber
&~& \braket{\nu | {\hat F} | \nu_0} 
\rightarrow \;{\rm QBA} \rightarrow \braket{\Phi_0 | [ \hQ_\nu, \hF ] | \Phi_0} \\
&=&
\sum_{\mu \mu'} f_{\mu \mu'}
\Big\{
\sum_{mi} X^\nu_{mi}  \braket{\Phi_0 | [ \ha^+_i \ha_m ,  \ha^+_\mu  \ha_{\mu'} ] | \Phi_0} 
- \sum_{mi} Y^\nu_{mi}  \braket{\Phi_0 | [ \ha^+_m \ha_i ,  \ha^+_\mu  \ha_{\mu'} ] | \Phi_0} 
\Big\}
.
\eeqn
The two matrix elements are
\beqn
\nonumber
\braket{\Phi_0 | [ \ha^+_i \ha_m ,  \ha^+_\mu  \ha_{\mu'} ] | \Phi_0} 
&=& \braket{\Phi_0 |  \ha^+_i \ha_m  \ha^+_\mu  \ha_{\mu'}  | \Phi_0} - 
\braket{\Phi_0 | \ha^+_\mu  \ha_{\mu'}  \ha^+_i \ha_m  | \Phi_0} = \delta_{m \mu} \delta_{i \mu'} - 0
,
\\ \nonumber 
\braket{\Phi_0 | [ \ha^+_m \ha_i ,  \ha^+_\mu  \ha_{\mu'} ] | \Phi_0} 
&=& \braket{\Phi_0 |  \ha^+_m \ha_i  \ha^+_\mu  \ha_{\mu'}  | \Phi_0} - 
\braket{\Phi_0 | \ha^+_\mu  \ha_{\mu'}  \ha^+_m \ha_i  | \Phi_0} = 0 - \delta_{m \mu'} \delta_{i \mu} 
;
\eeqn
therefore, 
\beq
\braket{\nu | \hF | \nu_0} \simeq 
\sum_{\mu \mu'} f_{\mu \mu'} 
\left( \sum_{mi} X^\nu_{mi} \delta_{m \mu} \delta_{i \mu'}  
+  \sum_{mi} Y^\nu_{mi}  \delta_{m \mu'} \delta_{i \mu} \right) 
=  \sum_{mi} \left( X^\nu_{mi} f_{mi} + Y^\nu_{mi} f_{im}
\right)
\label{eq:rpa.tprob}
.
\eeq
Also in RPA, the transition amplitude of a many-body system is expressed
as a linear combination of single-particle transitions.

\subsubsection{Sum Rules}
\label{sec:eom.srule}

We show  in Appendix \ref{sec:app.sr} that, in general, by indicating with  $\ket{\Psi_\nu}$ 
the eigenstates of the hamiltonian $\hH$ 
\[
\hH \ket{\Psi_\nu} = E_\nu \ket{\Psi_\nu}
,
\]
for an external operator $\hF$ inducing a transition of the system from the ground
state to the excited state one has that:
\beq
2 \sum_\nu (E_\nu - E_0) \left| \braket{\Psi_\nu | \hF | \Psi_0} \right|^2
= \braket{\Psi_0 | \Big[ \hF,  \big[ \hH, \hF \big] \Big] | \Psi_0} 
.
\label{eq:rpa:gensumrule}
\eeq
This expression puts a quantitative limit on the total value of the excitation strength 
of a many-body system. This value is determined only by the ground
state properties and the knowledge of the excited states structure is not required.
The validity of Equation~(\ref{eq:rpa:gensumrule}) is related to the fact that the  $\ket{\Psi_\nu}$ 
are eigenstates of  $\hH$. In actual calculations, states based on models or approximated 
solutions of the Schr\"odinger equations are used and Equation~(\ref{eq:rpa:gensumrule}) is
not properly satisfied.

On the other hand, for RPA theory, it has been shown \cite{tho61} that the following relation holds
\beq
2 \sum_\nu (E_\nu - E_0) \left| \braket{\nu | \hF | \nu_0} \right|^2
= \braket{\Phi_0 | \Big[ \hF,  \big[ \hH, \hF \big] \Big] | \Phi_0} 
.
\label{eq:rpa:rpasumrule}
\eeq
The above expression, formally speaking, is not a true sum rule since in the left hand side there are RPA states,
both ground and excited states, while in the right hand side there is an IPM ground state. These two
types of states are not eigenstates of the same hamiltonian.
When the residual interaction is neglected, one obtains mean-field excited states  
$\ket{\Phi_{ph}}$, 
i.e., single Slater determinants with a single 
particle--hole excitation. In this case, Equation (\ref{eq:rpa:gensumrule}) 
is verified since all these mean-field states are eigenstates of the unperturbed hamiltonian $\hH_0$
\beq
2 \sum_{ph} (\epsilon_p - \epsilon_h) \left| \braket{\Phi_{ph} | \hF | \Phi_0} \right|^2
= \braket{\Phi_0 | \Big[ \hF,  \big[ \hH_0, \hF \big] \Big] | \Phi_0}, 
\label{eq:rpa:HFsumrule}
\eeq
where the excitation energies of the full system are given by the difference between the 
single-particle energies of the particle--hole excitation. 

Since in RPA the full hamiltonian $\hH = \hH_0 + \hV_{\rm res}$ is considered, by inserting
this expression in Equation (\ref{eq:rpa:rpasumrule}) we obtain
\beq
2 \sum_\nu (E_\nu - E_0) \left| \braket{\nu | \hF | \nu_0} \right|^2
=
\braket{\Phi_0 | \Big[ \hF,  \big[ \hH_0, \hF \big] \Big] | \Phi_0} 
+\braket{\Phi_0 | \Big[ \hF,  \big[ \hV_{\rm res}, \hF \big] \Big] | \Phi_0} 
.
 \label{eq:rpa:sumrulehfrpa}
\eeq
For operators $\hF$ which commute with $\hV_{\rm res}$, the IPM and RPA 
sum rules coincide.

\subsubsection{ RPA Ground State}
\label{sec:RPA.ground}
\index{RPA ground state}
We have already indicated that RPA ground state is not an IPM state but it 
contains effects beyond it, correlations, expressed in terms of hole--particle excitations. 
A more precise representation of the RPA ground state comes from a theorem demonstrated
by  D.~J.~Thouless \cite{tho61} leading to an expression of the RPA ground state
of the type \cite{rin80}:
\beq
\ket{\nu_0} = {\cal N} e^{\hat S} \ket{\Phi_0}
\label{eq:rpa.thouless}
,
\eeq
where ${\cal N}$ is a normalization constant and the operator ${\hat S}$ is defined as
\beq
{\hat S} \equiv \half \sum_{\nu, m i n j} s^{(\nu)}_{minj} \ha^+_m \ha_i \ha^+_j \ha_n 
.
\eeq
The sum considers all the particle--hole,  $\ha^+_m \ha_i$, and hole--particle, 
$\ha^+_j \ha_n$, pairs and the index $\nu$ runs on all the possible angular momentum 
and parity combinations allowed by the particle--hole and hole--particle quantum numbers. 
We indicated with $s^{(\nu)}_{minj}$ an amplitude weighting the contribution of each 
couple of $ph$. 

Starting from the above expression, it is possible to calculate the $s^{(\nu)}_{minj}$ 
from the knowledge of RPA $X^\nu_{ph}$ and $Y^\nu_{ph}$ amplitudes \cite{suh07}. 
By using these expressions, the expectation value of 
a one-body operator with respect to the RPA ground state can be expressed as \cite{row68,ang01} 
\beqn
\nonumber
\braket{\nu_0 | \hF | \nu_0} &=& 
\braket{\nu_0 | \sum_{\mu \mu'}  f _{\mu\,\mu'} \ha^+_\mu \ha_{\mu'} | \nu_0} \\
&=&
\sum_h f_{h\,h}
\left[ 1 - \half \sum_\nu \sum_p |Y^\nu_{ph} |^2\right]
+
\sum_p f_{p,p}
\left[ \half \sum_\nu \sum_h |Y^\nu_{ph} |^2\right]
.
\eeqn
This clearly shows that the $Y^\nu_{ph}$ amplitudes
modify the expectation value of the operator with respect to the IPM result.
In TDA, the ground state is $\ket{\Phi_0}$ and the $Y$ amplitudes are zero; therefore,
the expectation value of $\hF$ is given by the sum of the s.p. expectation values 
of the states below the Fermi energy, as in the pure IPM. The TDA theory does not contain
ground state correlations as indicated in Equation~(\ref{eq:rpa:tdazero}).

\section {RPA with Green Function}
\label{sec:green}

\subsection{Field Operators and Pictures}
In this section, we use the field operators $\hpsi^+(\br)$, 
which creates a particle in the point $\br$. The hermitian
conjugate operator $\hpsi(\br)$  destroys a particle positioned in $\br$. 
These two operators are related to the creation and destruction operators
via the s.p.  wave functions $\phi_\nu(\br)$ generated by the solution 
of the IPM problem:
\beq
\label{eq:green.cdefa1}
\hpsi ({\bf r}) = \sum_{\nu} \ha_{\nu} \phi_{\nu} ({\bf r}) 
\;\;\;;\;\;\;
\hpsi^+({\bf r}) = \sum_{\nu} \ha^+_{\nu} \phi^{\ast}_{\nu} ({\bf r}) 
.
\eeq
These equations can be inverted to express the creation and
destruction operators in terms of field operator
\beq
\label{adefc}
\ha_{\nu} = \int d^3r \phi^{\ast}_{\nu}({\bf r}) \hpsi({\bf r}) 
\;\;\;;\;\;\;
\ha^+_{\nu} = \int d^3r \phi_{\nu}({\bf r}) \hpsi^+({\bf r})
.
\eeq
where we exploited the orthonormality of the $\phi_\nu$.
By using the anti-commutation relations of the creation 
and destruction operators, see Equation~(\ref{eq:app.anta}),
we obtain analogous relations for the field operators:
\beq
\label{eq:green.antc}
\left\{\hpsi^+({\bf r}), \hpsi^+({\bf r}')\right\}=0 \quad ; \quad
\left\{\hpsi({\bf r}), \hpsi({\bf r}')\right\}=0 \quad ;\quad
\left\{\hpsi^+({\bf r}), \hpsi({\bf r}')\right\}=\delta({\bf r}-{\bf r}')
.
\eeq

In the Heisenberg picture \cite{mes61,fet71}, the states are defined as
\beq
\label{eq:p.sheis}
|\Psi_{\rm H} (t)\rangle \equiv  e^{i{\frac{\hH t}{\hbar}}} |\Psi_{\rm S}(t)\rangle 
,
\eeq
with respect to those of the Schr\"odinger picture $ |\Psi_{\rm S}(t)\rangle$. 
In the above equations, $\hH$ is the full many-body hamiltonian.
The states in the Heisenberg picture are time-independent and the time evolution
of the system is described by the operators whose relation with the time-independent
operators of the Schr\"odinger picture is:
\beq
\label{eq:p.oheis}
\hO_{\rm H}\equiv e^{i{\frac{\hH t}{\hbar}}} \hO_S e^{-i{\frac{\hH t}{\hbar}}} 
,
\eeq
satisfying the equation:
\beq
\label{eq:p.theis}
i\hbar \frac{\partial}{\partial t} \hO_{\rm H}(t) = [\hO_{\rm H}(t), \hH] 
\eeq

By separating the hamiltonian in the Schr\"odinger picture into two parts
\beq
\label{eq:p.hint}
\hH = \hH_0 + \hH_1 
,
\eeq
it is possible to define an interaction picture whose states are defined as
\beq 
\label{eq:p.sint}
|\Psi_{\rm I}(t) \rangle \equiv e^{i{\frac{\hH_0 t}{\hbar}}} 
|\Psi_{\rm S}(t)\rangle
,
\eeq
and the operators 
\beq
\label{eq:p.oint}
\hO_{\rm I}(t) = e^{i{\frac{\hH_0 t}{\hbar}}} \hO_{\rm S} e^{-i{\frac{\hH_0 t}{\hbar}}}
.
\eeq

In the interaction picture \cite{fet71}, both states and operators evolve with the time
as indicated by the equations
\beq
\label{eq:p.timesint}
 i\hbar \frac{\partial}{\partial t} |\Psi_{\rm I}(t)\rangle = \hH_{1,{\rm I}}(t)|\Psi_{\rm I}(t)\rangle 
  ,
\eeq
and 
\beq
\label{eq:p.timeoint}
i\hbar \frac{\partial}{\partial t} \hO_{\rm I}(t)  =  \left[ \hO_{\rm I}(t), \hH_0\right] 
  .
\eeq

The fermionic field operators in the Heisenberg and interaction picture
are, respectively, defined as:
\beq
\hpsi_{\rm H}(\bx,t) = e^{\ih  \hH  t} \hpsi(\bx) e^{- \ih \hH t} 
\quad;\quad
\hpsi_{\rm I}(\bx,t) = e^{\ih  \hH_0  t} \hpsi(\bx) e^{- \ih \hH_0 t} 
\label{eq:psih}
\eeq
It can be shown that the anti-commutation relations 
(\ref{eq:green.antc}) of the field operators, as well as those
of the creation and destruction operators,  see Equations (\ref{eq:app.anta}),
are conserved in every representation \cite{fet71}.

\subsection{Two-Body Green Function and RPA}
\label{sec:green.tbgf}

The two-body Green function is defined as
\beq
(-i)^2 \rG(\bx_1,t_1,\bx_2,t_2,\bx_3,t_3,\bx_4,t_4)
\equiv
\frac{ \braket{
 \Psi_0 | \hT[\hpsi_{\rm H} (\bx_1,t_1)  \hpsi_{\rm H} (\bx_2,t_2) 
\hpsi^+_{\rm H} (\bx_3,t_3) \hpsi^+_{\rm H} (\bx_4,t_4) ] |\Psi_0 \rangle}
}
{ \braket {\Psi_0 | \Psi_0} }
\,\,\,.
\label{eq:gtbgf}
\eeq
In the above expression, $\ket{\Psi_0}$ indicates the ground state of the system in 
Heisenberg representation and $\hT$ is the time-ordering operator which arranges 
the field operators in decreasing time order. Because of the possible values that the 
times $t_i$ can assume, there are $4!=24$ cases, but, for the symmetry properties
\beq
\rG(1,2,3,4)=-\rG(2,1,3,4)=-\rG(1,2,4,3)=\rG(2,1,4,3)
\,\,\,,
\eeq
only six of them are independent. Out of these six cases, only three have physically 
interesting properties and between these latter cases we select that where 
 $t_1, t_3 > t_2 , t_4$ which implies
\beq
(-i)^2 \rG(\bx_1,t_1,\bx_2,t_2,\bx_3,t_3,\bx_4,t_4) = \frac{
- \langle \Psi_0 | \hpsi_{\rm H} (\bx_1,t_1) \hpsi^+_{\rm H} (\bx_3,t_3) 
 \hpsi_{\rm H} (\bx_2,t_2)  \hpsi^+_{\rm H} (\bx_4,t_4)  
|\Psi_0 \rangle
}
{ \braket {\Psi_0 | \Psi_0} }
\,\,\,,
\eeq
and describes the time evolution of a $ph$ pair.

Since we work in a non-relativistic framework, the creation and also the destruction,
of a particle--hole pair is instantaneous; therefore, we have that 
\beq
t_1 = t_3 = t' \,\,\,\,{\rm e}\,\,\,\, t_2 = t_4 =t
\,\,\,.
\eeq

For this case, we express the two-body Green function in terms of creation and destruction
operators as
\beqn
\nonumber
&~& \rG(\bx_1,t',\bx_2,t,\bx_3,t',\bx_4,t)  \\
\nonumber
&=& \sum_{\nu_1 \nu_2 \nu_3 \nu_4} 
\frac{
\phi_{\nu_1}(\bx_1)\phi^*_{\nu_3}(\bx_3) \phi_{\nu_2}(\bx_2)\phi^*_{\nu_4}(\bx_4)
\langle \Psi_0 | \hT[ \ha_{\nu_1}(t')  \ha^+_{\nu_3}(t')  \ha_{\nu_2}(t) \ha^+_{\nu_4}(t)]|\Psi_0 \rangle 
}
{ \braket {\Psi_0 | \Psi_0} }
\\
&=& \sum_{\nu_1 \nu_2 \nu_3 \nu_4} 
\phi_{\nu_1}(\bx_1) \phi_{\nu_2}(\bx_2)
\phi^*_{\nu_3}(\bx_3) \phi^*_{\nu_4}(\bx_4)
G(\nu_1,t' ,\nu_2,t, \nu_3,t', \nu_4,t)
\label{eq:tbgf4}
\,\,\,.
\eeqn
where it is understood that 
all the creation and destruction operators are expressed in the Heisenberg picture.
The previous equation defines a two-body Green function depending on the quantum
numbers $\nu$ characterizing the single-particle states.  

Since this Green function depends only on the time difference $\tau=t'-t$, 
we find it convenient to define the energy dependent two-body Green function as
\beq
{\tilde G}(\nu_1,\nu_2, \nu_3, \nu_4,E) 
 =  \int _{-\infty}^{\infty} d \tau \, G(\nu_1,\nu_2, \nu_3, \nu_4,\tau) \, e^{\ih E\tau}
\,\,\,,
\label{eq:green.fourier}
\eeq
For the case $\tau > 0$, by considering the expression of the creation and destruction
operators in the Heisenberg picture, see Equation~(\ref{eq:p.oheis}), and the fact that $\ket{\Psi_0}$
is eigenstate of $\hH$ whose eigenvalue is $E_0$, we obtain the expression
\beq
{\tilde G}_{\tau>0}(\nu_1,\nu_2, \nu_3, \nu_4,E) =
\frac{
\langle \Psi_0 |  
\ha_{\nu_1}  \ha^+_{\nu_3}  
\int_0^\infty d\tau \, e^{-\ih (\hH-E_0-E)\tau}  \ha_{\nu_2} \ha^+_{\nu_4} 
|\Psi_0 \rangle }
{\braket{\Psi_0 | \Psi_0}}
\,\,\,.
\eeq
We can express the value of the time integral as
\beq
\lim_{\eta \rightarrow 0} \int_0^\infty d\tau \, e^{-\ih (-\hH+E_0+E - i \eta)\tau} 
= \lim_{\eta \rightarrow 0}  \frac {-i \hbar}   {E-\hH+E_0 - i\eta} 
\,\,\,;
\eeq
therefore,
\beq
{\tilde G}_{\tau>0}(\nu_1,\nu_2, \nu_3, \nu_4,E) = \hbar
\langle \Psi_0 |  
\ha_{\nu_1}  \ha^+_{\nu_3}  
 \,  \frac {-i}   {E-\hH+E_0+i\eta} \, \ha_{\nu_2} \ha^+_{\nu_4} 
|\Psi_0 \rangle 
\frac{1}{\braket{\Psi_0 | \Psi_0}}
\,\,\,.
\eeq
With an analogous calculation for the $\tau < 0$ case, we obtain
\beqn
\nonumber 
\ih {\tilde G}(\nu_1,\nu_2, \nu_3, \nu_4,E) &=& 
\ih ({\tilde G}_{\tau>0} + {\tilde G}_{\tau<0})  
\\ \nonumber &=& 
\langle \Psi_0 |  
\ha_{\nu_1}  \ha^+_{\nu_3}  
 \,  \frac {1}   {E-\hH+E_0  - i\eta} \, \ha_{\nu_2} \ha^+_{\nu_4} 
|\Psi_0 \rangle 
\frac{1}{\braket{\Psi_0 | \Psi_0}}
\\
&-& 
\langle \Psi_0 |  
\ha_{\nu_2}  \ha^+_{\nu_4}  
 \,  \frac {1}   {E+\hH-E_0 + i\eta} \, \ha_{\nu_1} \ha^+_{\nu_3} 
|\Psi_0 \rangle 
\frac{1}{\braket{\Psi_0 | \Psi_0}}
\,\,\,.
\eeqn
By inserting the completeness of the eigenfunctions of $\hH$, 
$\sum_n |\Psi_n \rangle \langle \Psi_n | = 1$ and considering
$\hH  |\Psi_n \rangle = E_n  |\Psi_n \rangle$, we obtain the expression
\beqn
\nonumber &~& 
\ih {\tilde G}(\nu_1,\nu_2, \nu_3, \nu_4,E) = \frac{1}{\braket{\Psi_0 | \Psi_0}}\\
&~& 
\sum_n \left[
 \frac{ \langle \Psi_0 |  \ha_{\nu_1}  \ha^+_{\nu_3}  |\Psi_n \rangle 
          \langle \Psi_n | \, \ha_{\nu_2} \ha^+_{\nu_4} |\Psi_0 \rangle }
 {E-(E_n-E_0) - i\eta}  
-
\frac {\langle  \Psi_0 |  \ha_{\nu_2}  \ha^+_{\nu_4}  |\Psi_n \rangle 
        \langle \Psi_n |   \ha_{\nu_1}  \ha^+_{\nu_3} |\Psi_0 \rangle }
  {E+(E_n-E_0) + i\eta} 
\right]
\,\,\,.
\label{eq:gtbgfl}
\eeqn
In this expression, the states $ |\Psi_n \rangle $ have the same 
number of particles as the ground state.
The energy values related to the poles, 
$E = E_n - E_0$, represent the excitation energies of the $A$ particle
system.

The unperturbed two-body Green function is obtained by substituting 
in \linebreak\mbox{Equation~(\ref{eq:gtbgfl})} the eigenstates $\ket{\Psi}$ of the full hamiltonian
with $\ket{\Phi}$, the eigenstates of the IPM hamiltonian $\hH_0$. In this 
case, the action of the creation and destruction operators is well defined
and the energy eigenvalues are given by the s.p. energies of the 
$ph$ pair. Because of the properties of the creation and destruction 
operators we have that
\beq
{\tilde G^0}(m, i, j, n, E) = \hbar \frac {\delta_{ij} \delta_{mn}} {\epsilon_m - \epsilon_i - E - i\eta}
\quad , \quad 
{\tilde G^0}( i,m , n, j, E) = \hbar \frac {\delta_{ij} \delta_{mn}} {\epsilon_m - \epsilon_i + E - i\eta}
,
\label{eq:green.gnot1}
\eeq
and 
\beq
{\tilde G^0}( m , i , n, j, E) = {\tilde G^0}( i ,m , j , n , E) = 0
.
\label{eq:green.gnot2}
\eeq

We show in Appendix \ref{sec:rl} that the two-body Green function is strictly related
to the response of the system to an external probe. By using $\hF$, the 
one-body operator of Equation~(\ref{eq:rpa:opext}) 
describing the action of the external probe, we can write
the transition amplitude from the ground state to an excited state as
\beqn
\nonumber
S(E) &=&  
\sum_n | \langle \Psi_0  | \hF |\Psi_n \rangle |^2 \delta\big(E - (E_n - E_0)\big)
\\ &=& 
 \sum_{\nu_1 \nu_2} \sum_{\nu_3 \nu_4}  f_{\nu_1 \nu_2} f^*_{\nu_3 \nu_4}
\frac {\Im}{\pi} \left(   i \hbar {\tilde G}(\nu_1,\nu_3, \nu_2, \nu_4,E) \right)
\label{eq:transprob}
\,\,\,,
\eeqn
where $f_{\nu_1 \nu_2}$ indicates the matrix element between s.p. wave functions.

The expression (\ref{eq:transprob}) of the transition amplitude separates
the action of the external probe from the many-body response which is 
described by the two-body Green function. This latter quantity is related to the 
interaction between the particles composing the system and it is a universal
function independent of the kind of probe perturbing the system.
The knowledge of $S(E)$ allows a direct comparison with observable quantities
such as scattering cross sections. 

In the time-dependent perturbation theory, a theorem, called  Gell--Man and Low, 
indicates that the eigenvector $\ket{\Psi_0}$  of the full hamiltonian 
can be written as \cite{fet71}:
\beq
\frac{\hU (0,-\infty)|\Phi_0\rangle}{\langle \Phi_0|
\hU (0,-\infty)|\Phi_0\rangle} \equiv
\frac{|\Psi_0\rangle}{\langle\Phi_0| \Psi_0\rangle}
\,\,\,,
\label{eq:green.gellman}
\eeq
where the time evolution operator $\hU$ can be expressed in powers of the 
interaction $\hH_1$ expressed in the interaction picture
\beq
\hU(t,t_0) = \sum_{n=0}^{\infty} \left(\frac{-i}{\hbar}\right)^n \frac{1}{n!}
\int_{t_0}^t dt_1 \ldots \int_{t_0}^t dt_n \hT 
\left[ \hH_1(t_1) \ldots \hH_1(t_n)\right].
\label{eq:green.ut}
\eeq
In the above equation, we dropped the subindex ${\rm I}$ to simplify the 
writing.

We insert Equation~(\ref{eq:green.gellman}) into the expression (\ref{eq:gtbgf}) 
of the two-body Green function and we obtain a perturbative expression
of the full interacting Green function in powers of $\hH_1$ and of 
the unperturbed two-body Green function $\rG^0$.

\begin{figure}[ht]
\begin{center}
\captionsetup{margin=2cm}
\includegraphics [scale=0.45, angle=0] {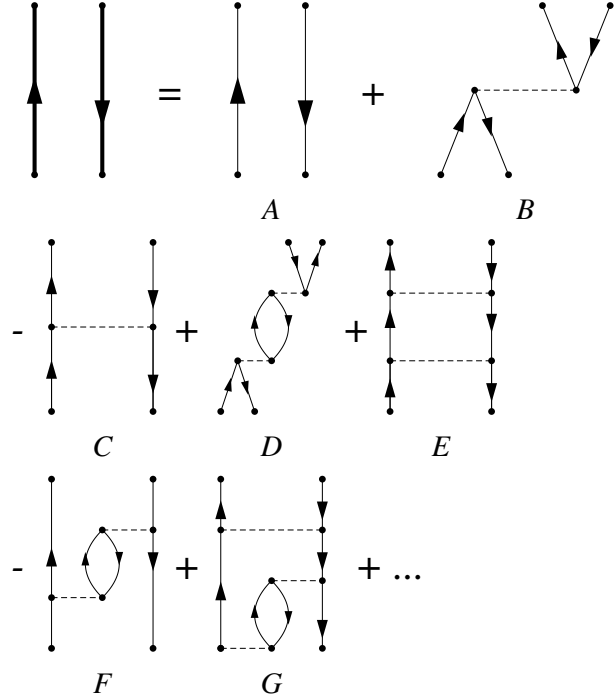}
\caption{Graphical representation of the perturbation expansion of the interacting Green 
function. The double thick lines represent $\rG$, the double thin lines $\rG^0$ and the dashed lines the 
two-body interaction $\intr$.   
}
\label{fig:gfed} 
\end{center}
\end{figure}

It is useful to consider a graphical representation of the Green function, as indicated in Figure~\ref{fig:gfed}.
The full two-body Green function is indicated by two continuous thick lines. The upward arrows stand for
the particle state and the downward arrows for the hole state. The continuous thin lines indicate
the unperturbed Green function $\rG^0$. In the figure, we consider only two-body  interactions, i.e., 
$\hH_1 = \intr(\rx_1, \rx_2)$ which is represented by a dashed line, with $\rx$ indicating both space and time
coordinates. 

Figure \ref{fig:gfed} shows some of the terms we obtain by carrying out the perturbation 
expansion of the two-body Green function. 
The explicit expressions of the various terms  are presented in Appendix \ref{sec:app.gfet}.
We observe that there are diagrams which, by cutting particle and hole lines,
can be separated into two diagrams already present in the expansion. 
This is the case, for example, of the diagram E of the figure
which is composed by two diagrams of C type and by the diagram G which is given by the product of a
diagram of C type and another one of F type. 
The contribution of these diagrams can be factorized in 
a term containing the four coordinates $\rx_1 \cdots \rx_4$ 
of the full Green function times, another term which
does not contain them. 
The sum of all the diagrams of this second type is identical to that of all the diagrams of the
denominator; therefore, these two contributions simplify the matter. Finally,
the calculation of $\rG$ can be carried out by considering only the remaining diagrams of the 
numerators which are called irreducible.

%

Formally, this can be expressed with an equation similar to the Dyson's equation for the 
one-body Green function \cite{fet71}:
\beqn
\nonumber
&~&\rG(\rx_1,\rx_2,\rx_3,x_4) = \rG^0(\rx_1,\rx_2,\rx_3,\rx_4) \\
&+& \int d^4\ry_1\, d^4\ry_2\, d^4\ry_3\, d^4\ry_4\,
 \rG^0(\rx_1,\rx_2,\ry_1,\ry_2) {\hat {\cal K}} (\ry_1,\ry_2,\ry_3,\ry_4) \rG(\ry_3,\ry_4,\rx_3,\rx_4)
\,\,\,.
\label{eq:green.dystbgf}
\eeqn

\begin{figure}[h]
\begin{center}
\includegraphics [scale=0.35, angle=0] {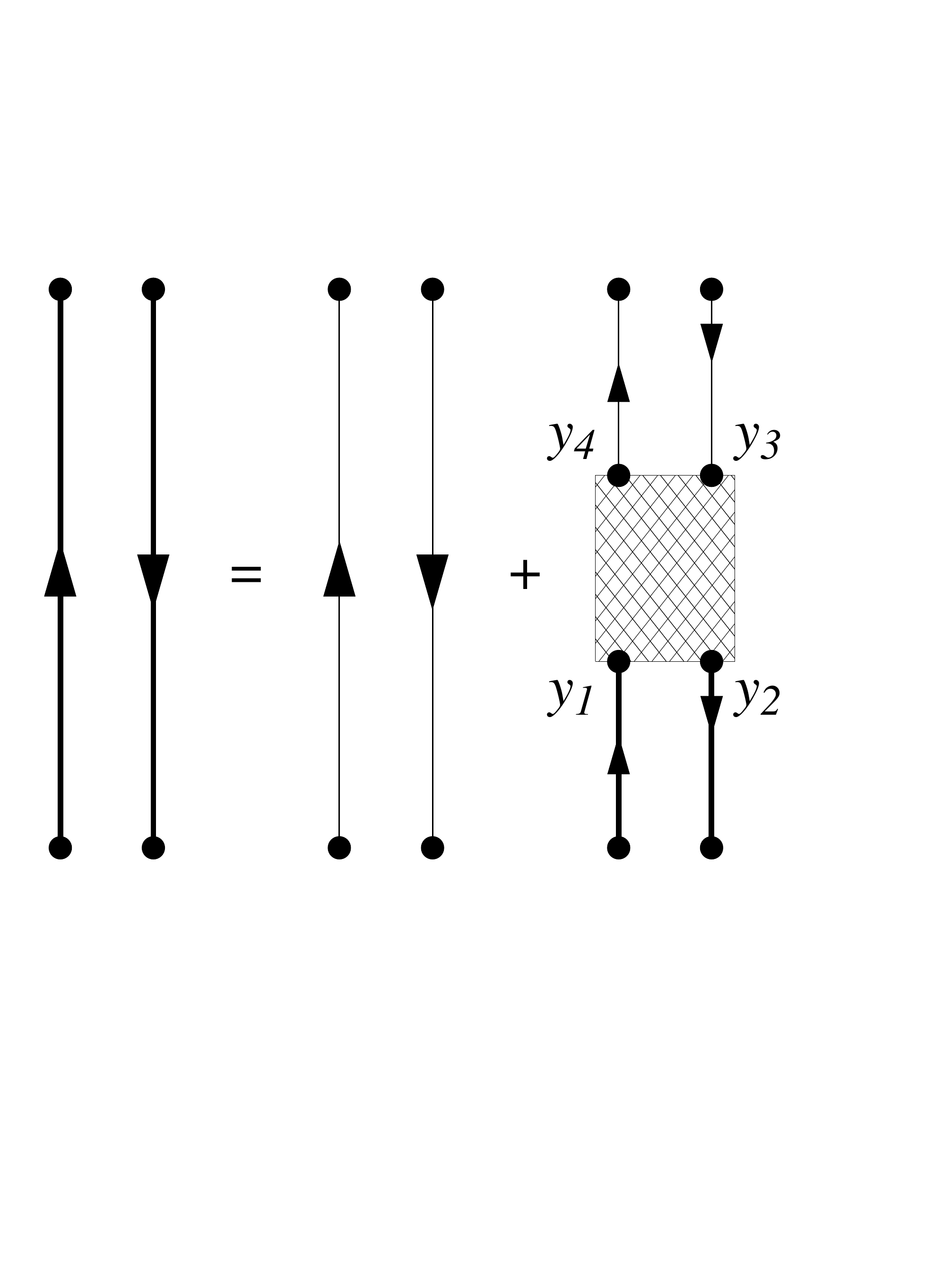}
\captionsetup{margin=2cm}
\caption{\small Graphical representation of Equation (\ref{eq:green.dystbgf}).
The criss-cross box represents the proper kernel ${\hat {\cal K}}$.
}
\label{fig:rpa3} 
\end{center}
\end{figure}

A graphical representation of Equation~(\ref{eq:green.dystbgf}) is shown in Figure~\ref{fig:rpa3}.
The dashed area indicates the kernel ${\hat {\cal K}}$ containing all the irreducible diagrams which can
be inserted between the four $y$ points.

%

\begin{figure}[h]
\begin{center}
\captionsetup{margin=2cm}
\includegraphics [scale=0.3, angle=90] {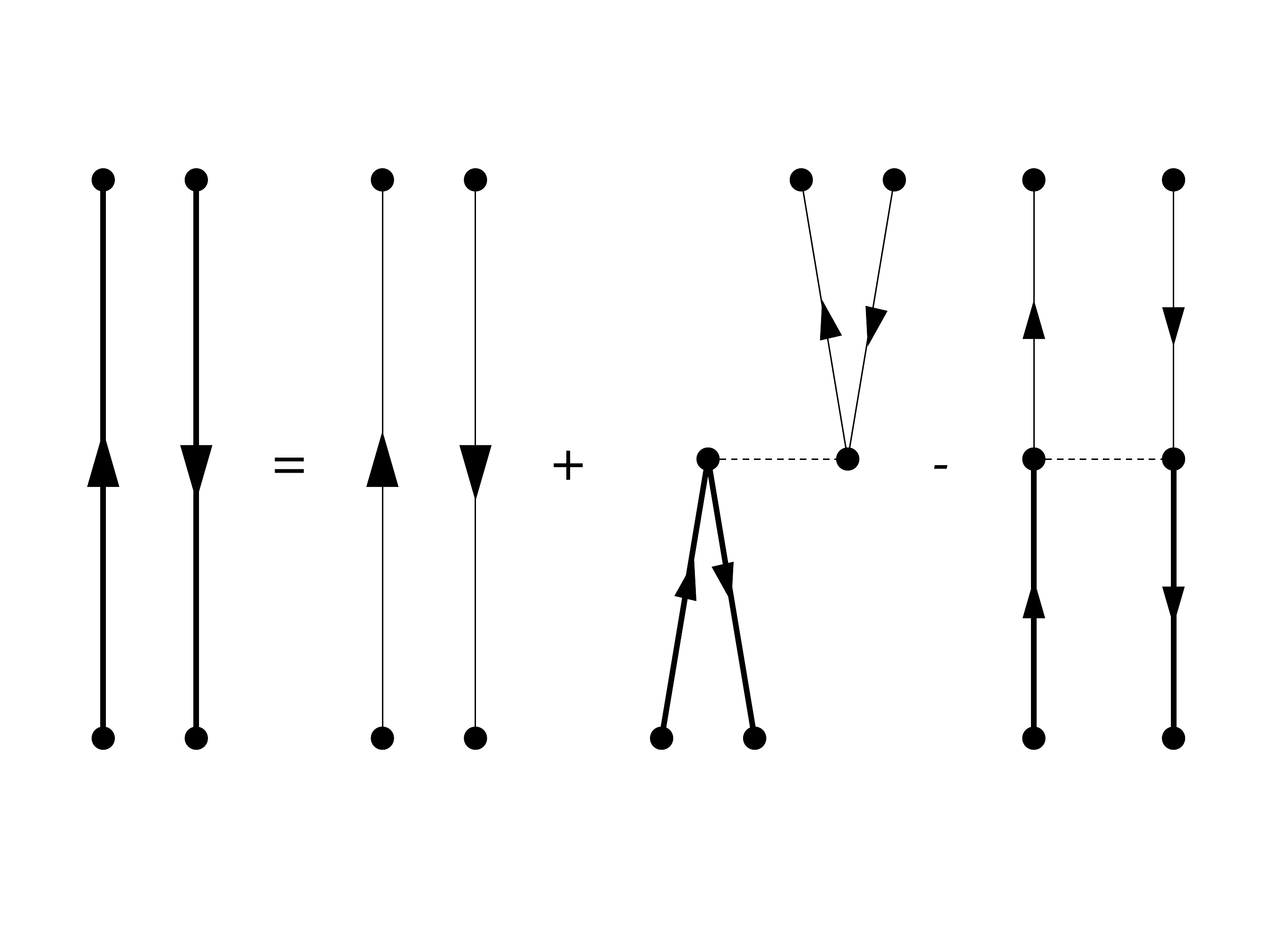}
\caption{\small Graphical representation of the two-body Green function in RPA.
} 
\label{fig:rpa4} 
\end{center}
\end{figure}

 RPA consists in considering, in the previous
equation, instead of the full kernel ${\hat {\cal K}}$, a single interaction $\intr$ 
depending only on two coordinates
\beq
{\hat {\cal K}}^{\rm RPA} (\ry_1,\ry_2,\ry_3,\ry_4) = 
\intr (\ry_1,\ry_4) \left[
\delta(\ry_1 - \ry_2) \delta(\ry_3 - \ry_4) - \delta(\ry_1 - \ry_3) \delta(\ry_2 - \ry_4)
\right]
\,\,\,;
\label{eq:rpa3}
\eeq
therefore,
\beqn
\nonumber 
&~& \rG^{\rm RPA}(\rx_1,\rx_2,\rx_3,\rx_4) = \rG^0(\rx_1,\rx_2,\rx_3,\rx_4) \\
\nonumber 
 &+& \int d^4\, \ry_1\, d^4\,\ry_2\, 
 \rG^0(\rx_1,\rx_2,\ry_1,\ry_1) \intr (\ry_1,\ry_2) \rG^{\rm RPA}(\ry_2,\ry_2,\rx_3,\rx_4) \\
 &-& \int d^4y_1\, d^4y_2\, 
 \rG^0(\rx_1,\rx_2,\ry_1,\ry_2) \intr (\ry_1,\ry_2) \rG^{\rm RPA}(\ry_1,\ry_2,\rx_3,\rx_4)
\,\,\,,
\label{eq:rpa4}
\eeqn
where we separated the direct and the exchange terms. The graphical representation
of the above equation is given in Figure \ref{fig:rpa4}.

In mixed representation, RPA equations are
\beqn
\nonumber &~& 
{\tilde G}^{\rm RPA}(\nu_1,\nu_2, \nu_3, \nu_4,E) = {\tilde G}^0(\nu_1,\nu_2, \nu_3, \nu_4,E) 
 \\ \nonumber &+& 
\sum_{\mu_1,\mu_2,\mu_3,\mu_4} {\tilde G}^0(\nu_1,\nu_2, \mu_1, \mu_2,E)
 \braket{\mu_1 \mu_3 | \hV | \mu_2 \mu_4} {\tilde G}^{\rm RPA}(\mu_3, \mu_4, \nu_3, \nu_4,E) \frac 1 \hbar
\\ \nonumber
&-& \sum_{\mu_1,\mu_2,\mu_3,\mu_4} {\tilde G}^0(\nu_1,\nu_2, \mu_1, \mu_2,E)
 \braket{\mu_1 \mu_2 | \hV | \mu_4 \mu_3} {\tilde G}^{\rm RPA}(\mu_3, \mu_4, \nu_3, \nu_4,E) \frac 1 \hbar
\\ \nonumber
&=&  \sum_{\mu_1,\mu_2,\mu_3,\mu_4} {\tilde G}^0(\nu_1,\nu_2, \mu_1, \mu_2,E) 
\Big\{ \delta_{\mu_1,\nu_3} \delta_{\mu_2,\nu_4}
\\ \nonumber
&+&  \frac 1 \hbar
\braket{\mu_1 \mu_3 | \hV | \mu_2 \mu_4} {\tilde G}^{\rm RPA}(\mu_3, \mu_4, \nu_3, \nu_4,E) 
 \\ 
&-& \frac 1 \hbar
\braket{\mu_1 \mu_2 | \hV | \mu_4 \mu_3} {\tilde G}^{\rm RPA}(\mu_3, \mu_4, \nu_3, \nu_4,E)
\Big\} 
\label{eq:RPA.grpa}
,
\eeqn
where we used $\intr = \hV / \hbar$. 

As indicated by Equations~(\ref{eq:green.gnot1}) and (\ref{eq:green.gnot2}),  
there are two possibilities of forming non-zero unperturbed
Green functions. By adopting 
the usual convention of indicating with $i,j,k,l$ the hole states and with 
$m,n,p,q$ the particle states, we express RPA Equation 
(\ref{eq:RPA.grpa}) as:
\beq
\sum_{q,l} \Big\{
\left[ A_{miql} - E\, \delta_{m,q} \, \delta_{i,l} \right] {\tilde G}^{\rm RPA}(q, l, j, n,E)
+ B_{miql}  {\tilde G}^{\rm RPA}(l,q, j, n,E)
\Big\} = \delta_{m,n} \delta_{i,j} 
,
\label{eq:RPA.grpa1}
\eeq
\beq
\sum_{q,l} \Big\{
\left[ A^*_{miql} + E \, \delta_{m,q} \delta_{i,l} \right] {\tilde G}^{\rm RPA}(l,q, j, n,E)
+ B^*_{miql}  {\tilde G}^{\rm RPA}(q, l, j, n,E)
\Big\} = 0
,
\label{eq:RPA.grpa2}
\eeq
\beq
\sum_{q,l} \Big\{ \left[ A_{miql}  - E) \delta_{m,q} \delta_{i,l}  \right] {\tilde G}^{\rm RPA}(q, l, n, j, E)
+ B_{miql}  {\tilde G}^{\rm RPA}(l,q, n, j, E) \Big\} = 0
,
\label{eq:RPA.grpa3}
\eeq
\beq
\sum_{q,l} \Big\{
\left[ A^*_{miql} + E \, \delta_{m,q} \delta_{i,l}  \right] {\tilde G}^{\rm RPA}(l,q, n, j,E)
+ B^*_{miql}  {\tilde G}^{\rm RPA}(q, l, n, j)
\Big\} = \delta_{m,n} \delta_{i,j} 
,
\label{eq:RPA.grpa4}
\eeq
where we have defined the matrices
\beqn
A_{miql} &=& (\epsilon_m - \epsilon_i) \delta_{m,q} \delta_{i,l}  + \barv_{iqml}
,
\\
B_{miql} &=& - \barv_{ilmq}
.
\eeqn
The  calculation is outlined in detail  in Appendix \ref{sec:app.gfmat}.
These equations can be expressed in matrix form. By defining
\beqn
\nonumber
&~& G_1(E) \equiv {\tilde G}^{\rm RPA}(m,i, j,n,E)\;\;;\;\;
G_2(E) \equiv {\tilde G}^{\rm RPA}(m,i,n,j,E) \;\;;\;\;
\\ &~&
G_3(E) \equiv {\tilde G}^{\rm RPA}(i,m, j,n,E)\;\;;\;\;
G_4(E) \equiv {\tilde G}^{\rm RPA}(i,m,n,j,E)
,
\eeqn
we obtain
\beq
\left(
\begin{array}{cc}
A - E\, {\mathbb I}& B \\
   & \\
B^*  & A^* + E \,{\mathbb I}
\end{array}
\right)
\left(  
\begin{array}{cc}
G_1(E) & G_2(E) \\
\\
G_3(E) & G_4(E) \\
\end{array}
\right) = 
\left(  
\begin{array}{cc}
{\mathbb I} & 0 \\
 & \\
0 & {\mathbb I}  \\
\end{array}
\right) 
\label{eq:arpa:matrx}
.
\eeq

The two-body Green functions depend on the energy $E$. The poles $\omega_n = E_n - E_0$
of these Green functions are the excitation energies of RPA excited states $\ket{\Psi_n}$.
When the value of the energy $E$ corresponds to that of a pole, the value of the
Green function goes to infinity; therefore, 
Equation (\ref{eq:arpa:matrx}) remains valid only if the matrix of the coefficients 
goes to zero. For this reason RPA 
excitation energies are those of the non-trivial 
solution of the homogeneous system of equations
\beq
\left(
\begin{array}{cc}
A -  \omega_n \, {\mathbb I}& B \\
   & \\
B^*  & A^* + \omega_n \,{\mathbb I}
\end{array}
\right)
\left(
\begin{array} {c} 
X_n \\
 \\
Y_n
\end{array}
 \right)  = 0
\ ,
 \eeq
which is the expression (\ref{eq:rpa:matrix}) of RPA equations. 

In Section \ref{sec:propeRPA}, we have shown that RPA equations for each 
positive eigenvalue $\omega_ n$ admit also a negative eigenvalue $-\omega_n$. 
The set of the vectors of the $X$ and $Y$ amplitudes is orthogonal
\[
\left( X^*_m , - Y^*_m \right) 
\left(
\begin{array} {c} 
X_n \\
 \\
 Y_n
\end{array}
 \right)  = \delta_{m,n}
,
\]
and complete
\[
\sum_{n>0}
\left(
\begin{array} {c} 
X_n \\
 \\
Y_n
\end{array} 
\right) 
\left( X^*_n ,  Y^*_n \right) 
-
\sum_{n<0}
\left(
\begin{array} {c} 
X^*_n \\
 \\
Y^*_n
\end{array} 
\right) 
\left( X_n ,  Y_n \right) 
= {\mathbb I}
,
\]
where $n>0$ indicates the sum on the positive $\omega_n$ and 
$n<0$ the sum on the negative values, as indicated by Equation~(\ref{eq:rpa:cmplx}).
By inserting the above expressions in Equation~(\ref{eq:arpa:matrx}), we identify the
solution as
\beqn
\nonumber
\left(  
\begin{array}{cc}
G_1(E) & G_2(E) \\
\\
G_3(E) & G_4(E) \\
\end{array}
\right) &=& 
\sum_{n}
\frac{1}{\omega_n -E}
\left(
\begin{array} {c} 
X_n \\
 \\
 Y_n
\end{array} 
 \right) 
\left( X^*_m ,  Y^*_m \right) 
\\ \nonumber 
&-&
\sum_n
\frac{1}{-|\omega_n| -E}
\left(
\begin{array} {c} 
Y^*_n \\
 \\
 X^*_n
\end{array} 
 \right) 
\left( Y_n ,  X_n \right) 
\\ \nonumber &~& \\ 
&=& \left(
\begin{array}{cc}
\sum_n \left( \frac{X_n X^*_n}{\omega_n -E} + \frac{Y_n Y^*_n}{\omega_n  + E}  \right) 
& 
\sum_n \left( \frac{X_n Y^*_n}{\omega_n -E} + \frac{X_n Y^*_n}{\omega_n  + E}  \right) \\
& \\
\sum_n \left( \frac{Y_n X^*_n}{\omega_n -E} + \frac{X^*_n Y_n}{\omega_n  + E}  \right) 
& 
\sum_n \left( \frac{Y_n Y^*_n}{\omega_n -E} + \frac{X_n X^*_n}{\omega_n  + E}  \right) 
\end{array}
\right)
,
\eeqn
where $\omega_n > 0$ always.
The comparison with the expression of the two-body Green function in the
representation of Equation~(\ref{eq:gtbgfl}) allows the identification of the $X$ and $Y$ 
amplitudes as 
\beqn
X_{mi} = \braket{\Psi_0 | \ha_m \ha^+_i | \Psi_n} &\;\;\;;\;\;\;& 
X^*_{mi} = \braket{\Psi_n | \ha_i \ha^+_m | \Psi_n} \;\;; \\
Y_{mi} = \braket{\Psi_0 | \ha_i \ha^+_m | \Psi_n} &\;\;\;;\;\;\;& 
Y^*_{mi} = \braket{\Psi_n | \ha_m \ha^+_i | \Psi_0} ,
\label{eq:perXY}
\eeqn 
where $\ket{\Psi_n}$ and $\ket{\Psi_0}$ are, respectively, RPA excited and ground
states, which we called $\ket{\nu}$ and $\ket{\nu_0}$ in Section~\ref{sec:RPA.rpa}.

\subsubsection*{Infinite Systems}
\label{sec.green.infinite}

In an infinite and homogeneous system with translational invariance, the s.p. wave functions 
are the plane waves (\ref{eq:mf.pw}) characterized by the modules of the wave vector
$k\equiv|\bk|$. 
If we use the representation of Equation (\ref{eq:gtbgfl}) of the unperturbed two-body Green
function, we obtain terms of the kind
\beq
\langle \Phi_0 |  \ha_{\nu_1}  \ha^+_{\nu_3}  |\Phi_n \rangle 
          \langle \Phi_n | \, \ha_{\nu_2} \ha^+_{\nu_4} |\Phi_0 \rangle = 
\delta_{k_1,k_4}\theta(k_1 - \kfermi) \,
\delta_{k_2,k_3}\theta(\kfermi- k_2) 
,
\eeq
since the action of the creation and destruction operators on the IPM states $\ket{\Phi}$
is well defined. 

We consider Green functions depending on the energy, as indicated by 
Equation~(\ref{eq:green.fourier}). In this representation, by inserting the plane wave function
in Equation~(\ref{eq:tbgf4}), we obtain for the unperturbed two-body Green function the expression
 \beq
\tilde{\rG}^0(\bx_1,\bx_2,\bx_3,\bx_4;E) = \frac{1}{(2 \pi)^6}
\int d \bk_1\,d \bk_2 e^{i \bk_1 \cdot (\bx_1 - \bx_4)} e^{-i\bk_2 \cdot (\bx_2 - \bx_3)}
\tg^0(k_1, k_2;E) 
,
\eeq
clearly indicating that there is a dependence only on the difference between the particle and 
hole coordinates. This is a consequence of the translational invariance of the system. 
The interacting Green function and the kernel of Equation (\ref{eq:green.dystbgf})
depend only on the coordinate difference. 
We can define the Fourier transforms of these quantities depending on 
the modules of two momenta
\beqn
\tilde{\rG}(\bx_1,\bx_2,\bx_3,\bx_4; E) &=& \frac{1}{(2 \pi)^6}
\int d \bk_1\,d \bk_2 e^{i\bk_1 \cdot (\bx_1 - \bx_4)} e^{i\bk_2 \cdot (\bx_2 - \bx_3)}
\tg(k_1,k_2; E) 
\,\,\,,\\
{\hat {\cal K}}(\bx_1,\bx_2,\bx_3,\bx_4) &=& \frac{1}{(2 \pi)^6}
\int d \bk_1\,d \bk_2 e^{i\bk_1 \cdot (\bx_1 - \bx_4)} e^{-i\bk_2 \cdot (\bx_2 - \bx_3)}
{\hat {\tilde {\cal K}}}(k_1,k_2) 
\,\,\,;
\label{eq:rpa5}
\eeqn
the kernel does not depend on the energy $E$.

\begin{figure}[h]
\begin{center}
\captionsetup{margin=2cm}
\includegraphics [scale=0.3, angle=90] {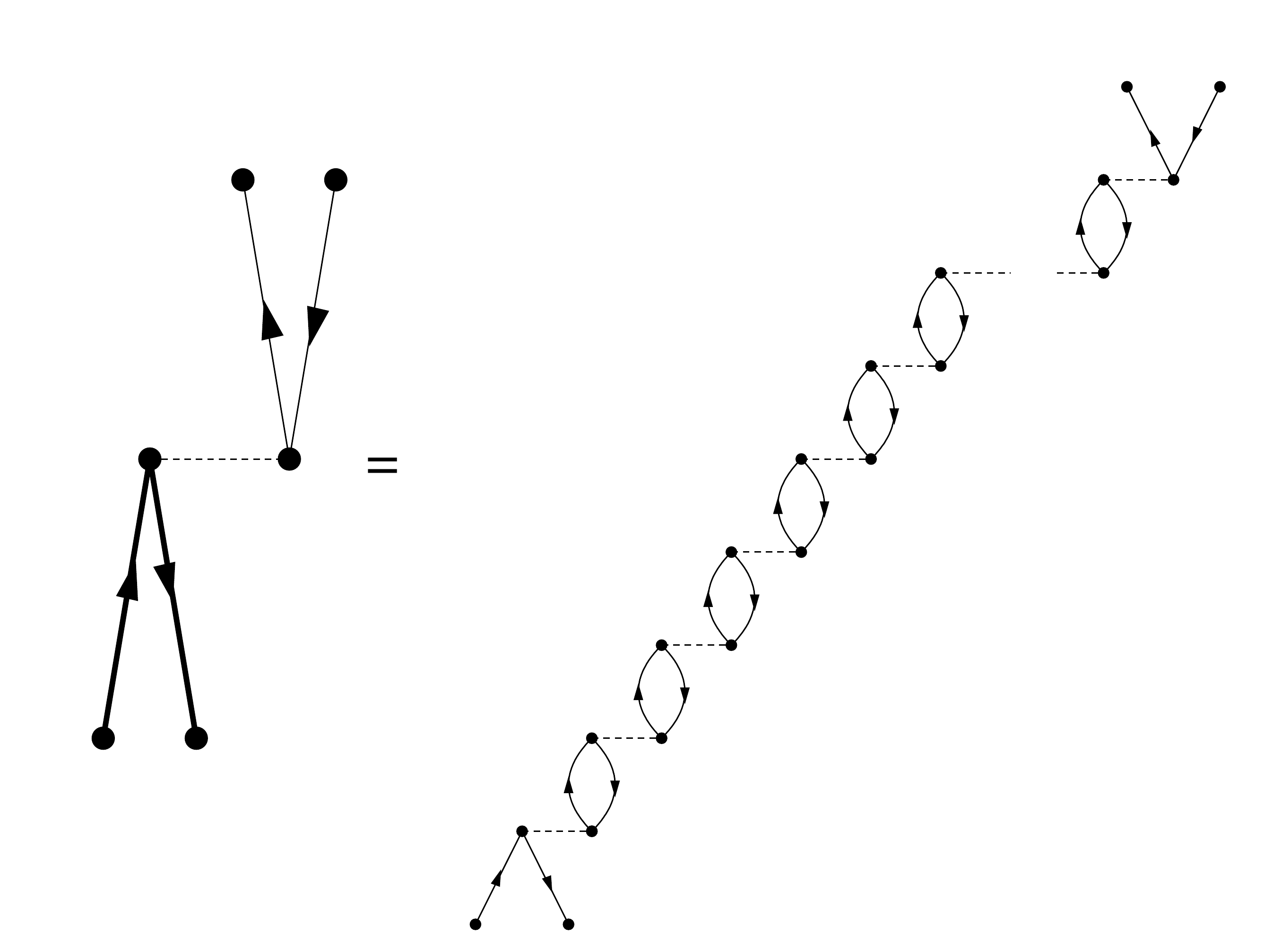}
\caption{\small Graphical representation of the ring diagram approximation  in RPA .
}
\label{fig:rpa5} 
\end{center}
\end{figure}

By inserting these definitions in RPA Equation (\ref{eq:rpa4}) and substituting 
${\hat {\cal K}}$ with $\intr$, which is RPA ansatz,  we obtain
\beqn
\nonumber
&~& \int d \bk_1\,d \bk_2 e^{i\bk_1 \cdot (\bx_1 - \bx_4)} 
e^{-i\bk_2 \cdot (\bx_2 - \bx_3)} \tg^{\rm RPA}(k_1,k_2; E) = \\
\nonumber
&~& \int d \bk_1\,d \bk_2 e^{i\bk_1 \cdot (\bx_1 - \bx_4)} e^{-i\bk_2 \cdot (\bx_2 - \bx_3)} \tg^0(k_1,k_2; E) \\
\nonumber
&+& \int d \by_1  d \by_2   d \by_3   d \by_4
\int d \bk_1\,d \bk_2 e^{i\bk_1 \cdot (\bx_1 - \by_2)} e^{-i\bk_2 \cdot (\bx_2 - \by_1)} \\
\nonumber
&~& \tg^0(k_1,k_2; E) 
\int d \bk_a\,d \bk_b e^{i\bk_a \cdot (\by_1 - \by_4)} e^{-i\bk_b \cdot (\by_2 - \by_3)}  
{\hat {\tilde {\cal U}}} (k_a,k_b) \\
\nonumber
&~& \int d \bk_3\,d \bk_4 e^{i\bk_3 \cdot (\by_3 - \bx_4)} e^{-i\bk_4 \cdot (\by_4 - \bx_3)} \tg^{\rm RPA}(k_3,k_4; E) 
\label{eq:rpa.diretto}\\
\nonumber
&+& \int d \by_1  d \by_2  d \by_3  d \by_4
\int d \bk_1\,d \bk_2 e^{i\bk_1 \cdot (\bx_1 - \by_2)} e^{-i\bk_2 \cdot (\bx_2 - \by_1)} \\
\nonumber
&~& \tg^0(k_1,k_2; E) 
\int d \bk_a\,d \bk_b e^{i\bk_a \cdot (\by_1 - \by_4)} e^{-i\bk_b \cdot (\by_2 - \by_3)}  
{\hat {\tilde {\cal U}}} (k_a,k_b) \\  
\nonumber
&~& \int d \bk_3\,d \bk_4 e^{i\bk_3 \cdot (\by_4 - \bx_4)} e^{-i\bk_4 \cdot (\by_3 - \bx_3)} \tg^{\rm RPA}(k_3, k_4; E) 
\label{eq:rpa.scambio}
,
\eeqn
where the second term of the right hand side is called direct 
and the third term is the exchange. 
The integration on the $\by$ coordinates in the direct term leads to the relations
\beq
- \bk_a = \bk_2 = \bk_4\;\;\;{\rm and}\;\;\; - \bk_b = \bk_1 =\bk_3
,
\eeq
while that of the exchange term leads to 
\beq
 \bk_a = \bk_2 = - \bk_3\;\;\;{\rm and}\;\;\;  \bk_b = \bk_1 = - \bk_4
.
\eeq
By considering the above conditions, we have 
\beqn
\nonumber
&~& \int d \bk_1\,d \bk_2 e^{i\bk_1 \cdot (\bx_1 - \bx_4)} e^{-i\bk_2 \cdot (\bx_2 - \bx_3)} 
\tg^{\rm RPA}(k_1,k_2; E) = \\
\nonumber
&~& \int d \bk_1\,d \bk_2 e^{i\bk_1 \cdot (\bx_1 - \bx_4)} e^{-i\bk_2 \cdot (\bx_2 - \bx_3)} \tg^0(k_1,k_2; E) \\
\nonumber
&+& \int d \bk_1\,d \bk_2 e^{i\bk_1 \cdot (\bx_1 - \bx_4)} e^{-i\bk_2 \cdot (\bx_2 - \bx_3)}
 \tg^0(k_1,k_2; E)  {\hat {\tilde {\cal U}}} (k_1,k_2) \tg^{\rm RPA}(k_1,k_2; E) 
\label{eq:rpa.diretto1}\\
\nonumber
&+& \int d \bk_1\,d \bk_2 e^{i\bk_1 \cdot (\bx_1 + \bx_3)} e^{-i\bk_2 \cdot (\bx_2 + \bx_4)}
 \tg^0( k_1, k_2; E)  {\hat {\tilde {\cal U}}}( k_2, k_1) \tg^{\rm RPA}(k_2, k_1; E) 
\label{eq:rpa.scambio1}
.
\eeqn
If we neglect the exchange term, we have a simple algebraic 
relation between the Green functions in momentum space
\beqn
&~& \nonumber
\tg^{\rm RPA,D} (k_1,k_2; E) = \tg^0(k_1,k_2; E) + \tg^0(k_1,k_2; E) 
{\tilde \intr}(k_1-k_2=q) \tg^{\rm RPA,D} (k_1,k_2; E)  ,
\\ \nonumber
&~& \tg^{\rm RPA,D}(k_1,k_1+q; E)  \left[1 -  \tg^0(k_1,k_1+q; E)  {\tilde \intr}(q) \right] 
= \tg^0 (k_1,k_1+q; E) ,
\\ 
&~& \tg^{\rm RPA,D} (k_1,k_1+q; E) = 
\frac {\tg^0(k_1,k_1+q; E)}
       {1 -  \tg^0(k_1,k_1+q; E)  {\tilde \intr}(q) }
.
\label{eq:ring}       
\eeqn
This  expression is commonly used to calculate the linear response
of infinite fermion systems to an external probe. 
The graphical representation of the above equation is given in Figure~\ref{fig:rpa5}. 
The equation represents an infinite sum of 
diagrams of ring form and it is, therefore, called ring approximation.

The exchange term cannot be factorized as the direct one. The inclusion of this term 
is sometimes treated by  using  approximated treatments. One of them is the 
continuous fraction technique \cite{dep98,mar08}.

\section{RPA with Time-Dependent Hartree--Fock}
\label{sec:TDHF}

Another way of obtaining RPA secular Equation (\ref{eq:rpa:matrix}) is that of using time-dependent
Hartree--Fock (TDHF) equations and the variational principle. 
We apply the variational principle to the time-dependent Schr\"odinger  equation 
\beq
\delta \braket{\Psi(t) | \left( \hH - i \hbar \frac{\partial}{\partial t}  \right) | \Psi(t)} = 0
\label{eq:D:var1}
.
\eeq
The search for the minimum of the above functional of $\Psi(t)$ is carried out
in the Hilbert subspace spanned by many-body wave functions of the form
\beq
\ket{\Psi(t)} = e^{\displaystyle{\sum_{mi} }C_{mi} (t) \ha^+_m \ha_i} 
\,\ket{\Phi_0(t)}
\label{eq:D:ket}
,
\eeq
where the time-dependent IPM state has been defined as
\beq
\ket{\Phi_0(t)} = e ^{-\ih \tene_0 t} \ket{\Phi_0}
.
\eeq
In the above equation,  $\ket{\Phi_0}$ is the stationary Hartree--Fock ground state of which $\tene_0$, \linebreak 
Equation~(\ref{eq:v.ezero}), is the energy eigenvalue. In Equation (\ref{eq:D:var1}), 
the variation of  the real and of the imaginary part of $\Psi(t)$ are independent.
The variation is carried on
the only time dependent terms which are $C_{mi}(t)$ and $C^*_{mi}(t)$.
We obtain a system composed by the equations
\beqn
\frac {\delta}{\delta C^*_{mi}(t)} 
\braket{\Psi(t) | \hH - i \hbar \frac{\partial}{\partial t} | \Psi(t)} &=& 0 
,
\label{eq:D:first}\\
\frac {\delta}{\delta C_{mi}(t)} 
\braket{\Psi(t) | \hH - i \hbar \frac{\partial}{\partial t} | \Psi(t)}
&=& 0
\label{eq:D.var2}
.
\eeqn
We consider Equation (\ref{eq:D:first}) and calculate the expectation value of operators
by using the power expansion of the exponential 
\beq
e^{\displaystyle{\sum_{mi} }C_{mi} (t) \ha^+_m \ha_i} 
= \hI + \sum_{mi} C_{mi} (t) \ha^+_m \ha_i
+ \half \sum_{minj} C_{mi} (t) \ha^+_m \ha_i C_{nj} (t) \ha^+_n \ha_j
+ \cdots
,
\eeq
For the hamiltonian expectation value, we obtain the expression
\beqn
\nonumber
\braket{\Psi(t) | \hH | \Psi(t)} &=& 
\braket{\Phi_0(t) | \hH | \Phi_0(t)} 
\\ \nonumber &+& 
\sum_{mi} C^*_{mi}(t) \braket{\Phi_0(t) | \ha^+_i \ha_m \hH | \Phi_0(t)} 
\\ \nonumber &+& 
\sum_{mi} C_{mi}(t) \braket{\Phi_0(t) | \hH \ha^+_m \ha_i | \Phi_0(t)} 
\\ \nonumber &+& 
\half \sum_{minj} C^*_{mi}(t) C^*_{nj}(t) 
\braket{\Phi_0(t) | \ha^+_j \ha_n \ha^+_i \ha_m \hH | \Phi_0(t)} 
\\ \nonumber &+& 
\half \sum_{minj} C_{mi}(t) C_{nj}(t) 
\braket{\Phi_0(t) | \hH \ha^+_m \ha_i \ha^+_n \ha_j | \Phi_0(t)} 
\\ &+& 
\sum_{minj} C^*_{mi}(t) C_{nj}(t) 
\braket{\Phi_0(t) | \ha^+_i \ha_m \hH  \ha^+_n \ha_j \hH | \Phi_0(t)} 
+ \cdots
.
\label{eq:D:ham1}
\eeqn

The first term of the above equation is the $\tene_0$ of Equation~(\ref{eq:v.ezero}).
The linear terms in $C_{mi}(t)$ are all zero since they  overlap with orthogonal
Slater determinants, or, in other words, since the number of $ph$ operators is odd. 

Let us calculate the matrix element of the fifth term by using the expression of the 
hamiltonian given in Equation~(\ref{eq:hf.ham4})
\beqn
\nonumber
&~&
\braket{\Phi_0(t) |\hH \ha^+_m \ha_i \ha^+_n \ha_j |  \Phi_0(t)} =
\sum_\nu \epsilon_\nu \braket{\Phi_0(t) | \ha^+_\nu \ha_{\nu} 
\ha^+_m \ha_i \ha^+_n \ha_j | \Phi_0(t)} 
\\ \nonumber 
&-& \half \sum_{k l} \barv_{k l k l}
\braket{\Phi_0(t) | \ha^+_m \ha_i \ha^+_n \ha_j | \Phi_0(t)} 
\\ 
&+&
\frac 1 4 \sum_{\mu \mu' \nu \nu'} \barv_{\nu \mu \nu' \mu'}
\braket{\Phi_0(t) |
\hN[\ha^+_\nu \ha^+_\mu \, \ha_{\mu'} \, \ha_{\nu'}] 
\ha^+_m \ha_i \ha^+_n \ha_j 
| \Phi_0(t)} 
\,\,\,.
\eeqn
The first and second terms are zero because of the orthogonality of the Slater determinants. 
With a calculation analogous to that leading to the interacting term of  $A_{minj}$ in
\mbox{Equation (\ref{eq:rpa:amnij})} (this calculation is  presented in Equation (\ref{eq:app:em1})), we obtain
\beq
\frac 1 4 \sum_{\mu \mu' \nu \nu'} \barv_{\nu \mu \nu' \mu'}
\braket{\Phi_0(t) |
\hN[\ha^+_\nu \ha^+_\mu \, \ha_{\mu'} \, \ha_{\nu'}] 
\ha^+_m \ha_i \ha^+_n \ha_j 
| \Phi_0(t)} 
= \barv_{i j m n}
.
\eeq
The fifth term of Equation (\ref{eq:D:ham1}) can be written as 
\beq
\half \sum_{minj} C_{mi}(t) C_{nj}(t) 
\braket{\Phi_0(t) | \hH \ha^+_m \ha_i \ha^+_n \ha_j | \Phi_0(t)} 
= \half \sum_{minj} C_{mi}(t) C_{nj}(t) \barv_{i j m n}
.
\eeq
By working in an analogous manner for the fourth term of Equation (\ref{eq:D:ham1}), we obtain
\beq
\half \sum_{minj} C^*_{mi}(t) C^*_{nj}(t) 
\braket{\Phi_0(t) | \ha^+_j \ha_n \ha^+_i \ha_m \hH  | \Phi_0(t)} 
= \half \sum_{minj} C^*_{mi}(t) C^*_{nj}(t) \barv_{m n i j}
.
\eeq
The expression of the last term of Equation~(\ref{eq:D:ham1}) is 
\beqn
\nonumber
&~& \sum_{minj} C^*_{mi}(t) C_{nj}(t) 
\braket{\Phi_0(t) | \ha^+_i \ha_m \hH  \ha^+_n \ha_j \hH | \Phi_0(t)} 
\\ \nonumber &=&
\sum_{mi} | C_{mi} |^2 \sum_k \epsilon_k +
\sum_{mi} | C_{mi} |^2 (\epsilon_m - \epsilon_i)
\\ \nonumber &~&
- \half \sum_{mi} | C_{mi} |^2  \sum_{kl }\barv_{k l k l}
+  \sum_{minj}  C_{mi} C^*_{nj} \barv_{m j i n}
\\ &\equiv&
\tene_0 \sum_{mi} | C_{mi}|^2 
+ \sum_{mi} | C_{mi} |^2 (\epsilon_m - \epsilon_i)
+   \sum_{minj}  C_{mi} C^*_{nj} \barv_{m j i n}
,
\eeqn
where we used the definition (\ref{eq:v.ezero}) of $\tene_0$. 
The final expression of Equation~(\ref{eq:D:ham1}) is then
\beqn
\nonumber
\braket{\Psi(t) | \hH | \Psi(t)} &=& 
\tene_0 \left( 1 + \sum_{mi} | C_{mi}|^2 \right) 
+ \sum_{mi} | C_{mi} |^2 (\epsilon_m - \epsilon_i)
+ \sum_{minj}  C_{mi} C^*_{nj} \barv_{m j i n}
\\ &~& 
+ \half \sum_{minj} C_{mi}(t) C_{nj}(t) \barv_{i j m n}
+ \half \sum_{minj} C^*_{mi}(t) C^*_{nj}(t) \barv_{m n i j}
.
\label{eq:D:ham2}
\eeqn

Let us calculate the second term of Equation~(\ref{eq:D:first}),
containing the time derivation. 
By considering the expression (\ref{eq:D:ket}) of $\ket{\Psi(t)}$,
we have
\beq
 i \hbar \braket{\Psi(t) |\frac{\partial}{\partial t} | \Psi(t)} =
\tene_0 \braket{\Psi(t) | \Psi(t)}
+ i \hbar \sum_{mi}
\frac{d}{dt} C_{mi}(t)  \braket{\Psi(t)| \ha^+_m \ha_i | \Psi(t)}
.
\label{eq:D:dert}
\eeq

We make a power expansion of the exponential function in Equation~(\ref{eq:D:ket}) and obtain
\beqn
\nonumber
\braket{\Psi(t) | \Psi(t)} &=&
\braket{\Phi_0(t) | \Phi_0(t)} \\
&+&
\sum_{minj} C^*_{mi}(t) C_{nj}(t) 
\braket{\Phi_0(t) | \ha^+_i \ha_m \ha^+_n \ha_j | \Phi_0(t)} 
+ \cdots
,
\eeqn
and after the application of the Wick's theorem
\beq
\braket{\Psi(t) | \Psi(t)} = 1 +
\sum_{mi} |C_{mi}(t)|^2 + \cdots
.
\eeq

The terms of first order in $C$ are zero because of the odd number of 
$ph$ excitation pairs. 
By using the power expansion of the exponential to calculate the second
term of Equation~(\ref{eq:D:dert}), we have 
\beq
\braket{\Psi(t)| \ha^+_m \ha_i | \Psi(t)} = 
\sum_{nj} C^*_{nj} 
\braket{\Phi_0(t) | \ha^+_j \ha_n \ha^+_m \ha_i | \Phi_0(t)} + \cdots
 =  C^*_{mi}  + \cdots
.
\eeq
The term related to the time derivative becomes
\beq
 i \hbar \braket{\Psi(t) |\frac{\partial}{\partial t} | \Psi(t)} =
\tene_0 \left( 1 + \sum_{mi} |C_{mi}(t)|^2  \right)  + 
i \hbar \sum_{mi} C^*_{mi}(t) \frac{d}{d t}  C_{mi}(t)
+ \cdots
.
\label{eq:D:dert1}
\eeq
We put together the results of Equations~(\ref{eq:D:ham2}) and (\ref{eq:D:dert1}); we
consider terms up to the second order in $C$ and obtain the expression
\beqn
\nonumber 
\braket{\Psi(t) | \hH - i \hbar \frac{\partial}{\partial t} | \Psi(t)} &\simeq&
\sum_{mi} |C_{mi}(t)|^2 (\epsilon_m - \epsilon_i)
 + \sum_{minj}  C_{mi} C^*_{nj} \barv_{m j i n}
\\ \nonumber &+&
 \half \sum_{minj} C_{mi}(t) C_{nj}(t) \barv_{i j m n}
+ \half \sum_{minj} C^*_{mi}(t) C^*_{nj}(t) \barv_{m n i j}
\\  &-&
 i \hbar \sum_{mi} C^*_{mi}(t) \frac{d}{d t} C_{mi}(t) 
 .
\eeqn

We have to impose the variational condition
\beq
\frac {\delta}{\delta C^*_{mi}(t)} 
\braket{\Psi(t) | \hH - i \hbar \frac{\partial}{\partial t} | \Psi(t)} 
= 
\frac {\partial}{\partial C^*_{mi}(t)} 
\braket{\Psi(t) | \hH - i \hbar \frac{\partial}{\partial t} | \Psi(t)} 
= 0 
,
\eeq
where the variational derivative has been changed in partial derivatives
since the $C$'s are the only terms depending on time. By working out the derivative, 
we obtain the expression
\beq
C_{mi}(t) (\epsilon_m - \epsilon_i)
+  \sum_{nj} C^*_{nj} \barv_{m n i j}
+  \sum_{nj} C_{nj} \barv_{m j i n}
=  i \hbar \sum_{mi} \frac{d}{d t} C_{mi}(t) 
\label{eq:D:main1}
.
\eeq

We consider harmonic oscillations around the ground state 
\beq
C_{mi}(t) = X_{mi} e^{-i \omega t} + Y_{mi} e^{ i \omega t}
.
\eeq
where $X$, $Y$ and $\omega$ are real numbers. 
By inserting this expression in Equation~(\ref{eq:D:main1}) and separating the
positive and negative frequencies, we obtain the system of equations
\beqn
X_{mi} (\epsilon_m - \epsilon_i) 
+ \sum_{nj} \barv_{m j i n} X_{nj} 
+ \sum_{nj} \barv_{m n i j} Y_{nj} &=& \hbar \omega X_{mi} 
,
\\
Y^*_{mi} (\epsilon_m - \epsilon_i) 
+ \sum_{nj} \barv_{m j i n} Y^*_{nj} 
+ \sum_{nj} \barv_{m n i j} X^*_{nj} &=& - \hbar \omega^* X_{mi} 
,
\eeqn
which is identical to Equation~(\ref{eq:rpa:matrix}) where the $A$
and $B$ matrices have been defined by Equations~(\ref{eq:rpa:amnij}) and 
(\ref{eq:rpa:bmnij}).

\section{Continuum RPA}
\label{sec:CRPA}

If the excitation energy $\omega$ of the system is larger than 
$|\epsilon_h|$, the particle lying on this 
state can be emitted and leave the system. In an atom, this effect produces
a positive ion, in a nucleus a new nucleus with $A-1$ nucleons. The 
RPA approach which explicitly considers the emission of a particle is called
Continuum RPA (CRPA), where continuum  refers to the fact that for $\epsilon_p > 0$
the IPM Schr\"odinger equation has a continuous spectrum.  
In this case, the s.p. wave function has an asymptotically oscillating behavior. 

In CRPA, the operator (\ref{eq:rpa:rpaq}) defining the excited state 
is written as
\beq
\hQ^\dag_\nu \, = \, \sum_{[p] h} \, \sumint_{\epsilon_p} \, 
\left[X^\nu_{ph}(\epsilon_p)\, \ha^\dag_p(\epsilon_p)\, \ha_h \,
     - \, Y^\nu_{ph}(\epsilon_p)\, \ha^\dag_h \, \ha_p(\epsilon_p) \right] \, ,
\label{eq:qnu}
\eeq
where we have introduced the symbol 
\beq 
\sumint_{\epsilon_p} \,
\equiv \, \sum_{\epsilon_{\rm F} \le \epsilon_p \le 0} \, + \,
\int_0^\infty {\rm d} \epsilon_p 
\label{eq:crpa.simbol}
\eeq
to indicate a sum on the discrete energies and an integral on 
the continuum part of the spectrum. The symbol $[p]$ indicates the set 
of quantum numbers characterizing the particle state with the exclusion
of the energy.

 RPA secular Equation (\ref{eq:rpa:matrix}) with the continuum can be written as 
\beqn
(\epsilon_p-\epsilon_h-\omega)\, X_{ph}^\nu(\epsilon_p)\,+ \nonumber   
&& \\  & &
\hspace*{-3.5cm}
\sum_{[p']h'} \, \sumint_{\epsilon_{p'}} \,
\left[ \barv_{p h' h p'}(\epsilon_p,\epsilon_{p'})\, X_{p'h'}^\nu(\epsilon_{p'})\,
   + \, \barv_{p p' h h'}(\epsilon_p,\epsilon_{p'}) \, Y_{p'h'}^\nu(\epsilon_{p'}) 
     \right] \, = \, 0 \,,
\label{eq:crpa.crpa1}
\\
(\epsilon_p-\epsilon_h+\omega) \, Y_{ph}^{\nu}(\epsilon_{p})\,+ \nonumber 
&& \\ & &
\hspace*{-3.5cm}
\sum_{[p']h'} \, \sumint_{\epsilon_{p'}}\, 
\left[
\barv_{h p p' h'}  (\epsilon_p,\epsilon_{p'})\ Y_{p'h'}^{\nu}(\epsilon_{p}) \,
 + \,  \barv_{h h' p p'} (\epsilon_p,\epsilon_{p'}) \, 
X_{p'h'}^{\nu}(\epsilon_{p'})
   \right] \, = \, 0
\,.
\label{eq:crpa.crpa2}
\eeqn
where we have explicitly written the dependences on the particle energies
which are now continuous variables. 

In order to discuss the implications related
to the fact that $\epsilon_p$ can assume a continuous set of values, 
it is useful to express the  $X$ and $Y$ RPA amplitudes as:
\beqn
X^\nu_{ph} (\epsilon_p) &=&  A^\nu_{ph} \delta(\epsilon_p - \epsilon_h -\omega)
+ {\cal P} \frac{B^{X,\nu}_{ph} (\epsilon_p)}{\epsilon_p - \epsilon_h -\omega} 
,
\\ 
Y^\nu_{ph} (\epsilon_p) &=& 
\frac{B^{Y,\nu}_{ph} (\epsilon_p)}{\epsilon_p - \epsilon_h + \omega} 
.
\eeqn
When 
$\epsilon_p$ assumes the value
$\epsilon_p = \epsilon_h + \omega$,
in the integrals of (\ref{eq:crpa.crpa1}) and  (\ref{eq:crpa.crpa2}), 
the $X$ amplitudes have a pole. In the above expression, the contribution
of the pole, multiplied by a constant $A^\nu_{ph}$, is separated 
 from the principal part, indicated by  ${\cal P}$. 

The CRPA secular equations in terms of the new unknown can be 
written as
%
\beqn
\nonumber   
 B^{X,\nu}_{ph}(\epsilon_p) &+&
\sum_{[p']h'} \, \sumint_{\epsilon_{p'}} \,
\left[ \barv_{p h' h p'}(\epsilon_p,\epsilon_{p'})\, 
  \frac{B^{X,\nu}_{p'h'}(\epsilon_{p'})}{\epsilon_{p'} - \epsilon_{h'} - \omega}
   + \, \barv_{p p' h h'}(\epsilon_p,\epsilon_{p'}) \, 
   \frac{B^{Y,\nu}_{p'h'}(\epsilon_{p'})}{\epsilon_{p'} - \epsilon_{h'} + \omega}
     \right] 
\\     &=& 
      - \sum_{[p']h'} \barv_{p h' h p'}(\epsilon_p,\epsilon_{p'}) A^\nu_{p'h'}
       \delta(\epsilon_{p'} - \epsilon_{h'} -\omega)
\;,
\label{eq:bcrpa1}
\eeqn
\beqn
\nonumber
B^{Y,\nu}_{ph}(\epsilon_p) &+&
\sum_{[p']h'} \, \sumint_{\epsilon_{p'}}\, 
\left[
\barv_{h p p' h'}  (\epsilon_p,\epsilon_{p'})
  \frac{B^{Y,\nu}_{p'h'}(\epsilon_{p'})}{\epsilon_{p'} - \epsilon_{h'} + \omega}
 + \,  \barv_{h h' p p'} (\epsilon_p,\epsilon_{p'}) \, 
  \frac{B^{X,\nu}_{p'h'}(\epsilon_{p'})}{\epsilon_{p'} - \epsilon_{h'} - \omega}
   \right] 
   \\ &=& 
      - \sum_{[p']h'} \barv_{h p p' h'}(\epsilon_p,\epsilon_{p'}) 
      A^\nu_{p'h'}
       \delta(\epsilon_{p'} - \epsilon_{h'} -\omega)
       \;.
\label{eq:bcrpa2}
\eeqn
The above equations indicate a system of linear equations whose unknowns are the $B$'s.

The continuum threshold, $\omega_{\rm thr}$, is the minimum value of the energy 
necessary to emit the particle,  i.e., the absolute value of the s.p. energy of the hole state closest to the 
Fermi surface.
For $\omega < \omega_{\rm thr}$, no particle can be emitted. In this case, all the $A^\nu_{p'h'}=0$;
 therefore, the system is homogeneous. 
Solutions, different from the trivial one, are obtained 
when the determinant of the known coefficients is zero. 
This happens for some specific values of the excitation energy $\omega$. 
Below the emission threshold, the CRPA equations predict a discrete spectrum of solutions. 
When $\omega > \omega_{\rm thr}$, some $ph$ pairs have enough
energy to put the particle in the continuum, i.e., with $\epsilon_p > 0$. In the CRPA jargon these
$ph$ pairs are called \emph{open channels}. Obviously, the other $ph$ pairs where $\epsilon_p < 0$
are called \emph{closed channels}. Every open channel generates a coefficient different from zero 
in the right hand side of Equations (\ref{eq:bcrpa1}) and (\ref{eq:bcrpa2}). 
The problem is defined by imposing boundary conditions, which is equivalent to saying  
that we have to select specific values of the $A^\nu_{ph}$ coefficients. 
The choice commonly adopted consists in imposing
that the particle is emitted in a specific open channel, called \emph{elastic channel}. 
This means
\beq
A^\nu_{ph} = \delta_{p,p_0} \delta_{h,h_0}
,
\eeq
where $p_0$ and $h_0$ are the quantum numbers characterizing the elastic channel. 
The sums in  terms of the right hand sides of Equations~(\ref{eq:bcrpa1}) and 
(\ref{eq:bcrpa2}) collapse to a single term. For each value of the excitation
energy $\omega$, the system has to 
be solved a number of times equal to the number of open channels. 

The solution of the CRPA system of equations can be obtained by solving directly 
the set of Equations~(\ref{eq:bcrpa1}) and (\ref{eq:bcrpa2}). The s.p. particle wave functions
in the continuum are obtained  by solving  the s.p. Schr\"odinger equation with 
asymptotically oscillating boundary conditions. This is the classical problem of a particle
elastically scattered by a potential. This problem has to be solved for a set of $\epsilon_p$
energy values mapping the continuum in such a way that the integral of Equation~(\ref{eq:crpa.simbol})
is numerically stable. This means that $\epsilon_p$ must reach 
values much larger than those of the  excitation energy region one wants to investigate. 
The selection of the $\epsilon_p$
values to obtain the s.p. wave function is not a simple problem to be solved. 
The various particles may have more or less sharp resonances and 
they have to be properly described by the choice of the $\epsilon_p$ values
mapping the continuum.

There is another technical problem in the direct approach to the solution of the 
 CRPA Equations  (\ref{eq:bcrpa1}) and (\ref{eq:bcrpa2}). The numerical stability 
of the interaction matrix elements $\barv$ is due to the fact that, in the integrals, 
hole wave functions, which asymptotically go to zero, are present. 
This works well for the direct matrix elements
\beq
\braket{p_1 h_2 | \hV | h_1 p_2} = 
\int d^3 r_1 \int d^3 r_2 \phi^*_{p_1} (\br_1) \phi^*_{h_2} (\br_2) V(\br_1,\br_2) 
\phi_{h_1} (\br_1) \phi_{p_2} (\br_2) 
,
\eeq
but it is a problem for the exchange matrix element
\beq
\braket{p_1 h_2 | \hV | p_2 h_1} = 
\int d^3 r_1 \int d^3 r_2 \phi^*_{p_1} (\br_1) \phi^*_{h_2} (\br_2) V(\br_1,\br_2) 
\phi_{p_2} (\br_1) \phi_{h_1} (\br_2) 
,
\eeq
where the two particle wave functions, both oscillating, are integrated together. 
The direct approach is suitable to be used with zero-range interactions
$V(\br_1,\br_2) = V_0 \delta(\br_1 - \br_2)$. In this case, direct and exchange
matrix elements are identical
\beq
\braket{p_1 h_2 | \hV | h_1 p_2} = \braket{p_1 h_2 | \hV | p_2 h_1} =
V_0 \int d^3 r_1 \phi^*_{p_1} (\br_1) \phi^*_{h_2} (\br_1)
\phi_{p_2} (\br_1) \phi_{h_1} (\br_1) 
,
\eeq
and the hole wave functions are always present in the integral. This 
ensures the numerical convergence. The direct approach is used, for example, 
in Refs. \cite{deh82,co85}, 
where the CRPA equations are expanded on a Fourier--Bessel basis. 

Another method of solving the CRPA equations consists in reformulating the
secular Equations (\ref{eq:crpa.crpa1}) and (\ref{eq:crpa.crpa2}) with new unknown
functions which do not have explicit dependence on the continuous
particle energy $\epsilon_p$. The new unknowns are the \emph{channel functions} 
$f$ and $g$ defined as:
\beq 
f^\nu_{ph}(\br) \, = \, \sumint_{\epsilon_p} \, 
X^\nu_{ph}(\epsilon_p) \, \phi_p(\br,\epsilon_p) \, ,
\label{eq:f}
\eeq 
and
\beq 
g^\nu_{ph}(\br) \, = \, \sumint_{\epsilon_p} \, 
Y^{\nu*}_{ph}(\epsilon_p) \, \phi_p(\br,\epsilon_p)  \, .
\label{eq:g}
\eeq 

In the first step of this procedure, we multiply 
Equations (\ref{eq:bcrpa1}) and (\ref{eq:bcrpa2}) by $ \phi_p(\br,\epsilon_p)$, 
which is the eigenfunction of the s.p. hamiltonian
\beq
\hh \phi_p(\br,\epsilon_p) = \epsilon_p \phi_p(\br,\epsilon_p)
,
\eeq
and we obtain
\beq 
(\epsilon_p-\epsilon_h-\omega) \phi_p(\br,\epsilon_p)\,X_{ph}^\nu(\epsilon_p) \, = \,
\hh \,\phi_p(\br,\epsilon_p) \, X_{ph}^\nu(\epsilon_p)    \,  - \,
(\epsilon_h+\omega) \, \phi_p(\br,\epsilon_p) \, X_{ph}^\nu(\epsilon_p) \, .  
\eeq
Since the s.p. hamiltonian $\hh$ does not depend on $\epsilon_p$, we can
write 
\beq
\sumint_{\epsilon_p}\, 
\hh \, \phi_p(\br,\epsilon_p) \, X_{ph}^\nu(\epsilon_p) 
= \hh \, f^\nu_{ph}(\br) 
\;\; .
\eeq 

We apply this procedure, i.e., multiplication of $\phi_p(\br)$ and 
integration on $\epsilon_p$, to all the terms of 
Equations (\ref{eq:crpa.crpa1}) and (\ref{eq:crpa.crpa2}).
By considering that 
\[
\int d \epsilon_p \phi^*_p(\br_1) \phi_p(\br_2) = \delta(\br_1 - \br_2)
,
\]
we obtain a new set of CRPA secular equations where the unknowns  
are the channel functions $f$ and $g$,  
\beqn 
\nonumber 
\hh \, f_{ph}(\br) 
 -\, (\epsilon_h\, +\, \omega)\, f_{ph}(\br) &=&
-\, {\cal F}_{ph}(\br) \, \\
&+&
\, \sum_{\epsilon_p<\epsilon_{\rm F}}  \, \phi_p(\br) \, 
\int {\rm d}^3 r_1 \phi^*_h(\br_1) \, {\cal F}_{ph}(\br_1) \, ,
\label{eq:feq}\\
\nonumber 
\hh \, g_{ph}(\br)  -
(\epsilon_h\, -\, \omega)\, g_{ph}(\br) &=& 
-\, {\cal G}_{ph}(\br) \, 
\\
&+&
\, \sum_{\epsilon_p<\epsilon_{\rm F}} \, \phi_p(\br) \, 
\int {\rm d}^3 r_1 \, \phi^*_h(\br_1) \, {\cal G}_{ph}(\br_1) \, ,
\label{eq:geq}
\eeqn
where we have defined 
\beqn
\nonumber
{\cal F}_{ph}(r) \, &=& \, \sum_{[p'] h'} \, \int {\rm d}^3 \,r_2 V(\br,\br_2)
\Bigg\{ \phi^*_{h'}(\br_2) \ \Bigg[
\, \phi_h(\br) \, f_{p'h'}(\br_2) - f_{p'h'}(\br) \, \phi_h (\br_2) \Bigg]
\\
&+& \, g^*_{p'h'}(\br_2) \, 
\Bigg[ \, \phi_h(\br)\, \phi_{h'}(\br_2)\, -  \phi_{h'}(\br) \, \phi_h(\br_2) \Bigg]  
\Bigg\} 
\, ,
\label{eq:Fcal}
\eeqn
and ${\cal G}_{ph}$ is obtained from the above equation by interchanging
the $f$ and $g$ channel functions.  The last terms of both 
Equations~(\ref{eq:feq}) and (\ref{eq:geq}) are the contributions of particle wave
functions $\phi_p$ which are not in the continuum.

We changed a set of algebraic equations with unknowns depending
on the continuous variable $\epsilon_p$ into a set of
integro-differential equations whose unknowns depend on $\br$.
In analogy to what we have discussed
above, for the direct solution of the CRPA secular equations, we
solve Equations (\ref{eq:feq}) and (\ref{eq:geq}) a number of times equal
to the number of the open channels, by imposing that  the particle 
is emitted only in the elastic channel.

In spherical systems, the boundary conditions 
are imposed on the radial parts of the $f$ and $g$ functions.
For an open $ph$ channel, the outgoing
asymptotic behavior of the channel function $f_{ph}^{p_0 h_0}$ is 
\beq
f_{ph}^{p_0 h_0}(r\to\infty)\, \to \,
R_{p_0}(r,\epsilon_p)\, \delta_{p,p_0}\, \delta_{h,h_0}\, +\, \lambda
\, H^-_p(\epsilon_h+\omega,r)
\,,
\label{eq:asintf} 
\eeq 
where $\lambda$ is a complex normalization constant and
$H^-_p(\epsilon_h+\omega,r)$ is an ingoing Coulomb function 
if the emitted particle is electrically charged or a Hankel function in cases of neutron.
The radial part of the s.p. wave function $R_{p}$ is the eigenfunction of the s.p. 
hamiltonian for positive energy.
In the case of a closed channel, the asymptotic behavior is given by 
a decreasing exponential function 
\beq
f_{ph}^{p_0 h_0}(r\to\infty) \, 
\rightarrow \, 
\frac{1}{r}\,
\exp\left[-r\left(\frac{2m|\epsilon_h+\omega|}{\hbar^2}\right)^{\frac{1}{2}}
\right]
\,,
\label{eq:asintf1}
\eeq
in analogy to the case of the channel functions
$g_{ph}^{p_0 h_0}$, 
\beq
g_{ph}^{p_0 h_0}(r\to\infty)
\, \rightarrow \,
\frac{1}{r} \,
\exp\left[-r\left(\frac{2m|\epsilon_h-\omega|}{\hbar^2}\right)^{\frac{1}{2}}
\right]
\,.
\label{eq:asintg}
\eeq
This approach solves the two technical problems of the direct approach 
indicated above, since the integration $\epsilon_p$ is formally done in
the definition of the two channel functions $f$ and $g$.

These CRPA secular equations can be solved 
by using a procedure similar to that presented in Refs. \cite{bub91,don11a}.
The channel functions $f$ and $g$ are expanded on the  basis of Sturm
functions $\Phi^{\mu}_p$ which obey the required boundary conditions
(\ref{eq:asintf})--(\ref{eq:asintg}).

In the IPM, the particle emission process is described by considering
that a particle lying on the hole state $h_0$ is emitted in the particle state $p_0$.
The CRPA considers this fact in the elastic channel and, in addition, 
takes care of the fact that the residual interaction
mixes this direct emission with all the other $ph$ pairs compatible 
with the total angular momentum of the excitation.

\section{Quasi-Particle RPA (QRPA)}
\label{sec:QRPA}

In the derivations  presented in the previous sections, we  considered that the IPM ground 
state is defined by a unique Slater determinant $\ket{\Phi_0}$, where all the s.p. states
below the Fermi energy are fully occupied and those above it are completely empty. 
This description does not consider the presence of an effect which is very important
in nuclei: the pairing. This effect couples two like-fermions to form
a unique bosonic system. In metals this  produces the effects of superconductivity. 
In nuclear physics this leads to the fact that all the even--even nuclei,  
without exceptions, have spin zero.

A convenient description of pairing effects is based on the Bardeen--Cooper--Schrieffer 
(BCS) theory of superconductivity \cite{bar57}. In this approach, the choice of $\ket{\Phi_0}$ 
for the description of the system ground state is abandoned.

Let us consider a finite fermion system and use the expression Equation~(\ref{eq:mf.spwf2}) for
the s.p. wave functions. We introduce a notation to indicate time-reversed s.p. 
wave functions
\beq
\ket{k} := \ket{nl \half jm} \;\;\;;\;\;\; \ket{-k} := \ket{nl \half j -m} 
. 
\eeq

The BCS ground state is defined as
\beq
\ket{{\rm BCS}} := \prod_{k>0}^\infty\left( u_k + v_k \ha^+_k \ha^+_{-k} \right) \ket{-} 
,
\label{eq:BCS.gs}
\eeq
where we have indicated with $\ket{-}$ the state describing the physical vacuum. 
The $v_k^2$ factor is the occupation probability of the $k$-th s.p. state,
and $u_k^2 = 1 - v_k^2$ the probability of being empty. When pairing effects
are negligible, for example, in doubly magic nuclei, $v_k=1$ for all the s.p. states
below the Fermi surface and $v_k=0$ for all the states above it;  
therefore, $\ket{\rm BCS} = \ket{\Phi_0}$.

A convenient manner of handling the $\ket{\rm BCS}$ states is to define quasi-particle
creation and destruction operators which are linear combinations of usual particle creation 
and destruction operators. The relations are known as \emph{Bogoliubov--Valatin 
transformations}
\beqn
\halpha_k &=& u_k \ha_k - v_k \ha^+_{-k} ,\\
\halpha^+_k &=& u_k \ha^+_k - v_k \ha_{-k}  .
\eeqn
Since the quasi-particle operators are linear combination of the creation and 
destruction operators, anti-commutation relations analogous to  (\ref{eq:app.anta})
are valid also for the $\halpha$ and $\halpha^+$ operators.
It is possible to show that \cite{suh07}
\beq
\halpha_k \ket{\rm BCS} = 0
\label{eq:BCS.zero}
,
\eeq
indicating that the $\ket{\rm BCS}$ states can be appropriately called 
quasi-particle vacuum.
The BCS ground state is not an eigenstate of the number operator 
\beq
{\hat N} = \sum_k \ha^+_k \ha_k
,
\label{eq:number1}
\eeq
and the number of particles is conserved only on average
\cite{rin80,suh07}
\beq
\braket {{\rm BCS} | {\hat N} |{\rm BCS}} = 2 \sum_{k>0}v^2_k = A
.
\eeq

The values of the $v_k$ coefficients, and consequently those of $u_k$, are
obtained by exploiting the variational principle. For this reason, it is common
practice to use a definition of the hamiltonian containing the Lagrange
multiplier $\lambda$, related to the total number of particle $A$ 
\beq
{\hat {\cal H}} = \hH - \lambda \hat N
.
\label{eq:H'}
\eeq

The hamiltonian $\hH$ is written by expressing in Equation~(\ref{eq:app.ham44}) 
the $\ha$ and $\ha^+$ operators in terms of the quasi-particle operators
$\alpha^+$ and $\alpha$. By observing the operator structure, it is
possible to identify four different terms (see Eq.~(13.32) of \cite{suh07}) 
\beq
\hat {\cal H}= \hH - \lambda {\hat N} = \hat {\cal H}_0 + 
\hat {\cal H}_{11} + \hat {\cal H}_{22} + \hH_{\rm int}
,
\label{eq:QRPA.fullh}
\eeq
where $\lambda$ is present only in the first three terms.
The various terms are defined as follows.
\begin{enumerate}
\item $\hat {\cal H}_0$ is purely scalar,
\beq
\hat {\cal H}_0 = \sum_{k}  \left[(\epsilon_k - \lambda - \mu_k) 2 v^2_k  +
u_k v_k \sum_{k'} \barv_{k,k',k,k'} u_{k'} v_{k'}  \right].
\eeq
\item $\hat {\cal H}_{11}$ depends on $\halpha^+_k \halpha_{k}$,
\beq
\hat {\cal H}_{11} = 
\sum_{k}  \left\{ \left[ \epsilon_k - \lambda  - \mu_k \right] (u^2_k-v^2_k) 
+ 2 u_k v_k \Delta_k \right\} \halpha^+_k \halpha_{k}
.
\label{eq:h11}
\eeq
\item $\hat {\cal H}_{22}$ depends on $\hN[\halpha^+_k \halpha^+_{k'} + \halpha_k \halpha_{k'} ]$.
\item $\hH_{\rm int} = \hH_{40} +  \hH_{31} + \hH_{22}$, where
\begin{itemize} 
 \item [] $\hH_{40}$ depends on 
 $[\halpha^+_{k_1} \halpha^+_{k_2} \halpha^+_{k_3} \halpha^+_{k_4} + h.c.]$,
 \item [] $\hH_{31}$ depends on 
 $[\halpha^+_{k_1} \halpha^+_{k_2} \halpha^+_{k_3} \halpha_{k_4} + h.c.]$,
 \item [] and finally,
 \beq
\hH_{22}= \frac 1 2 \sum_{abcd} V^{(22)}_{abcd} 
\halpha^+_{k_a} \halpha^+_{k_b} \halpha_{k_d} \halpha_{k_c}
,
\label{eq:h22}
 \eeq
with
\beq
V^{(22)}_{abcd} = (u_a u_b u_c u_d +  v_a v_b v_c v_d + 4 u_a v_b u_c v_d) \barv_{abcd} 
.
\label{eq:v22}
\eeq
\end{itemize}
\end{enumerate}
In the above equations, we used the following scalar quantities:
\beqn
\Delta_k &=& - \sum_{k'} \barv_{k,-k,k',-k'} u_{k'} v_{k'} 
, \\
\label{eq:deltak}
\epsilon_k &=& \int d^3r\, \phi^*_k(\br) \left( \frac {-\hbar^2 \nabla^2}{2 m}\right) \phi_k(\br) 
 + \half \sum_{k'} \barv_{k,k',k,k'} \, v^2_{k'} 
 , \\
\mu_k &=& -\half \sum_{k'} \barv_{k,k',k,k'} \, v^2_{k'} 
.
\eeqn
Because of Equation~(\ref{eq:BCS.zero}) the expectation value of $\hat {\cal H}$
with respect to the BCS ground state is
\beq
\braket{BCS | \hat {\cal H} | BCS} = 
\braket{BCS | \hat {\cal H}_0 | BCS} \equiv \tene^{\rm BCS}_A 
; 
\eeq
therefore, the application of the variational principle is 
\beq
\delta(\braket{BCS | \hat {\cal H}_0 | BCS}) = 0
,
\eeq
which implies the relation \cite{rin80,suh07}:
\beq
(u^2 - v^2) \Delta_k = 2 v_k u_k (\epsilon_k - \lambda + \mu_k) 
\Rightarrow
v_k u_k = \frac{\Delta_k}{2\sqrt{(\epsilon_k - \lambda-\mu_k)^2 + \Delta^2_k}}
.
\eeq
We insert the above result in Equation (\ref{eq:h11}) and obtain the BCS s.p. energies
\beq
\hat {\cal H}_{11} = 
\left\{ \sqrt{(\epsilon_k - \lambda-\mu_k)^2 + \Delta^2_k} \right\} \halpha^+_k \halpha_{k}
:= \epsilon^{\rm BCS}_k  \halpha^+_k \halpha_{k}
.
\eeq

In the BCS approach, the radial expressions of the 
s.p. wave functions are obtained by carrying out IPM calculations
and only the occupation amplitudes $v_k$ and $u_k$ are modified. There is a more fundamental 
approach, the Hartree--Fock--Bogolioubov theory, where s.p. wave functions, energies and
occupation probabilities are calculated in a unique theoretical framework whose only input
is the effective nucleon--nucleon interaction. 

After having defined a new ground state containing pairing effects, we can use it to develop
the theory describing the harmonic vibrations around it. 
The derivation of the QRPA secular equations is carried out by using the EOM method described
in Section \ref{sec:EOM}. In this case, the Slater determinant $\ket{\Phi_0}$ is 
substituted by the BCS ground state $\ket{\rm BCS}$ and the particle creation and destruction
operators $\ha_k$ and $\ha^+_k$ by the quasi-particle operators $\halpha_k$ and $\halpha^+_k$.
The QRPA excitation operator is given by 
\beq
Q^+_\nu \equiv \sum_{a \le b} X^\nu_{ab} \halpha^+_a \halpha^+_b 
                 - \sum_{a \le b} Y^\nu_{ab} \halpha_a \halpha_b 
\; \;.
\label{eq:QRPA.operator}                 
\eeq
The indexes
$a$ and $b$ containing all the quantum numbers which identify the quasi-particle states
are not, any more, referred to as 
particle or hole states. In this approach, the idea of Fermi surface
has disappeared. Each quasi-particle state can be partially occupied. For this reason, 
in the above equation, we had to impose restrictions on 
the sums in order to avoid double counting.

In the present case, the EOM (\ref{eq:rpa:moto}) assumes the expression
\beq
\braket{{\rm BCS}| \left[ \delta \hQ_\nu , [\hat {\cal H},\hQ^+_\nu] \right] |{\rm BCS}} 
= \omega \braket{{\rm BCS}| \left[ \delta \hQ_\nu, \hQ^+_\nu \right] | {\rm BCS}} 
,
\eeq
where we have substituted $\hat{\cal O}$ with $\delta \hQ_n$. 
Following the steps of the derivation of RPA equations, see Section \ref{sec:rpa:RPA},
and defining $A$ and $B$ matrices as
\beqn
A_{ab,cd} &\equiv& \braket{{\rm BCS}| \left[ \halpha_a \halpha_b , 
[\hat {\cal H},\halpha^+_c \halpha^+_d ] \right] |{\rm BCS}} 
,
\\
B_{ab,cd} &\equiv& -
\braket{{\rm BCS}| \left[ \halpha_a \halpha_b , 
[\hat {\cal H},\halpha_c \halpha_d ] \right] |{\rm BCS}} 
,
\eeqn
it is possible to obtain a set of linear equations analogous to those of RPA
\beqn
\sum_{c \le d} A_{ab,cd} X^\nu_{cd} + \sum_{c \le d} B_{ab,cd} Y^\nu_{cd} 
&=& \omega_\nu X^\nu_{ab}
,
\\ 
\sum_{c \le d} B^*_{cd,ab} X^\nu_{cd} + \sum_{c \le d} A_{cd,ab} Y^\nu_{cd}
 &=& - \omega_\nu Y^\nu_{ab}
 ,
\eeqn
which can be written in matrix form  analogously to  Equation~(\ref{eq:rpa:matrix}).
This calculation is explicitly carried out in Chapter 18 of Ref. \cite{suh07}
and it shows that only $\hat {\cal H}_0$, $\hat {\cal H}_{11}$ and $\hat {\cal H}_{22}$
contribute to the $A$ and $B$ matrices. These matrices contain, in addition to the 
particle--hole excitations present in the common RPA, also particle--particle and hole--hole
transitions, since each s.p. state is only partially occupied. 
The solution of the QRPA secular equations, for each excited state, provides the $X$ 
and $Y$ amplitudes which indicate the contribution of each quasi-particle excitation pair. 
 
The QRPA solutions have the same properties of those of RPA
solutions. The QRPA equations allow positive and negative eigenergies
with the same absolute value. 
Eigenvectors corresponding to different energy eigenvalues are orthogonal. 
The set of QRPA eigenstates is complete. 

The transition amplitudes from the QRPA ground state to an excited state, 
induced by an external one-body operator $\hat F$, Equation~(\ref{eq:rpa:opext}), is 
\beq
\braket{{\rm QRPA}; \nu | \hat F | {\rm QRPA};0} 
= \sum_{a \le b}  f_{ab} \left(v_a u_b + u_a v_b \right) (X^\nu_{ab} + Y^\nu_{ab} ) 
.
\eeq

For $ph$ transitions only, when $v_a=1, u_a=0$ and $v_b=0, u_b=1$ one recovers 
the ordinary RPA expression (\ref{eq:rpa.tprob}).

\section{Specific Applications}
\label{sec:applications}

In this section, I discuss some pragmatic issues arising in actual RPA calculations. 
The input of RPA calculations is composed by 
the s.p. energies and wave functions and also by the effective interaction
between the particles forming the system. 
There are various possible choices of these
quantities and they define different types of calculations. 

A fully \emph{phenomenological approach} is based on the Landau--Migdal theory of finite Fermi systems
\cite{spe77,rin78}. In this theory, the attention is concentrated on the small vibrations on top 
of the ground state, which is assumed to be perfectly known. 
The s.p. wave functions are generated by solving the MF Equation (\ref{eq:mf.hsp}) with a 
phenomenological potential whose parameters are chosen to reproduce
at best the empirical values of the s.p. energies of the levels around the Fermi surface.
In RPA calculations, these empirical values are used when available; otherwise,  
the s.p. energies obtained by solving the MF equation are considered.
The interaction is a zero-range density dependent Landau--Migdal force whose
parameters are selected to reproduce some empirical characteristics of the excitation spectrum. 

This approach has shown a great capacity to describe  features of the excited states and also
a remarkable predictive power. For example, the presence of a collective monopole excitation
 in $^{208}$Pb was predicted at the right energy \cite{rin74} 
 before it was identified by experiments
with $\alpha$ \cite{you77} and $^{3}$He scattering \cite{bue79}.
The drawback consists in the need for a continuous tuning of the 
MF potential and  the interaction
parameters, since the results strongly depend on the input. 
This means that there is a set of force parameters for each nucleus,  
and also in the same nucleus the values of these parameters change 
if the dimensions of the configuration space are modified.

An approach which avoids this continuous setting of the free parameters is the so-called 
\emph{self-consistent} approach. In this case, the s.p. wave functions and energies are generated
by solving HF or DFT equations. The parameters of the effective interaction are tuned to reproduce at best
experimental binding energies and charge radii, all along the isotope table. 
The same interaction, unique for all the nuclei, is used also in RPA calculations. 

The density dependent zero-range Skyrme force is probably the interaction
most used in this type of calculation  \cite{vau72}.
The zero-range characteristic allows great simplifications of  the expressions of the
interaction matrix elements and the numerical calculations are relatively fast. 
There are tens of different sets of parameters of the Skyrme force, each of them properly tuned 
to describe some specific characteristics of the nuclei. 
The zero-range feature of the Skyrme force is mitigated by the 
presence of momentum dependent terms. On the other hand, the sensitivity on the dimensions
of the s.p. configuration space is not negligible. For this reason, in BCS and QRPA calculations
it is necessary to use a different interaction to treat the pairing. 

These drawbacks are overcome by interactions which have finite range, a 
feature which clearly makes much more involved the numerical calculations.
A widely used finite-range interaction is that of Gogny \cite{dec80}.
Despite this difference, the philosophy of the calculations carried out with the two 
kinds of interaction is the same: a unique force, valid all along the nuclide chart, tuned to reproduce 
ground state properties with HF calculations and used in RPA. 
Discrete RPA calculations carried out
with Gogny force show a convergence of the results after certain dimensions of the configuration
space have been reached. 

The self-consistent approach does not provide an accurate 
description of the excited states obtained with the phenomenological approach. On the other
hand, by using self-consistent approaches it is possible  to make connections
between the properties of the ground 
and of the excited state and also between features appearing 
in different nuclei, everything described within a unique theoretical framework. 
This approach can make predictions on properties 
of nuclei far from stability where empirical quantities
have not yet been measured.

\begin{figure}[ht]
\begin{center} 
\includegraphics [scale=0.4,angle=90]{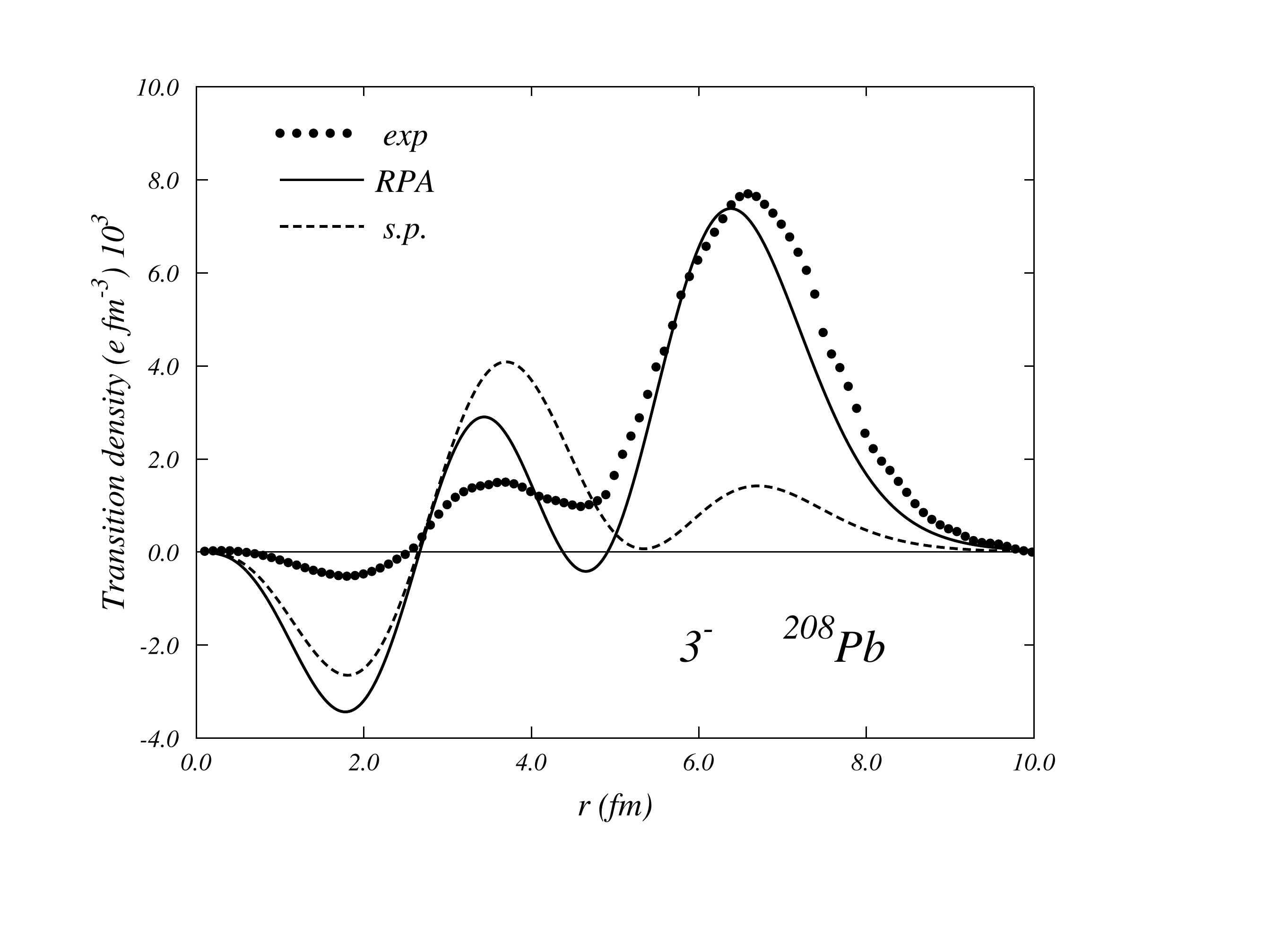} 
\captionsetup{margin=2cm}
\caption{\small 
Electron scattering
transition densities for the first $3^-$ excited state of $^{208}$Pb. The empirical
density, indicated by the dots, is extracted from the inelastic electron scattering
data of Ref.~\cite{gou80}. 
The dashed line shows the transition density calculated in an IPM particle model 
where the state is described by the s.p. proton transition from the $1s_{1/2}$ 
hole state to the $2f_{7/2}$ particle state. The full line shows the RPA result. 
} 
\label{fig:pb208} 
\end{center} 
\end{figure}

As an example of the RPA result, we consider the case of the $3^-$
state in $^{208}$Pb already mentioned in Section \ref{sec:IPM.excitation}.
We show in Figure~\ref{fig:pb208} a comparison between the transition 
densities calculated with an RPA 
theory (full line), that obtained by an s.p. transition (dashed lines) and 
the empirical transition density (dots) extracted from the 
inelastic electron scattering data of Ref.~\cite{gou80}. 
The s.p. excitation was obtained by considering the proton transition
from the $1s_{1/2}$ hole state to the $2f_{7/2}$ particle state with the 
excitation energy of 5.29 MeV.  RPA calculation was carried out
with the phenomenological approach and the excitation energy is of 
2.66 MeV to be compared with an experimental value of 2.63 MeV.

The s.p. transition, which is what the IPM at its best can provide, 
is unable to describe the large values of the transition density at 
the nuclear surface. This surface vibration is a characteristic feature
of this highly collective state.  RPA is able to reproduce the value of the 
excitation energy and also the correct behavior of the transition density.


\section{Extensions of RPA}
\label{sec:HRPA}

\subsection{Second RPA}
\label{sec:HRPA.srpa}

The main limitation of RPA theory
is due to the fact that the $\hQ^+_\nu$ operator
considers only $1p-1h$ and $1h-1p$ types of excitations,
see Equation~(\ref{eq:rpa:rpaq}).
The many-body system allows more complicated
excitation modes where $n$-particle and $n$-holes are created. The extension of $\hQ^+_\nu$ 
to consider also $2p-2h$ excitations is called Second RPA (SRPA) 
\cite{yan82,dro90,gam06}. In this theory, the operator which defines the excited states is 
\beqn
\nonumber
\hQ^+_\nu &\equiv& \sum_{m,i} 
\left(X^\nu_{mi} \ha^+_m \ha_i -  Y^\nu_{mi} \ha^+_i \ha_m \right)
\\
&+& \sum_{m<n, i<j} 
\left(X^\nu_{mnij} \ha^+_m \ha^+_n \ha_i \ha_j -  
Y^\nu_{mnij} \ha^+_i \ha^+_j \ha_m \ha_n \right)
+ \sum_{m,n, i,j} Z^\nu_{mjin} \ha^+_m \ha^+_j \ha_n \ha_i 
.
\label{eq:srpa:rpaq}
\eeqn
where the $X$, $Y$ and $Z$ factors are real numbers.

We insert this operator into the EOM Equation (\ref{eq:rpa:moto}) and substitute
${\hat {\cal O}}$ with $\delta \hQ_\nu$. Since $\delta \hQ_\nu$ implies variations
of the coefficients in (\ref{eq:srpa:rpaq}) and these variations are independent of
each other, we obtain five equations
\beqn
\braket{{\rm RPAII}| \left[\ha^+_i \ha_m , [\hH,\hQ^+_\nu] \right] | {\rm RPAII}}  &=& 
\omega_\nu \braket{{\rm RPAII}| \left[\ha^+_i \ha_m ,\hQ^+_\nu \right] | {\rm RPAII}} 
,
\label{eq:srpa.e1}
\\
\braket{{\rm RPAII}| \left[\ha^+_m \ha_i , [\hH,\hQ^+_\nu] \right] | {\rm RPAII}}  &=& 
\omega_\nu \braket{{\rm RPAII}| \left[\ha^+_m \ha_i ,\hQ^+_\nu \right] | {\rm RPAII}} 
,
\label{eq:srpa.e2} \\
 \braket{{\rm RPAII}| \left[\ha^+_i \ha^+_j \ha_n \ha_m , [\hH,\hQ^+_\nu] \right] | {\rm RPAII}}  &=& 
\omega_\nu \braket{{\rm RPAII}| \left[\ha^+_i \ha^+_j \ha_n \ha_m  ,\hQ^+_\nu \right] | {\rm RPAII}} 
,
\label{eq:srpa.e3}\\
\braket{{\rm RPAII}| \left[\ha^+_m \ha^+_n \ha_j \ha_i , [\hH,\hQ^+_\nu] \right] | {\rm RPAII}}  &=& 
\omega_\nu \braket{{\rm RPAII}| \left[\ha^+_m \ha^+_n \ha_j \ha_i  ,\hQ^+_\nu \right] | {\rm RPAII}} 
,
\label{eq:srpa.e4}\\
\braket{{\rm RPAII}| \left[\ha^+_i \ha^+_n \ha_j \ha_m , [\hH,\hQ^+_\nu] \right] | {\rm RPAII}}  &=& 
\omega_\nu \braket{{\rm RPAII}| \left[\ha^+_i \ha^+_n \ha_j \ha_m  ,\hQ^+_\nu \right] | {\rm RPAII}} 
,
\label{eq:srpa.e5}
\eeqn
where $\ket{\rm RPAII}$ is the SRPA ground state defined by the equation
\beq
Q_\nu \ket{\rm RPAII} = 0
.
\eeq
In analogy to what is presented in Section \ref{sec:RPA.rpa},  we use the QBA by assuming
\beq
\braket{{\rm RPA II} | \left[\hat {\cal O}_1 \, , \,  \hat {\cal O}_2 \right] | {\rm RPAII}} \simeq  
\braket{ \Phi_0 | \left[\hat {\cal O}_1 \, , \,  \hat {\cal O}_2 \right] | \Phi_0} 
,
\eeq
where  $\hat {\cal O}$ are two generic operators and $\ket{\Phi_0}$ indicates, as usual,
the IPM ground state. It is convenient to define the following matrix elements
\beqn
\braket{ \Phi_0 | \left[\ha^+_i \ha_m , [\hH, \ha^+_n \ha_j ] \right] | \Phi_0}&=& A_{mi,nj} , \\
\braket{ \Phi_0 | \left[\ha^+_i \ha_m , [\hH, \ha^+_j \ha_n ] \right] | \Phi_0}&=& - B_{mi,nj} , \\
\braket{ \Phi_0 | \left[\ha^+_i \ha_m , [\hH, \ha^+_n \ha^+_p \ha_l \ha_j ] \right] | \Phi_0}&=& A_{mi,npjl} 
, \\
\braket{ \Phi_0 | \left[\ha^+_i \ha_m , [\hH, \ha^+_j \ha^+_l \ha_p \ha_n ] \right] | \Phi_0}&=& - B_{mi,npjl} 
, \\
\braket{ \Phi_0 | \left[\ha^+_i \ha^+_j \ha_n \ha_m , 
[\hH, \ha^+_p \ha^+_q \ha_l \ha_k ] \right] | \Phi_0}&=& A_{mnij,pqkl} , \\
\braket{ \Phi_0 | \left[\ha^+_i \ha^+_j \ha_n \ha_m , 
[\hH, \ha^+_k \ha^+_l \ha_q \ha_p ] \right] | \Phi_0}&=& - B_{mnij,pqkl} 
.
\eeqn
The $A_{mi,nj}$ and $ B_{mi,nj}$ matrix elements are identical to
those defined in Equation~(\ref{eq:rpa:AB1}). 
Explicit expressions of the other matrix elements can be found in Ref. \cite{yan82}.
With the help of these definitions, 
Equations (\ref{eq:srpa.e1}--\ref{eq:srpa.e5}) can be expressed as:
\beqn
\sum_{pk} \left( A_{mi,pk} X^\nu_{pk} +  B_{mi,pk} Y^\nu_{pk} \right)
+ \sum_{p<q,k<l}  A_{mi,pqkl} X^\nu_{pqkl}  
&=& \omega_\nu X^\nu_{mi} 
,
\label{eq:srpa.eq1} \\
\sum_{pk} \left( B^+_{mi,pk} X^\nu_{pk} +  A^+_{mi,pk} Y^\nu_{pk} \right)
+ \sum_{p<q,k<l}  A^+_{mi,pqkl} Y^\nu_{pqkl}  
&=& - \omega_\nu Y^\nu_{mi} 
,
\label{eq:srpa.eq2} \\
\sum_{pk} A_{mnij,pk} X^\nu_{pk} + \sum_{p<q,k<l}  A_{mnij,pqkl} X^\nu_{pqkl}  
&=& \omega_\nu X^\nu_{mnij} 
,
\label{eq:srpa.eq3} \\
\sum_{pk} A^+_{mnij,pk} X^\nu_{pk} + \sum_{p<q,k<l}  A^+_{mnij,pqkl} Y^\nu_{pqkl} 
&=& - \omega_\nu Y^\nu_{mnij} 
,
\label{eq:srpa.eq4} 
\eeqn
where it appears evident that the $Z$ terms of Equation~(\ref{eq:srpa:rpaq}) do not contribute. 

The above equations form the complete set of SRPA secular equations. 
Usually, one does not search for the whole solution of these equations, but one
considers only the unknowns $X^\nu_{mi}$ and $Y^\nu_{mi}$. 
This is done by formally extracting $X^\nu_{mnij}$
and $Y^\nu_{mnij}$ from Equations (\ref{eq:srpa.eq3}) and  (\ref{eq:srpa.eq4}), respectively, 
and by inserting the obtained expressions into Equations (\ref{eq:srpa.eq1}) and (\ref{eq:srpa.eq2}).
In this way, we obtain two equations where the only unknowns are $X^\nu_{mi}$ and $Y^\nu_{mi}$ 
\beqn
\nonumber
 &~& \sum_{pk} \Bigg[ A_{mi,pk} - 
\sum_{p_1<q_1,k_1<l_1} \sum_{p_2<q_2,k_2<l_2}   \\
&~& \nonumber
 A_{mi,p_1q_1k_1l_1} \left(  A_{p_1q_1k_1l_1,p_2q_2k_2l_2} -
\omega_\nu \delta_{p_1,p_2} \delta_{q_1,q_2} \delta_{k_1,k_2} \delta_{l_1,l_2} \right)^{-1} 
A_{p_2q_2k_2l_2,pk} \Bigg] X^\nu_{pk} 
\\ &~&
+ \sum_{pk}  B_{mi,pk} Y^\nu_{pk} =  \omega_\nu X^\nu_{mi} 
,
\label{eq:srpa.eqn1}
\\ \nonumber
&~& \sum_{pk} \Bigg[ A^+_{mi,pk} - 
\sum_{p_1<q_1,k_1<l_1} \sum_{p_2<q_2,k_2<l_2}   
\\ &~& \nonumber
 A^+_{mi,p_1q_1k_1l_1} \left(  A^+_{p_1q_1k_1l_1,p_2q_2k_2l_2} -
\omega_\nu \delta_{p_1,p_2} \delta_{q_1,q_2} \delta_{k_1,k_2} \delta_{l_1,l_2} \right)^{-1} 
A^+_{p_2q_2k_2l_2,pk} \Bigg] Y^\nu_{pk} 
\\ &~&
+ \sum_{pk}  B^+_{mi,pk} X^\nu_{pk} =  -\omega_\nu Y^\nu_{mi} 
,
\label{eq:srpa.eqn2}
\eeqn
which, in matrix form, can be written in analogy to Equation (\ref{eq:rpa:matrix}) as
\beq
\left(
\begin{array}{cc}
{\cal A} & {\cal B} \\
   & \\
{\cal B}^+  & {\cal A}^+
\end{array}
\right)
\left(  
\begin{array}{c}
X^\nu \\
\\
Y^\nu
\end{array}
\right) = \omega_\nu
\left(
\begin{array}{c}
X^\nu \\
\\
- Y^\nu
\end{array}
\right) 
.
\label{eq:srpa:matrix}
\eeq
The second terms in square brackets of Equations (\ref{eq:srpa.eqn1}) and  (\ref{eq:srpa.eqn2}) 
couples $1p-1h$ excitations to $2p-2h$ excitations. If these terms are zero, RPA equations
are recovered.
The secular SRPA equations have the same properties as RPA equations.
\begin{enumerate}
\item Positive and negative energy eigenvalues with the same absolute value are allowed.
\item Eigenvectors of different eigenvalues are orthogonal.
\item The normalization between the excited states implies
\beq
\sum_{mi} \left(X^\nu_{mi} X^{\nu'}_{mi} - Y^\nu_{mi} Y^{\nu'}_{mi} \right) = \delta_{\nu,\nu'}
 .
\eeq
\end{enumerate}

The number of terms of the $A^+_{p_1q_1k_1l_1,p_2 q_2 k_2 l_2}$ and 
$A_{p_1q_1k_1l_1,p_2q_2k_2l_2}$ matrix elements
is quite large; for this reason, the so-called
diagonal approximation is often used. This approximation consists in considering in
$A_{p_1q_1k_1l_1,p_2q_2k_2l_2}$ only the diagonal part depending on the s.p. energies
involved in the $2p-2h$ excitations
\beq
A^+_{p_1q_1k_1l_1,p_2q_2k_2l_2} \longrightarrow
(\epsilon_{p_1} + \epsilon_{q_1} - \epsilon_{k_1} - \epsilon_{l_1} )
\delta_{p_1,p_2} \delta_{q_1,q_2} \delta_{k_1,k_2} \delta_{l_1,l_2} 
.
\eeq

The expression of the transition amplitude between the SRPA ground state and 
excited states can be calculated as indicated in Section \ref{sec.eom.trans}
and the same result, Equation~(\ref{eq:rpa.tprob}), is obtained. In this theoretical 
framework, the SRPA approach modifies the values of the 
$X$ and $Y$ RPA amplitudes by coupling them to the $2p-2h$ excitation space. 

\subsection{Particle-Vibration Coupling RPA}
\label{sec:HRPA.pvc}
The approach presented in the previous section
is general but rather difficult to implement because of the large number of $2p-2h$ pairs
to consider. Many of the $2p-2h$ matrix elements are relatively small with respect to the 
$1p-1h$ terms. Instead of evaluating many irrelevant matrix elements, it is more 
convenient to identify the important ones and calculate only them.

This is the basic idea of  Particle-Vibration Coupling RPA (PVCRPA) \cite{boh75},
also called Core Coupling RPA (CCRPA), where RPA excited states are coupled
to s.p. states. In this approach, the excited states have the expression
\beq
\ket{\cal R} = \sum_\nu \sum_{ph} \ket{ph} \otimes \ket{\nu}
,
\eeq
where $\ket{\nu}$ is an RPA excited state,  $\ket{ph}$ is $ph$ excitation pairs 
and $\otimes$ indicates a tensor coupling.

We define a set of operators which project the 
eigenstate $\Psi$ of the hamiltonian on 
IPM	eigenstates $\ket{\Phi_0}$, 
RPA states  $\ket {\nu}$ (composed by $1p-1h$ excitations) 
and particle-vibration coupled states $\ket{\cal R}$ (composed by $2p-2h$) 
excited pairs:
\beqn
\hP \ket{\Psi} &=& \ket{\Phi_0} \\
\hQ_1 \ket{\Psi} &=& \ket{\nu} \\
\hQ_2 \ket{\Psi} &=& \ket{\cal R} 
\eeqn
These operators have the properties
\beqn
\label{eq:srpa.p1}
&~& \hP^2=\hP;~~\hQ^2_1=\hQ_1;~~\hQ^2_2 = \hQ_2;\\
\label{eq:srpa.p2}
&~& \hP \hQ_1 = \hP \hQ_2 = \hP_1 \hP_2 =0; \\
\label{eq:srpa.p3}
&~& \hP+\hQ_1+\hQ_2={\hat {\mathbb I}} 
.
\eeqn
The latter property implies that $\ket{\Psi}$ does not contain excitations 
more complex than $2p-2h$ and automatically neglects some term of 
the many-body hamiltonian.

We can write the eigenvalue equation as
\beq
\hH  \ket{\Psi} = \hH (\hP+\hQ_1+\hQ_2) \ket{\Psi} = \omega (\hP+\hQ_1+\hQ_2) \ket{\Psi}
.
\eeq
We multiply both sides of the above equation, respectively,  by $\hP$, $\hQ_1$ and $\hQ_2$, 
and, by using the properties (\ref{eq:srpa.p1})--(\ref{eq:srpa.p3}), we obtain the following equations
\beqn
(\omega - \hP \hH \hP) \hP \ket{\Psi} &=&
\hP \hH \hQ_1 \ket{\Psi}  + \hP \hH \hQ_2 \ket{\Psi} \label{eq:pvc.p} 
\\
(\omega - \hQ_1 \hH \hQ_1) \hQ_1 \ket{\Psi} &=& 
\hQ_1 \hH \hP \ket{\Psi}  + \hQ_1 \hH \hQ_2 \ket{\Psi} \label{eq:pvc.q1}  \\
(\omega - \hQ_2 \hH \hQ_2) \hQ_2 \ket{\Psi} &=& 
\hQ_2 \hH \hP \ket{\Psi}  + \hQ_2 \hH \hQ_1 \ket{\Psi} \label{eq:pvc.q2} 
.
\eeqn
We formally obtain $\hP \ket{\Psi}$ from Equation~({\ref{eq:pvc.p}) and $\hQ_2 \ket{\Psi}$
from Equation~({\ref{eq:pvc.q2}) and we insert it into  Equation~({\ref{eq:pvc.q1}). This allows us
to express this latter equation as:
\beqn
\nonumber 
(\omega - \hQ_1 \hH \hQ_1) \hQ_1 \ket{\Psi} &=&
    \hQ_1 \hH \hP \frac{1}{\omega - \hP \hH \hP+i\eta} \hP \hH \hQ_1 \ket{\Psi}
 \\ \nonumber &+& 
  \hQ_1 \hH \hP \frac{1}{\omega - \hP \hH \hP+i\eta} \hP \hH \hQ_2 \ket{\Psi}
 \\ \nonumber  &+& 
 \hQ_1 \hH \hQ_2 \frac{1}{\omega - \hQ_2 \hH \hQ_2+i\eta} \hQ_2 \hH \hP \ket{\Psi}
  \\ 
 &+&  \hQ_1 \hH \hQ_2 \frac{1}{\omega - \hQ_2 \hH \hQ_2+i\eta} \hQ_2 \hH \hQ_1 \ket{\Psi}
,
\eeqn
where we inserted in the denominator a term  $i \eta$ to avoid divergences. In the two 
terms containing  $\hP \ket{\Psi}$ and $\hQ_2 \ket{\Psi}$,  we could insert again
the results of Equation~({\ref{eq:pvc.p}) and Equation~({\ref{eq:pvc.q2}) and we obtain terms 
with many denominator factors. We neglect these terms and obtain an eigenvalue
equation of the form \cite{she20}
\beqn
\nonumber 
{\hat {\cal H}}(\omega) \hQ_1 \ket{\Psi^{\cal R}}_N &=& 
( \Omega_N  - i \Gamma_N ) \hQ_1 \ket{\Psi}_N
\\ \nonumber 
= \Bigg[ \hQ_1 \hH \hQ_1 
&+& \hQ_1 \hH \hP \frac{1}{\omega - \hP \hH \hP+i\eta} \hP \hH \hQ_1 
\\ &+& 
\hQ_1 \hH \hQ_2 \frac{1}{\omega - \hQ_2 \hH \hQ_2+i\eta} \hQ_2 \hH \hQ_1
\Bigg] \hQ_1 \ket{\Psi^{\cal R}}_N
,
\label{eq:pvc.eigen}
\eeqn
where we distinguished the energy $\omega$ characterizing the effective hamiltonian $\hat {\cal H}$ 
from the energy eigenvalue which can be complex because of the
imaginary parts inserted in the denominators. 

Since the Hilbert subspace spanned by the $\hQ_1 \ket{\Psi^{\cal R}}$ states is composed by 
$1p-1h$ components
only, we can expand each  $\hQ_1 \ket{\Psi^{\cal R}}$ state in terms of RPA 
eigenstates $\ket{\nu}$ which form a basis
\beq
\hQ_1 \ket{\Psi^{\cal R}}_N = \sum_\nu {\cal F}_\nu^N \ket{\nu} 
,
\eeq
and write the eigenvalue Equation (\ref{eq:pvc.eigen}) in a matrix form
\beq
\sum_{\nu'} \braket{\nu | {\hat {\cal H}}(\omega) | \nu'} {\cal F}_{\nu'}^N = 
 ( \Omega_N  - i \Gamma_N ) {\cal F}_\nu^N
 .
\eeq 
The solution of the above eigenvalue problem provides the values of the 
$ {\cal F}_{\nu'}^N$ coefficients.
The transition probability of a transition from the ground state $\ket{\Psi^{\cal R}}_0$
to an excited state induced by a one-body operator $\hF$ is given by:
\beq
_N\bra{\Psi^{\cal R}} \hQ^+_1 \hF  \hQ^+_1 \ket{\Psi^{\cal R}}_0 = 
\sum_\nu {\cal F}_\nu^N   \braket {\nu| \hF | \nu_0 }
= \sum_\nu {\cal F}_{\nu}^N
\sum_{mi} \left( X^\nu_{mi} f_{mi} + Y^\nu_{mi} f_{im}  \right)
,
\eeq
where we used the result of Equation~(\ref{eq:rpa.tprob}).

In this approach, one has first to solve RPA equations for various multipoles which have to be 
inserted in the sums on $\nu$. The choice of RPA solutions to be inserted is an input of the method
and it is based on plausible physics hypotheses. 

\subsection{Renormalized RPA}
\label{sec:HRPA.renRPA}

The extensions of RPA theory presented in Sections \ref{sec:HRPA.srpa} and \ref{sec:HRPA.pvc}
aimed at including excitation modes more complicated than $1p-1h$. The renormalized RPA (r-RPA)
attacks another weak point of RPA theory: the QBA (\ref{eq:rpa.qba}). This approximation 
forces pairs of fermionic operators to work as  they would be bosonic operators. For this reason, 
in the literature, the QBA is associated to the statement that RPA violates the Pauli principle. 
The  r-RPA theory avoids the use of the QBA. 

As in the ordinary RPA, we indicated with $\ket{\nu_0}$ the ground state of the system and
with $\ket{\nu}$ the excited state which is a combination of $1p-1h$ and $1h -1p$ excitations.
We consider a $\hQ^+_\nu$ operator whose action is 
\beq
\ket{\nu} = \hQ^+_\nu \ket{\nu_0} \equiv \sum_{ph} 
\left( X^\nu_{ph} \hcB^+_{ph} - Y^\nu_{ph} \hcB_{ph} \right) \ket{\nu_0}
,
\eeq
where the renormalized $p-h$ operator is
\beq
\hcB^+_{ph} \equiv \sum_{p'h'} N_{ph,p'h'} \ha^+_p \ha_h
,
\eeq
and  $N_{ph,p'h'}$ is a number. The EOM method implies that the 
correlated ground state satisfies the equation
\[
\hQ_\nu \ket{\nu_0} = 0
.
\]
By using the anti-commutation relations (\ref{eq:app.anta}) of the creation and destruction 
operators, we express the orthonormality condition relating the excited states as 
\beqn
\nonumber
&~& \delta_{\nu \nu'} = \braket{\nu | \nu'} = \braket{\nu_0 | \left[ \hQ_{\nu'} , \hQ^+_{\nu} \right] | \nu_0} 
\\ \nonumber
&=&  
\sum_{ph,p'h'} \left(
X^{\nu'\,*}_{p'h'} X^{\nu}_{ph}
\braket{\nu_0 | \left[ \hcB_{p'h'} , \hcB^+_{ph} \right] | \nu_0} 
+ Y^{\nu'\,*}_{p'h'} Y^{\nu}_{ph}
\braket{\nu_0 | \left[ \hcB^+_{p'h'} , \hcB_{ph} \right] | \nu_0} 
\right)
\\ 
&=&  \nonumber
\sum_{ph,p'h'} \left( X^{\nu'\,*}_{p'h'} X^{\nu}_{ph} - Y^{\nu'\,*}_{p'h'} Y^{\nu}_{ph} \right)
\\ 
&~& \sum_{mi,nj} N^*_{p'h',mi} N_{ph,nj}
\left( \delta_{mn}  \braket{\nu_0 |  \ha^+_j \ha_i | \nu_0} 
- \delta_{ij}  \braket{\nu_0 |  \ha^+_n \ha_m | \nu_0} 
\right)
.
\eeqn
The above expression is simplified if we use the s.p. basis formed by the natural orbits.
By definition, this is the basis where the one-body density matrix 
is diagonal \cite{ari07}
\beq
 \braket{\nu_0 |  \ha^+_\alpha \ha_\beta | \nu_0} = n_\alpha \delta_{\alpha \beta}
 .
\eeq
If we assume that 
\beq
N_{ph,p'h'} = \frac {\delta_{p p'} \delta_{h,h'}} {\sqrt{n_h - n_p}}
,
\eeq
we obtain
\beq
\sum_{ph,p'h'} \left( X^{\nu'\,*}_{p'h'} X^{\nu}_{ph} - Y^{\nu'\,*}_{p'h'} Y^{\nu}_{ph} \right)
= \delta_{\nu \nu'}
,
\eeq
which is an expression analogous to that of the standard RPA, Equation (\ref{eq:RPA:normalization}).
It is worth  remarking that now the indexes $p$ and $h$ do not refer any more to s.p. states
which, in the ground state, are fully occupied or completely empty. 
The natural orbit s.p. states are partially occupied with probability $n_\alpha$; therefore,  
all the indexes in
the sums of the above equations run on the full configuration space. To avoid double counting, 
we assume that the $i,j,k,h$ indexes indicate natural orbits with energies smaller than
those of the states labelled with the $m,n,p,q$ indexes.

We proceed by using the EOM approach analogously to what was indicated in \mbox{Section  
\ref{sec:RPA.rpa}} where, now, the $\ha^+_p \ha_h$ operators are substituted by $\hcB^+_{ph}$,
and we define the following matrix elements
\beq
{\mathbb A}_{php'h'} \equiv 
\braket{\nu_0 | \left[ \hcB_{ph}, [\hH, \hcB^+_{p'h'}]\right] | \nu_0} 
,
\eeq
and
\beq
{\mathbb B}_{php'h'} \equiv 
- \braket{\nu_0 | \left[ \hcB_{ph}, [\hH, \hcB_{p'h'}]\right] | \nu_0} 
.
\eeq
We obtain a set of equations analogous to those of the usual RPA
\beq
\left(
\begin{array}{cc}
{\mathbb A} & {\mathbb B} \\
   & \\
{\mathbb B}^*  & {\mathbb A}^*
\end{array}
\right)
\left(  
\begin{array}{c}
X^\nu \\
\\
Y^\nu
\end{array}
\right) = 
\omega_\nu
\left(  
\begin{array}{c}
X^\nu \\
\\
- Y^\nu
\end{array}
\right) 
\label{eq:rrpa}
.
\eeq

The evaluation of the ${\mathbb A}$ and ${\mathbb B}$ matrix elements is carried out
by using the expressions of the $\hcB$ operators in terms of $\hQ$ operators
\beqn
\hcB^+_{ph}  &=& \sum_{\nu} 
\left(X^{\nu *}_{ph} \hQ^+_{\nu} + Y^\nu_{ph} \hQ_{\nu} \right) 
,
 \\
 \hcB_{ph}  &=& \sum_{\nu} 
\left(X^\nu_{ph} \hQ_{\nu} + Y^{\nu *}_{ph}  \hQ^+_{\nu} \right) 
,
\eeqn
and we obtain
\beqn
\nonumber
{\mathbb A}_{php'h'} &=&
\half \left( \sqrt{ \frac{n_h - n_p}{n_{h'} - n_p' } } +
\sqrt{ \frac{n_{h'} - n_{p'}}{n_h - n_p } }\right) 
({\tilde \epsilon}_{p p'} \delta_{h h'} - {\tilde \epsilon}_{h h'} \delta_{p p'})\\
&~& 
+ \sqrt{(n_h - n_p) (n_{h'} - n_{p'} )} \barv_{p h' h p'} 
\label{eq:rrpa.A}
,
\eeqn
and
\beq
{\mathbb B}_{php'h'} = \sqrt{(n_h - n_p) (n_{h'} - n_{p'} )} \barv_{h h' p p'} 
.
\label{eq:rrpa.B}
\eeq
In the above expressions, we  used the natural orbit energies defined as
\beq
{\tilde \epsilon}_{\alpha \alpha'} = \braket{\alpha |  \frac{- \hbar^ 2\nabla^2}{2 m} | \alpha'} 
+ \sum_\beta n_\beta \barv_{\alpha \beta \alpha' \beta}
.
\eeq

The key point of the r-RPA consists in expressing the occupation probabilities $n_\alpha$ 
in terms of $X$ and $Y$ amplitudes. In Ref. \cite{cat96}, by using a  method which iterates the 
anti-commutation relations of the creation and destruction operators,  it is shown that the
expressions of these occupation probabilities up to the fourth order in $Y$ are 
\beq
n_h \simeq 1- \sum_{p} \sum_{\nu \nu'} \Delta_{p h}^{\nu \nu'}  \; \; ; \; \;  
n_p \simeq \sum_{h} \sum_{\nu \nu'} \Delta_{p h}^{\nu \nu' }
,
\eeq
with
\beq
\Delta_{p h}^{\nu \nu' } = \left( \delta_{\nu \nu'} 
-\half \sum_{p'h'} (n_{h'} - n_{p'})  X^{\nu'}_{p'h'} X^{\nu *}_{p'h'} 
 \right) (n_h - n_p) Y^{\nu}_{ph}  Y^{\nu'\,*}_{ph} 
 .
\eeq

This result inserted in Equations (\ref{eq:rrpa.A}) and (\ref{eq:rrpa.B})
generates expressions of  the ${\mathbb A}$ and ${\mathbb B}$ matrix elements
in terms of $X$ and $Y$ amplitudes; therefore, 
Equation (\ref{eq:rrpa}) becomes  a system of nonlinear equations in the latter
unknowns. This is solved by using an iterative procedure. 
Starting from some initial guess for the $X$ and $Y$ amplitudes, 
obtained, for example, by solving the standard RPA equations, 
one calculates the  ${\mathbb A}$ and ${\mathbb B}$ matrix elements.
The solution of Equation (\ref{eq:rrpa}) provides new values of the 
$X$ and $Y$ amplitudes. The procedure is repeated until convergence 
is reached. A review of recent applications of the r-RPA theory is presented
in Ref. \cite{sch21}.


\section{Correlated RPA}
\label{sec:CCRPA}
Interactions built to describe systems composed of two particles are called \emph{microcospic}.
These interactions show similar features independently of the 
particles considered, nucleons, atoms or molecules. They are short ranged, meaning that 
they are zero after a certain value of the distance between the two particles. They 
have an attractive pocket at intermediate distances and a strongly repulsive core at short 
inter-particle distances. This latter feature inhibits the use of microscopic interactions in 
theories based on perturbation expansion such as RPA. The derivation of  the RPA with the 
Green function formalism clearly shows that RPA is the first term of a perturbative expansion
of the two-body Green functions. The presence of the strongly repulsive core would produce 
extremely large value of the interaction matrix elements with respect to the energy eigenvalues. 
This is because the s.p. wave functions obtained in the IPM would allow the particles to get too close
to each other. The traditional RPA requires the use of 
effective interactions, i.e., interactions which do not contain a short range repulsion. 

Microscopic many-body theories aim to describe many-particle systems by using
microscopic interactions. One method of handling the problem 
of  short range repulsion is to use a correlation function which modifies the IPM wave functions
in such a way that two particles do not get too close to each other. 
This is the basic idea of the Correlated Basis Function theory \cite{pan79,kro02,ari07}. 
The ansatz is that the ground state of the interacting particle system can be  expressed as
\beq
\ket{\Psi_0} = \frac { F \ket{\Phi_0} } {\braket{\Phi_0 | F^+ F | \Phi_0 }^\half}
,
\label{eq:ccrpa.psi0}
\eeq
where $\ket{\Phi_0}$ is the IPM Slater determinant and $F$ is a correlation function.
These two elements of the state are determined by minimizing the energy functional
\beq
\delta E[\Psi_0] = \delta \left[
\frac {\braket{\Phi_0 | F^+ \hH F | \Phi_0 }} {\braket{\Phi_0 | F^+ F | \Phi_0 }}
\right]  = 0
,
\eeq 
where the hamiltonian $\hH$ contains the microscopic interaction. The usual ansatz
on the expression of the correlation function $F$ is 
\beq
F ( \br_1 , \cdots , \br_A) = \prod_{i<j}^A f(r_{ij})
,
\eeq
where $f$ is a two-body correlation function depending only on the distance $r_{ij}$ between
the two interacting fermions. The need to keep finite the product of the interaction $\hV$
and the wave function $\ket{\Psi}$ requires that $f$ is almost 
zero for small values of $r_{ij}$ and it rapidly assumes the value of 1 when the distance becomes larger 
than that of the short range repulsive core. The minimization of the energy functional is carried out
by changing the parameters of $f$ and also the set of s.p. wave functions forming $\ket{\Phi_0}$.

After having solved the problem of finding the minimum of $E[\Psi_0]$, 
the correlated RPA aims to describe the excitations of the system 
 in this theoretical framework. 
There is an ambiguity in defining the expression of the excited state. 
If we consider the $\ket{\Psi_0}$ of Equation~(\ref{eq:ccrpa.psi0}) as ground state, 
the approach of the EOM (see Section~\ref{sec:EOM}) implies the calculation of 
matrix elements of the form
\beq
\braket{\Phi_0 | F^+ \ha^+_i \ha_m \hH \ha^+_n \ha_j F | \Phi_0}
,
\eeq
whose evaluation requires the knowledge of  the effects of creation and destruction
operators on $F \ket{\Phi_0}$. We attack the problem by considering the correlation function
acting on an exited IPM state. This implies that the
ansatz for the excited states is analogous to that of Equation~(\ref{eq:ccrpa.psi0})
\beq
\ket{\Psi} = \frac { F \ket{\Phi} } {\braket{\Phi | F^+ F | \Phi }^\half}
,
\label{eq:ccrpa.psi1}
\eeq
where $\ket{\Phi}$ is the Thouless variational ground state (\ref{eq:rpa.thouless}) 
\beq
\ket{\Phi} = e^{\displaystyle{\sum_{mi} C_{mi} \ha^+_m \ha_i }} \ket{\Phi_0}
,
\label{eq:ccrpa.phi}
\eeq
and the $C_{mi}$ coefficients are defined by using a variational procedure which
minimizes the energy functional
\beq
\delta E[ \Psi] = \delta \braket {\Psi | \hH | \Psi} = 0 
.
\eeq

In the expression  (\ref{eq:ccrpa.phi}) of $\ket{\Phi}$, we can consider $E$
as a function of the $C_{mi}$ coefficients and we make a power expansion
around the ground state energy value
\beq
H_{00} = \frac {\braket{\Phi_0 | F^+ \hH F | \Phi_0 }} {\braket{\Phi_0 | F^+ F | \Phi_0 }}
,
\eeq
which is obtained by considering all the $C_{mi}$ coefficients
equal to zero in Equation (\ref{eq:ccrpa.phi}). The power expansion can be written as
\beqn
\nonumber E[C_{mi},C^*_{mi}]
=&~& H_{00} + \sum_{mi} \left(  \left[ \frac {\delta E} {\delta C_{mi} } \right]_0 C_{mi} 
+ \left[ \frac {\delta E} {\delta C^*_{mi} } \right]_0 C^*_{mi} 
\right)
 \\ \nonumber &+&
\half \sum_{minj}  \left[ \frac {\delta^2 E} {\delta C_{mi} \delta C_{nj} } \right]_0 C_{mi}  C_{nj} 
+ \half \sum_{minj}  \left[ \frac {\delta^2 E} {\delta C^*_{mi} \delta C^*_{nj} } \right]_0 C^*_{mi}  C^*_{nj} 
\\ &+&
 \sum_{minj}  \left[ \frac {\delta^2 E} {\delta C^*_{mi} \delta C_{nj} } \right]_0 C^*_{mi}  C_{nj} 
+ \cdots
,
\label{eq:ccrpa.taylor}
\eeqn
where the subindex $0$ indicates that, after the evaluation of the variational derivative, all 
the $C$'s must be set equal to zero. 

The second term of the above equation is $\delta E$, the first variation of the energy functional. 
We obtain the minimum when this variation is zero and this implies that each variational term
must be zero. Let us consider the term with the variation about $C^*_{mi}$
\beq
 \frac {\delta E} {\delta C^*_{mi} } 
 = \frac{\partial}{\partial C^*_{mi}} \left[
\frac {\braket{\Phi | F^+ \hH F | \Phi }} {\braket{\Phi | F^+ F | \Phi }}
\right]
,
\eeq
where we have considered that, in this case, the functional derivative coincides with 
the partial derivative. By using the expression (\ref{eq:ccrpa.phi}), we obtain
\beq
 \frac {\delta E} {\delta C^*_{mi}} 
 =  \frac {\braket{\Phi | \ha^+_i \ha_m F^+ \hH F | \Phi }} {\braket{\Phi | F^+ F | \Phi }}
- {\braket{\Phi | \ha^+_i \ha_m F^+ F | \Phi }} 
 \frac {\braket{\Phi | F^+ \hH F | \Phi }} {\braket{\Phi | F^+ F | \Phi }^2}
 .
\eeq
After calculating the variation, we have to impose that all the $C$'s go to zero; this is equivalent to 
saying that in Equation~(\ref{eq:ccrpa.phi}) $\ket{\Phi} = \ket{\Phi_0}$, and we obtain a relation
\beq
\frac {\braket{\Phi_0 | \ha^+_i \ha_m F^+ \hH F | \Phi_0 }} {\braket{\Phi_0 | F^+ F | \Phi_0 }}
= H_{00} \frac {\braket{\Phi_0 | \ha^+_i \ha_m F^+ F | \Phi_0 }}  
                      {\braket{\Phi_0 | F^+ F | \Phi_ 0}}
                      .
\label{eq:ccrpa.brillouin}
\eeq
An analogous calculation carried out for the variation about $C_{mi}$ generates an expression
which is the complex conjugate of  (\ref{eq:ccrpa.brillouin}).

The fact that the value of $E$ is a minimum,
when the first variational derivatives are zero,  is ensured
if the sum of all the second order variational derivatives is positive. It is convenient to tackle this
problem in matrix form by defining the matrix elements
\beq
A_{minj} \equiv  \left[ \frac {\delta^2 E} {\delta C^*_{mi} \delta C_{nj}} \right]_0
\;\;\;\;\;\;{\rm and} \;\;\;\;\;\;
B_{minj} \equiv  \left[ \frac {\delta^2 E} {\delta C^*_{mi} \delta C^*_{nj}} \right]_0
.
\eeq
By carrying out calculations analogous to those carried out for the first variational
derivatives, i.e., by considering Equation~(\ref{eq:ccrpa.phi}) and making the limit for $C \rightarrow 0$,
we obtain the expressions
\beq
A_{minj} = 
\frac  { \braket{\Phi_0 | \ha^+_i \ha_m F^+ \hH F \ha^+_n \ha_j | \Phi_0 } }   
       {\braket{\Phi_0 | F^+ F | \Phi_0 } }
- H_{00} 
\frac {\braket{\Phi_0 | \ha^+_i \ha_m F^+ \hH F  \ha^+_n \ha_j | \Phi_0 }}  {\braket{\Phi_0 | F^+ F | \Phi_ 0}}
.
\label{eq:ccrpa:aminj}
\eeq
and 
\beq
B_{minj} = 
\frac  { \braket{\Phi_0 | \ha^+_i \ha_m \ha^+_j \ha_n  F^+ \hH F | \Phi_0 } }   
       {\braket{\Phi_0 | F^+ F | \Phi_0 } }
- H_{00} 
\frac {\braket{\Phi_0 |  \ha^+_i \ha_m \ha^+_j \ha_n F^+ \hH F | \Phi_0 }} 
 {\braket{\Phi_0 | F^+ F | \Phi_ 0}}
.
\label{eq:ccrpa:bminj}
\eeq

We consider the set of the $C$'s as a vector; therefore, we write
the condition that the sum of the second variational derivative is positive 
in matrix form as
\beq
\half (C^{*\,T}  C^T) 
\left( \begin{array}{cc}
A & B \\
   & \\
B^*  & A^*
\end{array} 
\right)
\left( \begin{array}{c}
C \\ \\ C^*
\end{array} 
\right)
>  0
.
\label{eq:ccrpa.mat1}
\eeq
This is equivalent to asking that in the eigenvalue problem
\beq
\left( \begin{array}{cc}
A & B \\
   & \\
B^*  & A^*
\end{array} 
\right)
\left( \begin{array}{c}
C \\ \\ C^*
\end{array} 
\right)
= \lambda
\left( \begin{array}{c}
C \\ \\ C^*
\end{array} 
\right)
,
\label{eq:ccrpa.mat2}
\eeq
the eigenvalues $\lambda$ are all positive. 
By inserting Equation~(\ref{eq:ccrpa.mat2}) into  Equation~(\ref{eq:ccrpa.mat1}), we obtain
\beq
\half (C^{*\,T} C + C^T \, C ) > 0
,
\eeq
which is satisfied for $\lambda>0$ since the part inside the round brackets is certainly
positive because it is the sum of squares moduli of complex numbers. 

The condition (\ref{eq:ccrpa.brillouin}) and its complex conjugate, together with 
(\ref{eq:ccrpa.mat1}) allows us to build equations for the Correlated RPA. 
 
We consider Equation~(\ref{eq:D.var2})
\beq
\braket{\delta \Psi(t) | \hH -
i \hbar \frac{\partial}{\partial t} | \Psi(t)} = 0
,
\label{eq:ccrpa.var1}
\eeq
where now the state $\ket{\Phi(t)}$ is 
\beq
\ket{\Psi(t)} = \frac { F \ket{\Phi(t)} } {\braket{\Phi(t) | F^+ F | \Phi(t)}^\half}
,
\label{eq:ccrpa.psit}
\eeq
In the above equation, 
the $\ket{\Phi(t)}$ states are defined  analogously to Equation~(\ref{eq:ccrpa.phi})
but now the $C$ coefficients are time-dependent 
\beq
\ket{\Phi(t)} = e^{\displaystyle{\sum_{mi} C_{mi}(t) \ha^+_m \ha_i }} \ket{\Phi_0(t)}
,
\label{eq:ccrpa.phit}
\eeq
and the time dependence of the ground state is defined as
\beq
\ket{\Phi_0(t)} = e^{ \frac i \hbar H_{00} t} \ket{\Phi_0}
,
\label{eq:ccrpa.phi0t}
\eeq
 analogously to the usual interaction picture (see Equation (\ref{eq:p.sint})). 

Since only the $C$ amplitudes can be varied, 
we can express Equation~(\ref{eq:ccrpa.var1}) as
\beqn
\nonumber
&~& 
\braket{\delta \Psi(t) | \hH -
i \hbar \frac{\partial}{\partial t} | \Psi(t)}  
\\ \nonumber &=&
\sum_{mi} \frac{\delta \bra{\Psi(t)} } {\delta C_{mi}} 
\left( \hH - i \hbar \frac{\partial}{\partial t} \right) \ket{ \Psi(t)}  \delta C_{mi} 
\\ \nonumber &+&
\sum_{mi} \frac{\delta \bra{\Psi(t)} } {\delta C^*_{mi}} 
\left( \hH - i \hbar \frac{\partial}{\partial t} \right) \ket{ \Psi(t)}  \delta C^*_{mi} 
\\
&\equiv&
\sum_{mi} \left( S_{mi} \delta C_{mi} + R_{mi} \delta C^*_{mi} \right) = 0
.
\label{eq:ccrpa.rmi}
\eeqn
The above equation is verified only if both the matrix elements $R_{mi}$ and $S_{mi}$ 
are zero for all the $m$ and $i$ particle--hole pairs and for all times $t$. 

The evaluation of $R_{mi}$ proceeds by considering the expressions (\ref{eq:ccrpa.psit}) for 
$\ket{\Psi(t)}$,  (\ref{eq:ccrpa.phit}) for $\ket{\Phi(t)}$ and (\ref{eq:ccrpa.phi0t}) for 
$\ket{\Phi_0(t)}$. We show in Appendix \ref{sec:app.tdhf}
the details of the calculation leading to the expression
\beq
0 = R_{mi} = \sum_{nj} A_{minj} C_{nj}(t) + \sum_{nj} B_{minj} C^*_{nj}(t)
- i \hbar \sum_{nj} \frac {d} {dt} C_{nj} M_{minj} 
,
\label{eq:ccrpa.rmi0}
\eeq
where the $A$ and $B$ matrix elements are those of Equations~(\ref{eq:ccrpa:aminj}) and
(\ref{eq:ccrpa:bminj}), respectively, and we defined
\beqn
\nonumber
M_{minj} &=& 
\frac  { \braket{\Phi_0 | \ha^+_i \ha_m  F^+ F \ha^+_n \ha_j | \Phi_0 } }   
       {\braket{\Phi_0 | F^+ F | \Phi_0 } }
\\ &-& 
\frac { \braket{\Phi_0 |  \ha^+_i \ha_m F^+ F | \Phi_0 } 
          \braket{\Phi_0 |  F^+ F \ha^+_n \ha_j | \Phi_0 } 
} 
 {\braket{\Phi_0 | F^+ F | \Phi_ 0}^2}
 .
\label{eq:ccrpa:mminj}
\eeqn

Analogously to what is presented in Section \ref{sec:TDHF},  we consider harmonic oscillations
of the $C$ amplitudes
\beq
C_{mi}(t) = X_{mi} e^{-i \omega t} + Y^*_{mi} e^{i \omega t} 
.
\label{eq:ccrpa.cmi}
\eeq
We insert Equation~(\ref{eq:ccrpa.cmi}) into  Equation~(\ref{eq:ccrpa.rmi0}); we separate the positive 
and negative frequency oscillations and obtain
\beq
\sum_{nj} A_{minj} X_{nj} + \sum_{nj} B_{minj} Y_{nj}(t)
=  \hbar \omega  \sum_{nj}  X_{nj} M_{minj} 
,
\eeq
and 
\beq
\sum_{nj} A_{minj} Y^*_{nj} + \sum_{nj} B_{minj} X^*_{nj}(t)
=  - \hbar \omega  \sum_{nj}  Y^*_{nj} M_{minj} 
.
\eeq

By considering the complex conjugated of the second equation we can cast 
the system in a matrix form
\beq
\left( \begin{array}{cc}
A & B \\
   & \\
B^*  & A^*
\end{array} 
\right)
\left( \begin{array}{c}
X^\nu \\ \\ Y^\nu
\end{array} 
\right)
= \hbar \omega_\nu
\left( \begin{array}{cc}
M & 0 \\
   & \\
0  & - M
\end{array} 
\right)
\left( \begin{array}{c}
X^\nu \\ \\ Y^\nu
\end{array} 
\right)
.
\label{eq:ccrpa.mat3}
\eeq

The structure of the standard RPA equations can be recovered by performing a transformation
on (\ref{eq:ccrpa.mat3}) such that the matrix on the right is converted into a unit-diagonal form
\beq
\left( \begin{array}{cc}
M & 0 \\
   & \\
0  & - M
\end{array} 
\right)
\longrightarrow
\left( \begin{array}{cc}
{\mathbb I} & 0 \\
   & \\
0  & - {\mathbb I} 
\end{array} 
\right)
.
\eeq
The properties of these equations have been studied \cite{row70} and they are similar to those
quoted in Section \ref{sec:propeRPA}. 

Obviously, we want to interpret the eigenvalues $\hbar \omega$ of Equation~(\ref{eq:ccrpa.mat3})
as excitation energies of the system. The question is if the amplitudes $X$ and $Y$ can be used
as  in Section~\ref{sec.eom.trans} to evaluate the transition probabilities. 
This is not straightforward since in the present approach we have worked with a hamiltonian of
the type $\hF^+ \hH \hF$. Consequently,  the one-body operators describing the external operator
should also be described as $\hF^+ {\hat {\cal O}} \hF$. 


\section{Summary and Conclusions}
\label{sec:conclusions}

In this article, I presented three different methods to obtain RPA secular equations.

The EOM approach emphasizes the fact that RPA considers only excitations of $1p-1h$ 
type and also that the RPA ground state is not the the IPM ground state, but it contains correlations.
These correlations are described in terms of $ph$ pairs;  therefore, RPA excited states 
contain also $hp$ excitations which are taken into account by the $Y$ amplitudes. 

 RPA secular equations are obtained by truncating at the first order the expansion
of the two-body Green function in powers of the interaction. As a consequence of this
truncation, RPA
requires the use of effective interactions, i.e., interactions without the strongly repulsive
core at short inter-particle distances, a feature which, instead,
characterizes the microscopic interactions. 

The derivation of RPA obtained with the TDHF approach clearly indicates that RPA
has to be used to describe harmonic oscillations around the many-body ground state, i.e.
excitations whose energies are relatively small with respect to the global binding energy of the 
system. 

RPA calculations require in input a set of s.p. wave functions and energies and also the 
effective particle--hole interaction. The solution of RPA secular equations 
provides not only the excitation spectrum, but for each excited state also the description
of the corresponding wave function in terms of $1p-1h$ and $1h-1p$ excitation pairs. 
The knowledge of RPA wave functions allows a rather straightforward evaluation of 
observable quantities because many-body transition amplitudes are expressed 
as linear combinations of s.p. transitions. 

 RPA is able to describe in a unique theoretical framework both single-particle and
collective excitations. This is particularly useful in atomic nuclei where these two types
of excitations are both present in the same energy range. 

 RPA is able to predict emergent phenomena which are unexpected in the IPM.
In the present article, I have considered, as an  illustrative example, the case of the $3^-$ state
of the $^{208}$Pb nucleus.  RPA has been widely used to investigate the giant 
resonances in nuclei~\cite{spe91}. The position of the peaks of the resonances and
the total strengths are well described. These latter quantities are related to RPA sum rules
whose values are rather different from those obtained in the IPM, as was pointed
out in Section \ref{sec:eom.srule}.
The accuracy of most modern data indicates that, even though RPA provides reasonable
values of the total excitation strengths, it fails in describing their energy distributions. This is the
main reason leading to an extension of the theory.

The main limitation of RPA is the fact that it considers $1p-1h$ excited pairs only.
The straightforward extension consists in considering, in addition, also $2p-2h$
excitations. The formulation of the SRPA has been presented in Section \ref{sec:HRPA.srpa}.
Applications of the SRPA are numerically very involved, but the obtained results
are rather promising. 
Another method of including $2p-2h$ excitations consists in coupling s.p. wave functions to 
RPA vibrational modes. 

It is possible to untie RPA theory from the use of effective interactions. The formulation
of a theory which uses microscopic interactions has been presented in \mbox{Section \ref{sec:CCRPA}.}
To the best of my knowledge, this formulation of RPA has never been used in actual 
calculations. Its validity in the description of observables remains an open question. 

 RPA is a milestone in many-body theories even though nowadays its role and
relevance is sometime overlooked because its relative simplicity in favour of theories
makes use of microscopic interactions.

\vspace{6pt}

\begin{appendices}
\appendix

\section{The Hartree--Fock Hamiltonian}
\label{sec:app:HF}
In this Appendix we obtain a useful expression of the hamiltonian for 
its use in HF and RPA calculations.
We consider the expression of the hamiltonian in ONR \cite{rin80,fet71}
\beq
\hH = \sum_{{\nu}{\nu}'}  T_{\nu,\nu'} 
\ha^+_{\nu} \ha_{\nu'}+\frac{1}{4}
\sum_{{\nu}{\nu}'{\mu}{\mu}'} \barv_{\nu\mu \nu'\mu'}
\ha^+_{\nu} \ha^+_{\mu} \ha_{\mu'} \ha_{\nu'} 
,
\label{eq:app.ham1}
\eeq
where we have defined
\beq
T_{\nu,\nu'} = \langle\nu|{\hat T}|\nu'\rangle 
,
\eeq
and $\barv_{\nu\mu \nu'\mu'}$ is the antisymmetric matrix element of Equation~(\ref{eq:v.avu}). 
We indicate with
$\ha^+_\nu$ and $\ha_\nu$ the usual fermion creation and destruction operators
satisfying the anti-commutation relations
\beq
\label{eq:app.anta}
\left\{\ha_{\nu}, \ha^+_{\nu'}\right\} = \delta_{\nu \nu'} \quad\quad 
\left\{\ha_{\nu}, \ha_{\nu'}\right\} = 0 \quad\quad
\left\{\ha^+_{\nu}, \ha^+_{\nu'}\right\} = 0
,
\eeq
From the definition of contraction (see \cite{fet71}) we have that
\beq
\acontraction{}{\ha^+_\nu}{}{\ha_{\nu'}}
\ha^+_\nu \, \ha_{\nu'} = \delta_{\nu \nu'} \delta_{\nu' i} \,\,\,\,;\,\,\,\,
\acontraction{}{\ha_\nu}{}{\ha^+_{\nu'}}
\ha_\nu \, \ha^+_{\nu'} = 0 \,\,\,\,;\,\,\,\,
\acontraction{}{\ha_\nu}{}{\ha_{\nu'}}
\ha_\nu \, \ha_{\nu'} = 0 \,\,\,\,;\,\,\,\,
\acontraction{}{\ha^+_\nu}{}{\ha^+_{\nu'}}
\ha^+_{\nu} \, \ha^+_{\nu'} = 0 
\,\,\,.
\eeq
where $i$ indicates a state below the Fermi surface.
By considering the definition of normal ordered product $\hN$ we obtain
\beq
\ha^+_\nu \, \ha_{\nu'} = \hN[ \ha^+_\nu \, \ha_{\nu'} ] +
\acontraction{}{\ha^+_\nu}{}{\ha_{\nu'}} 
\ha^+_\nu \, \ha_{\nu'} 
\label{eq:app1.enne}
\,\,\,,
\eeq
and, for the Wick's theorem,
\beqn
\nonumber
\ha^+_\nu \ha^+_\mu \, \ha_{\mu'} \, \ha_{\nu'} &=& 
\hN[\ha^+_\nu \ha^+_\mu \, \ha_{\mu'} \, \ha_{\nu'}] \\
\nonumber
&+& 
\hN[\ha^+_\mu \ha_{\mu'}] 
\acontraction{}{\ha^+_\nu}{}{\ha_{\nu'}} 
 \ha^+_{\nu} \, \ha_{\nu'} 
+
\hN[\ha^+_\nu \ha_{\nu'}] 
\acontraction{}{\ha^+_\mu}{}{\ha_{\mu'}} 
 \ha^+_{\mu} \, \ha_{\mu'} \\
 \nonumber
&-&
\hN[\ha^+_\mu \ha_{\nu'}] 
\acontraction{}{\ha^+_\nu}{}{\ha_{\mu'}} 
 \ha^+_{\nu} \, \ha_{\mu'} 
-
\hN[\ha^+_\nu \ha_{\mu'} ]
\acontraction{}{\ha^+_\mu}{}{\ha_{\nu'}} 
 \ha^+_{\mu} \, \ha_{\nu'} \\
 &+&
 \acontraction{}{\ha^+_\mu}{}{\ha_{\mu'}} 
 \ha^+_\mu \ha_{\mu'}
\acontraction{}{\ha^+_\nu}{}{\ha_{\nu'}} 
 \ha^+_{\nu} \, \ha_{\nu'} 
-
\acontraction{}{\ha^+_\nu}{} {\ha_{\mu'}}
\ha^+_\nu \ha_{\mu'}
\acontraction{}{\ha^+_\mu}{}{\ha_{\nu'}} 
 \ha^+_{\mu} \,\ha_{\nu'} 
\,\,\,.
\eeqn

We insert the above expression in Equation (\ref{eq:app.ham1}) 
\beqn
\nonumber
\hH &=& \sum_{\nu \nu'} T_{\nu \nu'} \ha^+_\nu \ha_{\nu'} 
+ \frac 1 4 \sum_{\mu \mu' \nu \nu'} \barv_{\nu \mu \nu' \mu'}
\Big\{\hN[\ha^+_\nu \ha^+_\mu \, \ha_{\mu'} \, \ha_{\nu'}] \\
\nonumber
&+&\hN[\ha^+_\mu \ha_{\mu'} ] \delta_{\nu \nu'} \delta_{\nu i}
+\hN[\ha^+_\nu \ha_{\nu'} ] \delta_{\mu \mu'} \delta_{\mu i}
-\hN[\ha^+_\mu \ha_{\nu'} ] \delta_{\nu \mu'} \delta_{\nu i}
-\hN[\ha^+_\nu \ha_{\mu'} ] \delta_{\mu \nu'} \delta_{\mu i} \\
&+& \delta_{\nu \nu'} \delta_{\nu i} \delta_{\mu \mu'} \delta_{\mu j}
- \delta_{\nu \mu'} \delta_{\nu i} \delta_{\mu \nu'} \delta_{\mu j} 
\Big\}
,
\eeqn
where we have already considered the fact that a contraction is different from 
zero only if the single-particle state is of hole type, i.e., if its energy is below
the Fermi surface.  

By considering the restrictions imposed by the Kronecker's  $\delta$, we obtain
\beqn
\nonumber
\hH &=& \sum_{\nu \nu'} T_{\nu \nu'} \ha^+_\nu \ha_{\nu'} 
+ \frac 1 4 \sum_{\mu \mu' \nu \nu'} \barv_{\nu \mu \nu' \mu'}
\hN[\ha^+_\nu \ha^+_\mu \, \ha_{\mu'} \, \ha_{\nu'}] \\
\nonumber
&+&
\frac 1 4 \sum_{\mu \mu' i} \barv_{\mu i \mu' i} \hN[\ha^+_\mu \ha_{\mu'} ] 
+
\frac 1 4 \sum_{\nu \nu' i} \barv_{i \nu i \nu'} \hN[\ha^+_\nu \ha_{\nu'} ]  \\
\nonumber
&-& \frac 1 4 
\sum_{\mu \nu' i} \barv_{i \mu \nu' i} \hN[\ha^+_\mu \ha_{\nu'} ] 
- \frac 1 4 \sum_{\nu \mu' i} \barv_{\nu i i \mu'} \hN[\ha^+_\nu \ha_{\mu'} ] 
\\
&+& \frac 1 4 \sum_{i j} \barv_{i j i j} 
- \frac 1 4 \sum_{i j} \barv_{i j j i} 
\,\,\,.
\eeqn

The definition  (\ref{eq:v.avu}) of the antisymmetric matrix element 
implies the following relations:
\beq
\barv_{\nu \mu \nu' \mu'} = - \barv_{\mu \nu \nu' \mu'}
=  \barv_{\mu \nu \mu' \nu' } = - \barv_{\nu \mu \mu' \nu'}
\,\,\,,
\eeq
therefore
\beqn
\nonumber
\hH &=& \sum_{\nu \nu'} T_{\nu \nu'} \ha^+_\nu \ha_{\nu'} 
+\frac 1 4 \sum_{\mu \mu' \nu \nu'} \barv_{\nu \mu \nu' \mu'}
\hN[\ha^+_\nu \ha^+_\mu \, \ha_{\mu'} \, \ha_{\nu'}] \\
&+&
\sum_{\nu \nu' i} \barv_{\nu i \nu' i} \hN[\ha^+_\nu \, \ha_{\nu'}] 
+\half \sum_{ij} \barv_{i j i j}
\label{eq:v.ham2}
\,\,\,.
\eeqn
We use the definition (\ref{eq:app1.enne}) of $\hN$ and we obtain the following 
expression for the hamiltonian
\beqn
\nonumber
\hH &=& \sum_{\nu \nu'} 
\left(T_{\nu \nu'}  + \sum_{i} \barv_{\nu i \nu' i} \right) \ha^+_\nu \ha_{\nu'} \\
&+&
\frac 1 4 \sum_{\mu \mu' \nu \nu'} \barv_{\nu \mu \nu' \mu'}
\hN[\ha^+_\nu \ha^+_\mu \, \ha_{\mu'} \, \ha_{\nu'}] 
 - \half \sum_{ij} \barv_{i j i j}
\,\,\,.
\label{eq:v.ham3a}
\eeqn
This expression makes evident the presence of a one-body hamiltonian
operator, the term multiplying $\ha^+_\nu \ha_{\nu'}$. 

Up to now, we did not make any assumption on the structure of the basis
of single-particle wave functions composing the Slater determinant
on which the creation and destruction operators are acting. 
We choose the single-particle basis which diagonalizes the
one-body term of Equation (\ref{eq:v.ham3a})
\beq
\left( T_{\nu \nu'}  + \sum_{i} \barv_{\nu i \nu' i} \right) \delta_{\nu,\nu'} 
\equiv h_{\nu \nu'} \delta _{\nu , \nu'} = \epsilon_\nu \delta_{\nu,\nu'} 
\label{eq:app.hdiag}
\,\,\,.
\eeq

In this basis, the expression of the hamiltonian is 
\beq
\hH = \sum_\nu \epsilon_\nu \ha^+_\nu \ha_{\nu}
- \half \sum_{ij} \barv_{i j i j}
+
\frac 1 4 \sum_{\mu \mu' \nu \nu'} \barv_{\nu \mu \nu' \mu'}
\hN[\ha^+_\nu \ha^+_\mu \, \ha_{\mu'} \, \ha_{\nu'}] 
\equiv \hH_0 +\hV_{\rm res}
\label{eq:app.ham44}
\,\,\,.
\eeq
where $\hH_0$ is the sum of the first two terms. 

\section{RPA Double Commutators}
\label{sec:app.doublec}

In this Appendix, we calculate the double commutator
\[
\braket { \Phi_0 | \Big[ \ha^+_i \ha_m , [\hH, \ha^+_n \ha_j] \Big] | \Phi_0}
,
\]
of Equation~(\ref{eq:rpa:tda2}) 
by considering the hamiltonian expressed as in Equation~(\ref{eq:app.ham44}).
The second term of the hamiltonian (\ref{eq:app.ham44}) is a number,
therefore commuting with every operator. 
By considering the anti-commutation rules (\ref{eq:app.anta}) 
of the creation and destruction operators,  we have that
\[
[\ha^+_\alpha \ha_\beta , \ha^+_n \ha_j] = 
\delta_{n \beta} \ha^+_\alpha \ha_j - \delta_{j \alpha} \ha^+_n \ha_\beta
,
\]
therefore, the commutator of the hamiltonian can be written as
\beqn
\nonumber
[\hH,\ha^+_n \ha_j] &=&  \sum_{\alpha \beta} h_{\alpha \beta} 
\left( \delta_{n \beta} \ha^+_\alpha \ha_j - \delta_{j \alpha} \ha^+_n \ha_\beta \right) \\
\nonumber &+& 
\frac{1}{4} \sum_{\alpha \alpha'  \beta \beta'} \barv_{\alpha \beta \alpha' \beta'}
\Big[ \hN[\ha^+_\alpha \ha^+_\beta \, \ha_{\beta'} \, \ha_{\alpha'}] , \ha^+_n \ha_j \Big]
.
\eeqn

The double commutator of the first term of the hamiltonian can be rewritten as
\beqn
\nonumber
&~& h_{\alpha \beta}
\braket{\Phi_0 | 
\Big[ \ha^+_i \ha_m , \left( \delta_{n \beta} \ha^+_\alpha \ha_j 
- \delta_{j \alpha} \ha^+_n \ha_\beta \right) \Big]
| \Phi_0} \\
\nonumber &=&
 h_{\alpha \beta} \braket{\Phi_0 | 
 \left( \ha^+_i \ha_m \delta_{n \beta} \ha^+_\alpha \ha_j - \ha^+_i \ha_m  \delta_{j \alpha} 
 \ha^+_n \ha_\beta \right)
| \Phi_0} \\
\nonumber &=&
h_{\alpha \beta} \braket{\Phi_0 | \ha^+_i \ha_m \ha^+_\alpha \ha_j | \Phi_0}  \delta_{n \beta} 
-h_{\alpha \beta} \braket{\Phi_0 | \ha^+_i \ha_m \ha^+_n \ha_\beta | \Phi_0}  \delta_{j \alpha} \\
\nonumber
&=& h_{\alpha \beta} \delta_{ij} \delta_{m \alpha} \delta_{n \beta} 
-  h_{\alpha \beta} \delta_{i \beta} \delta_{m n} \delta_{j \alpha} \\
&=& 
(\epsilon_m - \epsilon_i) \delta_{ij} \delta_{mn} 
,
\label{eq:app.pdfilon}
\eeqn
where in the last step we considered the diagonal expression of  $h_{\alpha,\beta}$, Equation~(\ref{eq:app.hdiag}).

For the calculation of the second term of the hamiltonian we have that
\[
\Big[ \hN[\ha^+_\alpha \ha^+_\beta \, \ha_{\beta'} \, \ha_{\alpha'}] , \ha^+_n \ha_j  \Big]
= 
\hN[\ha^+_\alpha \ha^+_\beta \, \ha_{\beta'} \, \ha_{\alpha'}]  \ha^+_n \ha_j 
-  \ha^+_n \ha_j  \hN[\ha^+_\alpha \ha^+_\beta \, \ha_{\beta'} \, \ha_{\alpha'}] 
,
\]
therefore
\beqn
&~& \nonumber
\braket{ \Phi_0 | \Big[
   \ha^+_i \ha_m , \hN[\ha^+_\alpha \ha^+_\beta \, \ha_{\beta'} \, \ha_{\alpha'}]  \ha^+_n \ha_j 
   \Big]
-  \Big[ \ha^+_n \ha_j  \hN[\ha^+_\alpha \ha^+_\beta \, \ha_{\beta'} \, \ha_{\alpha'}] ,   \ha^+_i \ha_m 
\Big] | \Phi_0}\\
&=&
\label{eq:rpa:con1}
\braket{ \Phi_0 | \ha^+_i \ha_m  \hN[\ha^+_\alpha \ha^+_\beta \, 
\ha_{\beta'} \, \ha_{\alpha'}]  \ha^+_n \ha_j | \Phi_0}\\
&-&
\label{eq:rpa:con2}
\braket{ \Phi_0 | \hN[\ha^+_\alpha \ha^+_\beta \, \ha_{\beta'} \, \ha_{\alpha'}]  
\ha^+_n \ha_j \ha^+_i \ha_m  | \Phi_0}\\
&+&
\label{eq:rpa:con3}
\braket{ \Phi_0 | \ha^+_i \ha_m \ha^+_n \ha_j  
\hN[\ha^+_\alpha \ha^+_\beta \, \ha_{\beta'} \, \ha_{\alpha'}] | \Phi_0}\\
&-&
\label{eq:rpa:con4}
\braket{ \Phi_0 |  \ha^+_n \ha_j  
\hN[\ha^+_\alpha \ha^+_\beta \, \ha_{\beta'} \, \ha_{\alpha'}] \ha^+_i \ha_m | \Phi_0}
.
\eeqn
The terms (\ref{eq:rpa:con2}) and (\ref{eq:rpa:con4}) are zero since $ a_m \ket{\Phi_0} =0 $.
The situation for the term (\ref{eq:rpa:con3}) is more involved. In the application of the Wick's
theorem one can see that in all the possible set of contractions there are always terms where
$\ha^+_n$ is contracted with $\ha_{\alpha'}$ or $\ha_{\beta'}$ and these contractions are zero.
Only the term (\ref{eq:rpa:con1}) is different from zero and, by applying the Wick's theorem 
we have to consider all the possible contractions and we obtain
\beqn
\nonumber
\braket{ \Phi_0 | a^+_i a_m  \hN[a^+_\alpha a^+_\beta \, a_{\beta'} \, a_{\alpha'}]  a^+_n a_j | \Phi_0}
&=& \delta_{i \alpha'} \delta_{m \alpha} \delta_{\beta' n} \delta_{\beta j} 
-  \delta_{i \alpha'} \delta_{m \beta} \delta_{\beta' n} \delta_{\alpha j} \\
&-&  \delta_{i \beta'} \delta_{m \alpha} \delta_{\alpha' n} \delta_{\beta j} 
+  \delta_{i \beta'} \delta_{m \beta} \delta_{\alpha' n} \delta_{\alpha j} 
.
\label{eq:app:em1}
\eeqn

This expression is used to obtain the TDA equation~(\ref{eq:rpa:tda}) 
whose terms are  equivalent to the 
$A_{minj}$ matrix elements (\ref{eq:rpa:AB1}) of RPA equation. 

For the term $B_{minj}$ of Equation (\ref{eq:rpa:AB1}) we use again the expression (\ref{eq:app.ham44})
of the hamiltonian. In this case also the contribution
of the one-body term is zero. By considering the anti-commutation properties
of the creation and destruction operators we obtain
\[
[\ha^+_\alpha \ha_\beta, \ha^+_j \ha_n] = 
\delta_{\beta j} \ha^+_\alpha \ha_n - \delta_{n \alpha}  \ha^+_j \ha_\beta 
,
\]
therefore
\beqn
\nonumber
&~& \braket{\Phi_0 | \Big[  \ha^+_i \ha_m  , \big[ \ha^+_\alpha \ha_\beta, \ha^+_j a_n \big]  \Big] | \Phi_0}  \\
\nonumber &=& \braket{\Phi_0 | \ha^+_i \ha_m  \ha^+_\alpha \ha_n  | \Phi_0} \rightarrow 0 \\
\nonumber &-& \braket{\Phi_0 | \ha^+_\alpha \ha_n \ha^+_i \ha_m | \Phi_0} \rightarrow 0 \\
\nonumber &-& \braket{\Phi_0 | \ha^+_i \ha_m  \ha^+_j \ha_\beta  | \Phi_0} = \delta_{j \beta} \delta_{im} \rightarrow 0 \\
\nonumber &+& \braket{\Phi_0 | \ha^+_j \ha_\beta \ha^+_i \ha_m | \Phi_0} \rightarrow 0
.
\eeqn
For the two-body term we have to evaluate
\[
\braket{\Phi_0 | \Big[  a^+_i a_m  , \big[ \hN[a^+_\alpha a^+_\beta a_{\beta'} a_{\alpha'}, a^+_j a_n \big]  \Big] | \Phi_0} 
.
\]
Three terms of the double commutators are zero since they contain $a_m \ket{\Phi_0}=0$. 
Only the term
\[
- \braket{\Phi_0 | \ha^+_i \ha_m  \ha^+_j \ha_n \ha^+_\alpha \ha^+_\beta \ha_{\beta'} \ha_{\alpha'} | \Phi_0} 
,
\]
is different from zero, therefore
\beq
B_{minj} = \frac 1 4 \sum_{\alpha \beta \alpha' \beta'} \barv_{\alpha \beta \alpha' \beta'} 
\braket{\Phi_0 | \ha^+_i \ha_m  \ha^+_j \ha_n \ha^+_\alpha \ha^+_\beta \ha_{\beta'} \ha_{\alpha'} | \Phi_0} 
\;.
\eeq
By considering the symmetry properties of  $\barv$ and all the possible contractions we obtain 
Equation~(\ref{eq:rpa:bmnij}).
%

\section{Sum Rules}
\label{sec:app.sr}

We derive here the expression of the sum rule (\ref{eq:rpa:gensumrule}).

\beqn
\nonumber  
\braket{\Psi_0 | \Big[ \hF, [ \hH, \hF] \Big] | \Psi_0} 
&=& \braket{\Psi_0 | \Big[ \hF \hH \hF - \hF \hF \hH - \hH \hF \hF + \hF \hH \hF)  \Big] | \Psi_0} 
\\ &=& \nonumber
\Big[ 2 \braket{\Psi_0 | \hF \hH \hF | \Psi_0} 
- \braket{\Psi_0 | \hF \hF | \Psi_0} E_0  - E_0 \braket{\Psi_0 | \hF \hF | \Psi_0}   \Big] 
\\ &=& \nonumber
2 \braket{\Psi_0 | \hF (\hH - E_0) \hF | \Psi_0} 
.
\eeqn
We insert the completeness $ \sum_\nu \ket{ \Psi_\nu}  \bra{\Psi_\nu} = {\mathbb I}  $
\beqn
&~& \nonumber
2 \braket{\Psi_0 | \hF \sum_\nu | \Psi_\nu} \braket{\Psi_\nu | (\hH - E_0) \hF | \Psi_0} 
\\ &=& \nonumber
2 \braket{\Psi_0 | \hF \sum_\nu | \Psi_\nu} \braket{\Psi_\nu | (E_\nu - E_0) \hF | \Psi_0} 
= 2  \sum_\nu (E_\nu - E_0) \braket{\Psi_0 | \hF | \Psi_\nu} 
\braket{\Psi_\nu |  \hF | \Psi_0} 
.
\eeqn
%

\section{Linear Response} 
\label{sec:rl}
\index{Response, linear}
Let us consider the situation where the many-body system is subject
to an external perturbation. We express the total hamiltonian describing 
the perturbed system as sum of the hamiltonian $\hH$ describing the system in absence 
of the perturbation, whose eigenstates are $| \Psi \rangle$, plus a time-dependent
term $\hH^{\rm ext}(t)$:
\beq
\hH^{\rm tot} = \hH + \hH^{\rm ext}(t) = \hH + \hF A(t)
\,\,\,,
\eeq
where $\hF$ is the operator describing the action of the external 
perturbation on the system. The function
$A(t)$ describes the time evolution of the perturbation 
and is defined such as $A(t) =0$ for $t < t_0 =0 $. 
This means that the perturbation is switched on after a specific
time, $t_0$ which we define as zero time.

We assume that, under the action of the external perturbation, 
the reaction times of the many-body system are
much faster than those needed to the perturbation to switch on and off. Then, 
when the perturbation is totally switched on the hamiltonian is 
$\hH^{\rm tot} = \hH + \hF$. In this case we can treat  $\hF$ as a 
perturbative term of the total time-dependent hamiltonian. For this
reason, we can consider the equation of 
motion (\ref{eq:p.timesint}) in the interaction picture 
\beq
i \hbar \frac{\partial}{\partial t}  | \Psi_{\rm I}(t) \rangle = \hF_{\rm I}(t)  | \Psi_{\rm I} (t) \rangle 
\,\,\,,
\label{eq:gris1}
\eeq
where
\beq
 \hF_{\rm I}(t) = e^{\ih \hH t} \hF  e^{-\ih \hH t} \;\;\; {\rm e } \;\;\; 
  | \Psi_{\rm I}(t) \rangle =  e^{\ih \hH t}  | \Psi(t) \rangle 
\,\,\,.
\eeq
In this section, we use the convention that states and operators without
sub-indexes are expressed in the Schr\"odinger picture. 
We formally integrate Equation (\ref{eq:gris1}) 
\beq
i \hbar \int_{-\infty}^t dt' \,\frac{\partial}{\partial t'}  | \Psi_{\rm I}(t') \rangle =
\int_{-\infty}^t dt' \, \hF_{\rm I}(t')  | \Psi_{\rm I} (t') \rangle 
,
\eeq
and obtain the expression 
\beq
 | \Psi_{\rm I}(t) \rangle =  | \Psi_{\rm I}(-\infty) \rangle 
 -\ih \int_{-\infty}^t dt' \, \hF_{\rm I}(t')  | \Psi_{\rm I} (t') \rangle 
\,\,\,.
\eeq
Since the perturbation is switched off when $t=-\infty$ we have that
$ | \Psi_{\rm I}(-\infty) \rangle = | \Psi_0 \rangle $ which is the ground
state of the system.
We can express the above equation as perturbative expansion
by iterating $| \Psi_{\rm I}(t) \rangle $
\beq
 | \Psi_{\rm I}(t) \rangle =  | \Psi_0 \rangle 
 - \ih \int_{-\infty}^t dt' \, \hF_{\rm I}(t')  | \Psi_0 \rangle \;\; +  \;\; \cdots 
\eeq

We call ${\hat D}$ the operator which describe how the system reacts to 
the external perturbation induced by the operator $\hF$. The expectation
value of this operator is given by
%
\beqn
\nonumber &~&
\langle \Psi_{\rm I}(t) | {\hat D}_{\rm I}(t)  | \Psi_{\rm I}(t) \rangle \\
\nonumber
&=&
\left\{  \langle \Psi_0 |
 + \ih \int_{-\infty}^t dt' \, \hF_{\rm I}(t')  \langle \Psi_0 | \;\; + \;\; \cdots
 \right\}  {\hat D}_{\rm I}(t)  
 \left\{
  | \Psi_0 \rangle 
 - \ih \int_{-\infty}^t dt' \, \hF_{\rm I}(t')  | \Psi_0 \rangle \;\; + \;\; \cdots
 \right\} \\
 &=&
 \langle \Psi_0 |  {\hat D}_{\rm I}(t)   | \Psi_0 \rangle 
 + \ih \int_{-\infty}^t dt'  \langle \Psi_0 |  [ \hF_{\rm I}(t'),{\hat D}_{\rm I}(t)]   | \Psi_0 \rangle 
 \;\;\;+ \cdots
\eeqn
We define the response function as
\beq
R(t'-t) = 
\left\{  \begin{array}{ll}
              0      &   \; \; t '< t \\
              \displaystyle
         \ih \frac{ \langle \Psi_0 |  [ \hF_{\rm I}(t'), {\hat D}_{\rm I}(t)]   | \Psi_0 \rangle } 
         {\braket{\Psi_0 | \Psi_0}}  &   
           \;\; t' > t 
       \end{array} 
\right.
\,\,\,.
\eeq
This definition implies causality. The system cannot respond before that the 
perturbation is switched on.

By making explicit the time dependence of $\hF_{\rm I}(t')$ 
and ${\hat D}_{\rm I}(t)$, 
\beq
\hF_{\rm I}(t') = e^{\ih \hH t'} \hF e^{- \ih \hH t'} 
\,\,\,\,;\,\,\,\,
{\hat D}_{\rm I}(t) = e^{\ih \hH t} {\hat D} e^{- \ih \hH t} 
\,\,\,,
\eeq
we can express the response as
\beq
R(t'-t) = \ih \frac{\langle \Psi_0 | \hB e^{\ih (\hH - E_0)(t-t')} {\hat D}   | \Psi_0 \rangle}
  {\braket{\Psi_0 | \Psi_0}} 
- \ih \frac{\langle \Psi_0 | {\hat D} e^{-\ih (\hH - E_0)(t-t')} {\hat B}   | \Psi_0 \rangle }
  {\braket{\Psi_0 | \Psi_0}} 
\,\,\,,
\eeq
and, since it depends only on the time difference $\tau=t-t'$,
by using the definition of Fourier transform, we obtain
\beqn
\nonumber
&~& {\tilde R}(E) = \int_{-\infty}^\infty d\tau \, R(\tau) \, e^{\ih E \tau}\\
\nonumber
&=& \ih \left(
\langle \Psi_0 | \hF \int_{-\infty}^\infty d\tau e^{\ih (\hH - E_0 + E)\tau} {\hat D}   | \Psi_0 \rangle 
- \ih \langle \Psi_0 | {\hat D} \int_{-\infty}^\infty d\tau e^{-\ih (\hH - E_0- E)\tau} \hF   | \Psi_0 \rangle \right) 
\frac{1}   {\braket{\Psi_0 | \Psi_0}} \\
&=& 
 - \frac{\langle \Psi_0 | \hF (\hH - E_0 +E + i \eta)^{-1} {\hat D}   | \Psi_0 \rangle}  {\braket{\Psi_0 | \Psi_0}}  
 - \frac{\langle \Psi_0 | {\hat D} (\hH - E_0 - E - i \eta)^{-1} \hF   | \Psi_0 \rangle}  {\braket{\Psi_0 | \Psi_0}} 
 \,\,\,.
\eeqn
We insert the completeness $ \sum_n |\Psi_n \rangle \langle \Psi_n | =1$
and obtain
\beq
{\tilde R}(E) = \sum_n
\left[
\frac{\langle \Psi_0  | {\hat D} |\Psi_n \rangle \langle \Psi_n | \hF  | \Psi_0 \rangle } 
{E - (E_n -E_0) + i\eta} 
- 
\frac{\langle \Psi_0  | \hF |\Psi_n \rangle \langle \Psi_n | {\hat D} | \Psi_0 \rangle } 
{E + (E_n -E_0) + i\eta} 
\right] \frac{1}   {\braket{\Psi_0 | \Psi_0}}
\,\,\,.
\label{eq:gres2}
\eeq
The poles of $\tilde R(E)$ correspond to the excitation energies of the system.
For each positive pole there is a negative pole, equal in absolute value
to the positive one.

We consider the  Dirac expression
\beq
\frac{1}{x'-x \pm i \eta} = {\cal P} \frac{1}{x'-x} \mp i \pi \delta(x-x') 
\,\,\,,
\eeq
where ${\cal P}$ indicates the principal part, therefore
\beq
\delta(x-x') = - \frac{1}{\pi} \Im \left( \frac{1}{x'-x \pm i \eta} \right)
\,\,\,,
\eeq
with the symbol $\Im$ indicating the imaginary part.
 
We assume ${\hat D}=\hF$, as it usually happens and consider
only positive energies. The transition probability from the ground state to 
an excited state is given by
\beq
S(E) =  - \frac{1}{\pi} \Im \big( R(E) \big) = 
\sum_n | \langle \Psi_0  | \hF |\Psi_n \rangle |^2 \delta\big(E - (E_n - E_0)\big)
\,\,\,.
\eeq
This is the traditional expression obtained by applying the time-dependent
perturbation theory \cite{mes61}.
Assuming that  $\hF$ is a one-body operator 
\beq
\hF= \sum_{\nu_1 \nu_2} f_{\nu_1 \nu_2} \ha_{\nu_1} \ha^+_{\nu_2}
\;\; {\rm and}\;\;
f_{\nu_1 \nu_2} = \int d^3r\, \phi^*_{\nu_1}(\br) \, f(\br) \, \phi_{\nu_2}(\br) 
\,\,\,,
\eeq
we obtain
\beqn
\nonumber
{\tilde R}(E) = \sum_{\nu_1 \nu_2} \sum_{\nu_3 \nu_4} \sum_n
&~&
\Big[ f_{\nu_1 \nu_2} f^*_{\nu_3 \nu_4}
\frac{\langle \Psi_0  | \ha_{\nu_1} \ha^+_{\nu_2} |\Psi_n \rangle 
        \langle \Psi_n |  \ha_{\nu_3} \ha^+_{\nu_4}  | \Psi_0 \rangle } 
{E - (E_n -E_0) + i\eta} \\
&-&  f_{\nu_3 \nu_4} f^*_{\nu_1 \nu_2}
\frac{\langle \Psi_0  |  \ha_{\nu_3} \ha^+_{\nu_4}  |\Psi_n \rangle 
        \langle \Psi_n |  \ha_{\nu_1} \ha^+_{\nu_2} | \Psi_0 \rangle } 
{E + (E_n -E_0) + i\eta} 
\Big] \frac{1}   {\braket{\Psi_0 | \Psi_0}}
\,\,\,,
\eeqn
Since $\hF$ is hermitian, $f_{\nu_1 \nu_2} = f^*_{\nu_2 \nu_1}$
and the indexes $\nu$ are dummy, we can write
\beqn
\nonumber
{\tilde R}(E) &=& \sum_{\nu_1 \nu_2} \sum_{\nu_3 \nu_4} 
 f_{\nu_1 \nu_2} f^*_{\nu_3 \nu_4} \\
\nonumber
&~& 
\sum_n
\Big[
\frac{\langle \Psi_0  | \ha_{\nu_1} \ha^+_{\nu_2} |\Psi_n \rangle 
        \langle \Psi_n | \ha_{\nu_3} \ha^+_{\nu_4}  | \Psi_0 \rangle } 
{E - (E_n -E_0) + i\eta} -
\frac{\langle \Psi_0  |  \ha_{\nu_3} \ha^+_{\nu_4}  |\Psi_n \rangle 
        \langle \Psi_n |  \ha_{\nu_1} \ha^+_{\nu_2} | \Psi_0 \rangle } 
{E + (E_n -E_0) + i\eta} 
\Big] \frac{1}   {\braket{\Psi_0 | \Psi_0}} \\
 &=& \sum_{\nu_1 \nu_2} \sum_{\nu_3 \nu_4} 
 f_{\nu_1 \nu_2} f^*_{\nu_3 \nu_4} (-i) {\tilde G}(\nu_1,\nu_3, \nu_2, \nu_4,E) 
\,\,\,,
\eeqn
where, in the last step, we considered the expression (\ref{eq:gtbgfl}) 
of the two-body Green function. The transition probability is given by 
\beq
S(E) =  - \frac{1}{\pi} \Im \big( R(E) \big) = 
 \sum_{\nu_1 \nu_2} \sum_{\nu_3 \nu_4}  f_{\nu_1 \nu_2} f^*_{\nu_3 \nu_4}
\frac {\Im}{\pi} \left(   i \hbar {\tilde G}(\nu_1,\nu_3, \nu_2, \nu_4,E) \right)
\,\,\,.
\eeq

\section{Green Function Expansion Terms}
\label{sec:app.gfet}
In this appendix we show the explicit expressions of the diagrams of Figure~\ref{fig:gfed}. 
The $\rx$ and $\ry$ labels
indicate both space and time coordinates. The integration on all the $\ry$ coordinates is  understood.
The symbol $\intr$ indicates the two-body interaction. 
\beq
A \equiv \rG^0(\rx_1,\rx_2,\rx_3,\rx_4)
,
\eeq
\beq
B \equiv \rG^0(\rx_1,\rx_2,\ry_1,\ry_1) \intr (\ry_1,\ry_2) \rG^0(\ry_2,\ry_2,\rx_3,\rx_4) 
,
\eeq
\beq
C \equiv \rG^0(\rx_1,\rx_2,\ry_1,\ry_2) \intr (\ry_1,\ry_2) \rG^0(\ry_1,\ry_2,\rx_3,\rx_4)
,
\eeq
\beq
D \equiv \rG^0(\rx_1,\rx_2,\ry_1,\ry_1) \intr (\ry_1,\ry_2) \rG^0(\ry_2,\ry_2,\ry_3,\ry_3)
\intr (\ry_3,\ry_4) \rG^0(\ry_4,\ry_4,\rx_3,\rx_4)
,
\eeq
\beq
E \equiv \rG^0(\rx_1,\rx_2,\ry_1,\ry_2) \intr (\ry_1,\ry_2) \rG^0(\ry_1,\ry_2,\ry_3,\ry_4)
\intr (\ry_3,\ry_4) \rG^0(\ry_3,\ry_4,\rx_3,\rx_4)
,
\eeq
\beq
F \equiv \rG^0(\rx_1,\rx_2,\ry_1,\ry_3) \intr (\ry_1,\ry_2) \rG^0(\ry_2,\ry_2,\ry_3,\ry_2)
\intr (\ry_3,\ry_4) \rG^0(\ry_1,\ry_4,\rx_3,\rx_4)
,
\eeq
\beqn
\nonumber
G &\equiv& \rG^0(\rx_1,\rx_2,\ry_1,\ry_3) \intr (\ry_1,\ry_2) \rG^0(\ry_2,\ry_2,\ry_4,\ry_4)
\intr (\ry_3,\ry_4) \rG^0(\ry_1,\ry_3,\ry_5,\ry_6) \\
&~&
\intr (\ry_5,\ry_6) \rG^0(\ry_5,\ry_6,\rx_3,\rx_4) 
.
\eeqn

\section{RPA Green Function in Matrix Form}
\label{sec:app.gfmat}

We consider Equation~(\ref{eq:RPA.grpa}) and we calculate first
\beqn
\nonumber &~& 
\hbar {\tilde G}^{\rm RPA}(m, i, j, n ,E) 
\\ \nonumber
&=&  
\sum_{\mu_1, \mu_2 ,\mu_3,\mu_4} {\tilde G}^0(m, i, \mu_1, \mu_2,E) 
\Big\{ \delta_{\mu_1, j} \delta_{\mu_2, n}
+ \braket{\mu_1 \mu_3 | \hV | \mu_2 \mu_4} {\tilde G}^{\rm RPA}(\mu_3, \mu_4, j, n ,E) 
\\ \nonumber
&-& \braket{\mu_1 \mu_2 | \hV | \mu_4 \mu_3} {\tilde G}^{\rm RPA}(\mu_3, \mu_4, j , n , E)
\Big\} 
.
\eeqn

Because of Equations~(\ref{eq:green.gnot1}) and (\ref{eq:green.gnot2}) we have that

\beqn
\nonumber &~& 
\hbar {\tilde G}^{\rm RPA}(m, i, j, n ,E) = \frac{ \delta_{i, j} \delta_{m, n}}{\epsilon_m - \epsilon_i -E} \Big\{ 1 + 
\\ \nonumber
&+&  
\sum_{\mu_3,\mu_4} \Big[
+ \braket{i \mu_3 | \hV | m \mu_4} {\tilde G}^{\rm RPA}(\mu_3, \mu_4, j, n ,E) 
- \braket{i m | \hV | \mu_4 \mu_3} {\tilde G}^{\rm RPA}(\mu_3, \mu_4, j , n , E)
\Big]
\Big\} 
.
\eeqn

Making explicit the sum on $\mu_3$ and $\mu_4$ and considering that, for the 
conservation of the number of particles, one of the indexes must indicate a particle state
and the other one a hole state, we can rewrite the above expression as:
\vspace{-6pt}
\beqn
\nonumber 
\hbar (\epsilon_m - \epsilon_i -E) {\tilde G}^{\rm RPA}(m, i, j, n ,E) -
\sum_{lq} &\Big[& 
\braket{i l | \hV | m q} {\tilde G}^{\rm RPA}(l , q, j, n ,E)  
\\ \nonumber &+& \braket{i q | \hV | m l} {\tilde G}^{\rm RPA}(q , l, j, n ,E) 
\\ \nonumber &-&  \braket{i m | \hV | l q} {\tilde G}^{\rm RPA}(q , l, j, n ,E) 
\\ \nonumber &-&  \braket{i m | \hV | q l} {\tilde G}^{\rm RPA}(l , q, j, n ,E) 
\Big]
=  \delta_{i, j} \delta_{m, n}
.
\eeqn

By considering the antisymmetrized matrix element (\ref{eq:v.avu})
we can express the above equation as
\beqn
\nonumber 
\sum_{lq} \Big\{ \Big[
\hbar (\epsilon_m - \epsilon_i -E) \delta_{i, l} \delta_{m, q}
&+&  \barv_{iqml} {\tilde G}^{\rm RPA}(q , l, j, n ,E)  
\\ \nonumber &+&  \barv_{ilmq} {\tilde G}^{\rm RPA}(l , q, j, n ,E) 
\Big] =  \delta_{i, j} \delta_{m, n}
,
\eeqn
which is Equation~(\ref{eq:RPA.grpa1}). The evaluation of 
Equations~(\ref{eq:RPA.grpa2})--(\ref{eq:RPA.grpa4}) is carried out in 
analogous manner.

\section{Correlated TDHF}
\label{sec:app.tdhf}

In this section, we obtain the explicit expression of the $R_{mi}$ and $S_{mi}$ factors defined in Equation~(\ref{eq:ccrpa.rmi}) as 
\beq
R_{mi} \equiv
\frac{\delta \bra{\Psi(t)} } {\delta C^*_{mi}(t)} 
\left( \hH - i \hbar \frac{\partial}{\partial t} \right) \ket{ \Psi(t)}
,
\eeq
and 
\beq
S_{mi} \equiv
\frac{\delta \bra{\Psi(t)} } {\delta C_{mi}(t)} 
\left( \hH - i \hbar \frac{\partial}{\partial t} \right) \ket{ \Psi(t)}
.
\eeq

To simplify the writing we define
\beq
\deno = \braket{\Phi(t)| F^+ F | \Phi(t)}
,
\eeq
By considering Equation~(\ref{eq:ccrpa.phit}) for $\ket{\Phi(t)}$ we have
\beq
\frac {\delta }{\delta C^*_{mi}(t)} \deno^\half = \half \frac{\braket{\Phi(t) |\ha^+_i \ha_m  F^+ F | \Phi(t)} } {\deno^\half}  
,
\eeq
and 
\beqn
\nonumber
\frac {\partial }{\partial t} \deno  &=& 
\sum_{nj} \frac {d }{d t} C^*_{nj}(t) \braket{\Phi(t) |\ha^+_j \ha_n  F^+ F | \Phi(t)} \\
&+&
\sum_{nj} \frac {d }{d t} C_{nj}(t) \braket{\Phi(t) | F^+ F \ha^+_n \ha_j | \Phi(t)} 
.
\eeqn
From the time dependence (\ref{eq:ccrpa.phit}) and (\ref{eq:ccrpa.phi0t}) of $\ket{\Phi(t)}$
we have
\beq
i \hbar \frac {\partial }{\partial t} \ket{\Phi(t)} =
H_{00} \ket{\Phi(t)} + i \hbar \sum_{nj} \frac {d }{d t} C_{nj}(t) \ha^+_n \ha_j \ket{\Phi(t)} 
.
\eeq

We use the above expressions and Equation~(\ref{eq:ccrpa.psit}) of
$\ket{\Psi(t)}$ and obtain the following expressions
\beqn
\frac {\delta \bra{\Psi(t)}} {\delta C^*_{mi}(t)}
&=& 
\frac{\bra{\Phi(t)} \ha^+_i \ha_m F^+}{\deno^\half}
- \half \frac {\braket{\Phi(t) |\ha^+_i \ha_m  F^+ F | \Phi(t)} } { \deno } 
\bra{\Psi(t)}
, \\
\label{eq:app.tdhf1}
\frac {\delta \bra{\Psi(t)}} {\delta C_{mi}(t)}
&= &
- \half \frac {\braket{\Phi(t) | F^+ F \ha^+_m \ha_i | \Phi(t)} } { \deno } 
\bra{\Psi(t)}
,
\label{eq:app.tdhf1}
\eeqn
\beq
\left( \hH - i \hbar \frac{\partial}{\partial t} \right) \ket{ \Psi(t)}  =
\frac {(\hH - i \hbar \frac{\partial}{\partial t}) F \ket{\Phi(t)}} { \deno^\half } 
+ \frac {i \hbar}{2} 
\frac{F \ket{\Phi(t)}}{\deno^{3/2} }
\frac{\partial}{\partial t} \deno
.
\label{eq:app.tdhf2}
\eeq

Putting together the above equations, we obtain
\beqn
\nonumber
0 = R_{mi}  &=& \frac {1} {\deno} \braket{\Phi(t) | \ha^+_i \ha_m F^+ \hH F | \Phi(t)}
\\ \nonumber &-& 
\half \frac {1} {\deno} \braket{\Phi(t) | \ha^+_i \ha_m F^+ F | \Phi(t)} H_{00}
\\ \nonumber &-& 
\half \frac{1}{\deno^2} \braket{\Phi(t) | \ha^+_i \ha_m F^+ F | \Phi(t)} \braket{\Phi(t) |F^+ \hH F | \Phi(t)}
\\ \nonumber &-& 
 \frac{i \hbar}{\deno} \sum_{nj} \frac {d}{dt} C_{nj}(t) 
\braket{\Phi(t) | \ha^+_i \ha_m F^+ F \ha^+_n \ha_j | \Phi(t)} 
\\ \nonumber &+& 
\frac 3 4 \frac{i \hbar}{\deno^2} \sum_{nj} \frac {d}{dt} C_{nj}(t) 
\braket{\Phi(t) | \ha^+_i \ha_m F^+ F | \Phi(t)} \braket{\Phi(t) | F^+  F  \ha^+_n \ha_j | \Phi(t)} 
\\ &+& 
\frac 1 4 \frac{i \hbar}{\deno^2} \sum_{nj} \frac {d}{dt} C^*_{nj}(t) 
\braket{\Phi(t) | \ha^+_i \ha_m F^+ F | \Phi(t)} \braket{\Phi(t) | \ha^+_j \ha_n F^+  F | \Phi(t)}                         
,
\label{eq:app.rmi2}
\eeqn 
and
\beqn
\nonumber
0 = -2 S_{mi}  &=& \frac {1} {\deno^2} 
\braket{\Phi(t) | F^+ F \ha^+_m \ha_i | \Phi(t)}
\braket{\Phi(t) | F^+ \hH F | \Phi(t)}
\\ \nonumber &-& 
\frac {1} {\deno} \braket{\Phi(t) | F^+ F  \ha^+_m \ha_i | \Phi(t)} H_{00}
\\ \nonumber &-& 
 \frac{i \hbar}{\deno^2}  \braket{\Phi(t) | F^+ F  \ha^+_m \ha_i | \Phi(t)} \sum_{nj} \frac {d}{dt} C_{nj}(t) 
\braket{\Phi(t) | F^+ F \ha^+_n \ha_j | \Phi(t)} 
\\ \nonumber &+& 
 \frac{i \hbar}{2} \frac{1} {\deno^2} \braket{\Phi(t) | F^+ F  \ha^+_m \ha_i | \Phi(t)} 
\\ \nonumber &\;\;\;& 
\Big[ \sum_{nj} \frac {d}{dt} C^*_{nj}(t) \braket{\Phi(t) | \ha^+_j \ha_n F^+  F | \Phi(t)} 
 \\ &\;\;\;&
+ \sum_{nj} \frac {d}{dt} C_{nj}(t)  \braket{\Phi(t) |  F^+  F \ha^+_n \ha_j | \Phi(t)}   
\Big]                      
.
\label{eq:app.smi2}
\eeqn  

We carry out a power expansion of Equation~(\ref{eq:ccrpa.phit}) 
\beq
\bra{\Phi(t)} = \bra{\Phi_0(t)} \left[ 1 + \sum_{mi} C^*_{mi}(t) \ha^+_i \ha_m + \cdots \right] 
,
\label{eq:app.braexp}
\eeq
and
\beq
\ket{\Phi(t)} = \left[ 1 + \sum_{mi} C_{mi}(t) \ha^+_m \ha_i + \cdots \right] \ket{\Phi_0(t)} 
.
\label{eq:app.ketexp}
\eeq

Since $H_{00}$ is a number, the expectation values of an operator between $\ket{\Phi_0(t)}$ 
states are identical to those obtained between those calculated between time-independent
$\ket{\Phi_0}$ states. This means that the expression (\ref{eq:ccrpa.brillouin}) is 
also valid in the form
\beq
\frac {\braket{\Phi_0 (t) | \ha^+_i \ha_m F^+ \hH F | \Phi_0(t) }} {\braket{\Phi_0(t) | F^+ F | \Phi_0(t) }}
= H_{00} \frac {\braket{\Phi_0(t) | \ha^+_i \ha_m F^+ F | \Phi_0(t) }}  
                      {\braket{\Phi_0(t) | F^+ F | \Phi_ 0(t)}}
.
\label{eq:app.brillouin}
\eeq

We use the approximation
\beq
\frac {\braket{\Phi (t) | F^+ \hH F | \Phi(t) }} {\braket{\Phi(t) | F^+ F | \Phi(t) }} \simeq
\frac {\braket{\Phi_0(t) | F^+ \hH F | \Phi_0(t) }} {\braket{\Phi_0(t) | F^+ F | \Phi_0(t) }} 
= H_{00} 
.
\label{eq:app.approximation}
\eeq
This expression has been obtained by using the expansions (\ref{eq:app.braexp}) 
and (\ref{eq:app.ketexp}) in both numerator and denominator and by neglecting all the terms
containing $C$'s. 

By using the expansions (\ref{eq:app.braexp}) and (\ref{eq:app.ketexp}) and the
approximation (\ref{eq:app.approximation}) in \mbox{Equation (\ref{eq:app.smi2})} we obtain
the relation
\beqn
\nonumber
0 &=& \sum_{nj} \frac {d}{dt} C_{nj}(t) 
\frac {\braket{\Phi_0 | F^+ F \ha^+_n \ha_j | \Phi_0 } } {\braket{\Phi_0| F^+ F | \Phi_0 }}               
\\ &-&
\sum_{nj} \frac {d}{dt} C^*_{nj}(t) 
\frac {\braket{\Phi_0 | \ha^+_j \ha_n F^+ F  | \Phi_0 } } {\braket{\Phi_0 | F^+ F | \Phi_0 }}
.
\label{eq:app.relation}
\eeqn

We consider the expansions (\ref{eq:app.braexp}) and (\ref{eq:app.ketexp}) and the
approximation (\ref{eq:app.approximation}) in Equation (\ref{eq:app.rmi2}) and obtain
\beqn
\nonumber
&~& R_{mi}  = 
\\ \nonumber &~& 
\frac {1} {\deno} 
\bra{\Phi_0(t)} \left[ 1 + \sum_{nj} C^*_{nj}(t) \ha^+_j \ha_n + \cdots \right] 
\ha^+_i \ha_m F^+ \hH F
\left[ 1 + \sum_{nj} C^*_{mi}(t) \ha^+_n \ha_j + \cdots \right] \ket{\Phi_0(t)} 
\\ \nonumber &-& 
\half \frac {1} {\deno} 
\bra{\Phi_0(t)} \left[ 1 + \sum_{nj} C^*_{nj}(t) \ha^+_j \ha_n + \cdots \right] 
\ha^+_i \ha_m F^+ F
\left[ 1 + \sum_{mi} C^*_{nj}(t) \ha^+_n \ha_j + \cdots \right] \ket{\Phi_0(t)} 
\\ \nonumber &~& \;\;\;\;\;\;\;\;\;\;\;\;\;\;\;\;\;\; 
\left[ H_{00} + \frac {\braket{\Phi (t) | F^+ \hH F | \Phi(t) }} {\braket{\Phi(t) | F^+ F | \Phi(t) }} 
\right] 
\\  &+& {\cal K}\left[ \frac {d C} {d t} \right] 
,
\label{eq:app.rmi3}
\eeqn 
where we have indicated with ${\cal K}$ the terms depending on the derivative of $C$'s. 
We retain only the terms containing a single $C$, we use 
the approximation (\ref{eq:app.approximation}) and obtain

\beqn
\nonumber
R_{mi}  &=& 
\frac {1} {\deno} 
\bra{\Phi_0(t)} \ha^+_i \ha_m F^+ \hH F \ket{\Phi_0(t)} 
\\ \nonumber &+& 
\frac {1} {\deno} \sum_{nj} C^*_{nj}(t) 
\bra{\Phi_0(t)} \ha^+_j \ha_n \ha^+_i \ha_m F^+ \hH F \ket{\Phi_0(t)} 
\\ \nonumber &+& 
\frac {1} {\deno} \sum_{nj} C_{nj}(t) 
\bra{\Phi_0(t)} \ha^+_i \ha_m F^+ \hH F \ha^+_n \ha_j \ket{\Phi_0(t)} 
\\ \nonumber &-& 
\frac {1} {\deno}  \bra{\Phi_0(t)} \ha^+_i \ha_m F^+ F \ket{\Phi_0(t)} H_{00}
\\ \nonumber &-& 
\frac {1} {\deno}  \sum_{nj} C^*_{nj}(t)
\bra{\Phi_0(t)}  \ha^+_j \ha_n \ha^+_i \ha_m F^+ F \ket{\Phi_0(t)} H_{00}
\\ \nonumber &-& 
\frac {1} {\deno} \sum_{nj} C_{nj}(t) 
\bra{\Phi_0(t)}  \ha^+_i \ha_m F^+ F \ha^+_n \ha_j \ket{\Phi_0(t)} H_{00}
\\  &+& {\cal K}\left[ \frac {d C} {d t} \right] 
.
\label{eq:app.rmi4}
\eeqn 

By applying the condition (\ref{eq:app.brillouin}) the terms without $C$'s cancels 
and we obtain
\beq
R_{mi} = \sum_{nj} A_{minj} + \sum_{nj} B_{minj} +  {\cal K}\left[ \frac {d C} {d t} \right] =0
.
\eeq
where we used the definitions (\ref{eq:ccrpa:aminj}) and (\ref{eq:ccrpa:bminj}).

We use the relation (\ref{eq:app.relation}) and
we can write the term dependent on the derivative
of the $C$'s as:
\beqn
\nonumber
{\cal K}\left[ \frac {d C} {d t} \right] &=&
-i \hbar \sum_{nj} \sum_{nj} \frac { d C_{nj}(t) }{d t}  
\Bigg[ \frac{\bra{\Phi_0(t)}  \ha^+_i \ha_m F^+ F \ha^+_n \ha_j \ket{\Phi_0(t)} }
{\bra{\Phi_0(t)} F^+ F \ket{\Phi_0(t)} }
\\ \nonumber &-&
\frac{
\bra{\Phi_0(t)}  \ha^+_i \ha_m F^+ F  \ket{\Phi_0(t)} 
\bra{\Phi_0(t)}  F^+ F \ha^+_n \ha_j \ket{\Phi_0(t)} 
}
{\bra{\Phi_0(t)} F^+ F \ket{\Phi_0(t)}^2 }
\Bigg] 
\\ & \equiv & 
-i \hbar \sum_{nj} \sum_{nj} \frac { d C_{nj}(t) }{d t}  M_{minj}
.
\eeqn

\end{appendices}

\begin{table}[h]
\begin{center}
\caption{ABBREVIATIONS}
\begin{tabular}{l l}
CCRPA & Core coupling RPA \\
DFT & Density Functional Theory \\
EOM & Equation of Motion \\
HF & Hartree--Fock \\
$hp$ & Hole--particle \\
IPM & Independent Particle Model \\
KS & Khon-Sham \\
MF & Mean-Field \\
ONR & Occupation Number Representation \\
$ph$ & Particle--hole \\
PVCRPA & Particle-vibration coupling RPA \\
QRPA & Quasi-particle RPA \\
RPA & Random Phase Approximation \\
r-RPA & Renormalized RPA \\
s.p. & Single particle \\
SRPA & Second RPA \\
TDHF & Time-dependent Hartree--Fock 
\end{tabular}
\end{center} 
\label{tab:acronyms}
\end{table}




\newpage \clearpage

\end{document}